\documentclass[ejs, preprint]{imsart}

\RequirePackage[numbers]{natbib}
\RequirePackage[colorlinks,citecolor=blue,urlcolor=blue]{hyperref}
\usepackage{graphicx}
\usepackage{amsmath, amsfonts, amssymb, amsthm}
\usepackage[english]{babel}
\usepackage{changepage}   
\usepackage{algorithm}
\usepackage[noend]{algpseudocode}
\usepackage{booktabs}

\startlocaldefs
\numberwithin{equation}{section}
\theoremstyle{plain}
\newtheorem{theorem}{Theorem}[section]
\newtheorem{lemma}[theorem]{Lemma}

\newtheorem{corr}[theorem]{Corollary}
\theoremstyle{definition}

\newtheorem{ass}{Assumption}

\newenvironment{remark}[1][Remark]{\begin{trivlist}
\item[\hskip \labelsep {\bfseries #1}]}{\end{trivlist}}
\newcommand{\bs}[1]{\boldsymbol{#1}}
\newcommand{\var}{\text{Var}}
\newcommand{\cov}{\text{Cov}}
\newcommand{\cum}{\text{cum}}
\newcommand{\te}[1]{\text{#1}}
\newcommand{\nn}{\nonumber \\}
\newcommand{\tr}{\text{tr}}
\endlocaldefs

\begin{document}

\begin{frontmatter}

\title{Functional mixed effects wavelet estimation for spectra of replicated time series}
\runtitle{Mixed effects spectral estimation}

\begin{aug}
\author{\fnms{Joris} \snm{Chau}\corref{J. Chau} \ead[label=e1]{j.chau@uclouvain.be}}
\and
\author{\fnms{Rainer} \snm{von Sachs}\ead[label=e2]{rvs@uclouvain.be}}
\address{Institut de statistique, biostatistique et sciences actuarielles\\ Universit\'e catholique de Louvain\\ Voie du Roman Pays, 20, B-1348 \\
Louvain-la-Neuve, Belgium \\ \printead{e1,e2}}
\runauthor{J. Chau and R. von Sachs}
\affiliation{Universit\'e catholique de Louvain}
\end{aug}

\begin{abstract}
Motivated by spectral analysis of replicated brain signal time series, we propose a functional mixed effects approach to model replicate-specific spectral densities as random curves varying about a deterministic population-mean spectrum. In contrast to existing work, we do not assume the replicate-specific spectral curves to be independent, i.e. there may exist explicit correlation between different replicates in the population. By projecting the replicate-specific curves onto an orthonormal wavelet basis, estimation and prediction is carried out under an equivalent linear mixed effects model in the wavelet coefficient domain. To cope with potentially very localized features of the spectral curves, we develop estimators and predictors based on a combination of generalized least squares estimation and nonlinear wavelet thresholding, including asymptotic confidence sets for the population-mean curve. We derive $L_2$-risk bounds for the nonlinear wavelet estimator of the population-mean curve --a result that reflects the influence of correlation between different curves in the replicate population-- and consistency of the estimators of the inter- and intra-curve correlation structure in an appropriate sparseness class of functions. To illustrate the proposed functional mixed effects model and our estimation and prediction procedures, we present several simulated time series data examples and we analyze a motivating brain signal dataset recorded during an associative learning experiment.
\end{abstract}



\end{frontmatter}

\section{Introduction}\label{sec:1}
Spectral analysis of replicated time series has recently gained growing interest, in particular in the field of brain data analysis, where it is common to collect time series data (such as EEG or local field potential data) from multiple subjects, or over multiple trials in an experiment, and the inferential focus is not on the mean responses of the time series but on the stochastic variation of the time series about their means. Other applications can be found, for instance, in biomedical experiments, geophysical and financial data analysis, or speech modeling. While there is an extensive literature on spectral analysis and inference of individual time series, this is not necessarily the case for replicated time series, and existing
approaches mostly work under simplifying assumptions such as independent or at least uncorrelated time series replications, which if not satisfied can lead to statistically inefficient estimators
or even give misleading inferences. \\
In this paper we address the specific problem of analyzing spectra of replicated time series showing potentially very localized features, allowing for explicit correlation between the time series replicates.  To illustrate, one can think of subject-replicated time series data collected from multiple subjects in an experiment with possible correlation between subjects due to unknown covariates (age, gender, etc.), or data collected over multiple trials of an experiment, where the spectral characteristics of the trial-replicated time series evolve over the course of the experiment. A particular motivating example for the latter is spectral analysis of brain data trial-replicated time series in the context of learning experiments, such a dataset is analyzed in Section \ref{sec:7}. As pointed out by \cite{M14} and \cite{FO16} there is a strong need to generalize existing approaches into this direction, however only few modifications to the assumption of independent time series replicates have been developed by now.\\
In the context of second-order spectral analysis for independent stationary replicated time series, \cite{DA97} introduced a log-linear mixed effects model, which was later generalized by \cite{HAS99} and \cite{IC01} by considering nonparametric estimation of the fixed effects curve. Also in the case of independent replicated time series, \cite{FOvS10} developed a tree-structed wavelet method for log-spectral estimation, whereas \cite{KHG11} introduced a more general mixed-effects approach based on spline smoothing of empirical log-spectra handling two-level nested designs with replicated time series for a number of independent subjects, here different time series replicates within a subject are allowed to be correlated based on known covariates. \cite{QGL09} considered a covariate-indexed functional fixed effects model for time-varying spectra of independent replicated nonstationary time series, and \cite{M13} applied the Bayesian wavelet-based mixed effects approach developed by \cite{MC06} to model time-varying spectra of replicated nonstationary time series, allowing for potential correlation between the time series replicates induced by the experimental design. More recently, in the context of learning experiments, \cite{FO16} model log-spectra of replicated nonstationary time series trials by including a replicate-time effect that evolves over the course of the experiment. In a general functional data analysis context, not aimed at spectral analysis of time series in particular, nonparametric functional mixed effects models have been considered among others by \cite{G02}, and \cite{QG06} using smoothing-spline approaches, and in \cite{A10} using functional principal components. In order to avoid the modelling of functional data by inherently smooth curves, wavelet-based approaches have been considered by \cite{MC06} and \cite{M08} using Bayesian wavelet shrinkage methods, by \cite{G13} using nonlinear wavelet thresholding, and by \cite{AS07} focusing on inference in a wavelet-based functional mixed effects model (see \cite{M14} for a comprehensive overview).\\
In this work we introduce an additive two-layer functional mixed effects model in the frequency domain for a collection $\{ X_s(t) \}_{s=1,\ldots,S}$ of $S$ individual time series, each with discrete observations over time. The time series replicates are modeled to have random replicate-specific log-spectra, which consist of a fixed effects curve on the first layer (population-average or -mean log-spectrum), additional to replicate-specific random effects curves on the second layer. We model explicit correlation between the random effects curves and do this in an appropriate way to allow for its fully nonparametric estimation, disposing of only a single realization for each of the $S$ time series replicates. As we observe only the noisy replicate-specific log-periodograms, we face a denoising problem of log-periodogram curves in the presence of potentially very localized structure for the underlying log-spectra, this problem is addressed by nonlinear wavelet thresholding. By projection onto an orthonormal wavelet basis, we obtain an equivalent finite-dimensional linear mixed effects model in the coefficient domain. This allows us to apply traditional linear mixed model estimation methods combined with nonlinear wavelet thresholding in a unified framework for both the fixed- and random effects empirical wavelet coefficients. To achieve simultaneous estimation of the fixed effects curve and the correlation structure between different random effects curves, we propose an easy-to-implement iterative generalized least squares estimation algorithm. We complete our methodology by proposing predictors of the individual replicate-specific log-spectra, as well as asymptotic confidence regions for the population-mean log-spectrum, which is interesting in its own as the literature on inference in the context of nonlinear wavelet thresholding estimators is relatively sparse.\\
The structure of the paper is as follows. In Section \ref{sec:2} we introduce the model set-up in both the frequency and wavelet coefficient domain with an appropriate combined $\ell_0$-sparseness constraint for the fixed- and random effects that allows for general inhomogeneous functional behavior over frequency. Some conditions on the variance-covariance structure of the random effects allow for its consistent estimation. In Section \ref{sec:3} we present estimators for the different components in the model, and we also propose predictors for the replicate-specific log-spectra. Section \ref{sec:4} provides consistency results for the estimators of the fixed effects curve and the variance-covariance-structure of the random effects curves, where we consider asymptotics in both the time series length $T$ and the replicate sample size $S$. In particular, we derive bounds on the $L_2$-risk of the nonlinear wavelet estimator of the fixed effects curve in an appropriate $\ell_0$-sparseness class, a result that reflects the influence of correlation between different curves in the replicate population. In Section \ref{sec:5} we derive asymptotic confidence regions for the population-mean log-spectrum based on the nonlinear wavelet estimator. Section \ref{sec:6} presents numerical results on the performance of the estimation and inference procedures for simulated time series data, and in Section \ref{sec:7} we analyze a motivating data example consisting of brain signal time series data recorded over the course of an associative learning experiment. The technical proofs are deferred to the Appendix section, which can be found in the supplementary material.

\section{Methodology} \label{sec:2}

\subsection{Model setup} \label{sec:2.1}
Let $\{ X_s(t) \}_{t > 0}$ be a collection of mean-zero second-order stationary univariate time series for replicates $s=1,\ldots,S$. We assume that the replicated time series are weakly dependent, as detailed in Section \ref{sec:2.1.1} below, in order to ensure that the power spectra are well-defined as the Fourier transform of the replicate-specific autocovariance functions. If we observe a collection of discretely sampled time series $\{ X_s(t),\ t = 1,\ldots 2T\}$, their raw bias-corrected log-periodograms at frequencies $\omega_\ell = \ell/(2T) \in [0,1)$ are computed as,
\begin{equation}\label{eq:1}
Y^f_s(\omega_\ell) = \log \frac{1}{2T}\left| \sum_{t=1}^{2T} X_{s}(t) \exp(-2\pi i \omega_\ell t) \right|^2 + \gamma
\end{equation}
where $\gamma \approx 0.577$ is a bias-correction equal to the Euler-Mascheroni constant (see \cite{W80}). For convenience, we consider the time series length to be dyadic $2T = 2^{J}$ in order to avoid additional complications in the subsequent wavelet estimation. We also note that it suffices to consider the log-periodograms only over the range of frequencies $\omega_\ell \in [0, 1/2)$, i.e. indices $\ell = 0, \ldots, T-1$, since the log-spectra are $[0,1]$-periodic and symmetric in $\omega_\ell = 1/2$.  

\subsubsection{Frequency domain functional mixed model}\label{sec:2.1.1}
We model the replicate-specific log-spectra as random curves varying about a deterministic population-mean log-spectrum, which is common to all replicates, see Figure \ref{fig:1} for a simulated example. Similar approaches are considered in \cite{DA97}, \cite{FOvS10}, and \cite{KHG11} to model the (log-)spectra of stationary replicated time series. We express the raw log-periodograms in terms of the following functional mixed effects model in the frequency domain:
\begin{eqnarray}
Y^f_s(\omega_\ell) &=& H^f_s(\omega_\ell) + E^f_{s}(\omega_\ell), \quad \quad \quad s=1,\ldots, S, \quad \ell = 0,\ldots, T-1 \nn
&=& h^f(\omega_\ell) + U^f_s(\omega_\ell) + E^f_{s}(\omega_\ell) \label{eq:2}
\end{eqnarray}
where,
\begin{enumerate}
\item $h^f \in L_2([0,1/2])$ is a population-mean log-spectrum (functional fixed effect). Hereafter,  $L_p(X)$ always denotes the $L_p$-space of measurable functions on $X$ with respect to the Lebesgue measure. 
\item $\{ U^f_s, s=1, \ldots, S \}$ are mean-zero random processes (functional random effects) with realizations in $L_2([0,1/2])$ for each replicate $s$. The distributional assumptions and variance-covariance structure of the functional random effects are detailed in Section \ref{sec:2.1.2} and Section \ref{sec:2.2} respectively.
\item $E^f_{s}(\omega_\ell) \overset{d}{\to} \log(\chi_2^2/2) + \gamma$ are asymptotically independent noise terms, with $\mathbb{E}[E^f_{s}(\omega_\ell)] = o_T(1)$ and $\var(E^f_{s}(\omega_\ell)) = \sigma_e^2 + o_T(1)$, where $\sigma_e^2 := \pi^2/6$ as shown in \cite{W80}. Note that for $\omega_0=0$, $E_{s}^f(\omega_0) \overset{d}{\to} \log(\chi^2_1) + \gamma$, but since the influence of this term is negligible for $T$ large enough, we consider the error term $E_{s}^f(\omega_0)$ to have the same asymptotic distribution as the other error terms in our subsequent analysis (as in \cite{M94}, \cite{KHG11},  and \cite{FOvS10}). The errors  $E_{s}^f(\omega_\ell)$ are assumed to be independent between different replicates and independent of the functional random effects $U_s^f$ for all $s, \ell$.
\end{enumerate}
Since our main interest lies in the analysis of the spectral characteristics of the replicated time series, we have introduced a functional mixed model on the level of the (log)-spectra in the frequency domain. It is nonetheless important to examine the implications of this model in the time domain, as the frequency domain model does not map to an additive functional mixed model for the replicated time series in the time domain. Each stationary mean-zero time series replicate $\{X_s(t)\}_{t>0}$ has a Cram\'er representation of the form: 
\begin{equation}
X_s(t) = \int_{0}^{1} A^f_s(\omega) \exp(2\pi i \omega t)\ dZ_s(\omega) \nonumber
\end{equation}
where the replicate-specific transfer functions $A^f_s(\omega)$ are $[0,1]$-periodic random Hermitian functions, i.e. $A^f_s(-\omega) = A^{f*}_s(\omega)$ (here $^*$ denotes the complex conjugate). The random processes $Z_s(\omega)$ are orthogonal increment processes that are independent between replicates and independent of the random transfer functions $A^f_s(\omega)$, such that:
\begin{equation}
\mathbb{E}\left[ dZ_s(\omega) dZ^*_s(\nu) \right] = \left\{ \begin{array}{ll}
1 & \te{if } \omega = \nu \\
0 & \te{if } \omega \neq \nu
\end{array} \right. \nonumber
\end{equation}
This is related to the stochastic transfer function models in \cite{KHG11} and \cite{K16b} for replicated time series organized in multiple groups or units, whereas in our case we dispose only of a single time series replicate per group. Conditional on the functional random effects $U^f_s(\omega) = u^f_s(\omega)$ in the frequency domain, for each $s=1,\ldots,S$, the time series replicate $\{X_s(t)\}_{t > 0}$ has a replicate-specific spectrum: 
\begin{eqnarray}
|A^f_s(\omega)|^2 &=& \exp\left(H^f_s(\omega) | u^f_s(\omega) \right) \nn
&=& \exp\left(h^f(\omega) \right) \exp\left( u^f_s(\omega) \right) \label{eq:3}
\end{eqnarray}
where the realized replicate-specific spectra are non-negative by construction. Furthermore, conditional on the random effects $U^f_s(\omega)$ in the frequency domain, we assume that the time series replicates are weakly dependent in the sense that $\sum_{h=-\infty}^\infty |\cov(X_s(t), X_s(t+h)| < \infty$ for all $s=1,\ldots, S$. This ensures that the realized replicate-specific spectra above are well-defined as the Fourier transforms of the realized replicate-specific autocovariance functions, and the reverse for the inverse Fourier transform. 

\subsubsection{Wavelet domain linear mixed model}\label{sec:2.1.2}
Since the realizations of the random replicate-specific log-spectra are $L_2([0,1])$-periodic functions, we consider a periodized orthonormal wavelet basis of $L_2([0,1])$, denoted by $\mathcal{B} = \{ \psi_k \}_{k=0}^\infty$, constructed from the translated and dilated versions of a sufficiently smooth father and mother wavelet function, compactly supported on $[0,1]$. Here, for ease of notation we compress the usual scale and location indices $(j,m)$ into a single scale-location index $k$, using classical lexicographical ordering. Since the log-periodograms $Y^f_s$ are sampled over a discrete grid of frequencies, instead of true wavelet coefficients (projections of the replicate-specific log-spectra), we compute the empirical wavelet coefficients:
\begin{equation}
Y_{sk} = \langle Y^f_s, \psi_k \rangle_T = \frac{1}{T} \sum_{\ell=0}^T Y^f_s(\omega_\ell)\ \psi_k(\omega_\ell) = \int Y^f_s(\omega)\ \psi_k(\omega)\ d\omega + o_T(1) \nonumber
\end{equation} 
More specifically, projecting the discrete sampled frequency domain model in eq.(\ref{eq:2}) onto the wavelet basis $\mathcal{B}$ via its discrete wavelet transform (we denote the discrete wavelet transform-matrix by $\bs{W}_\mathcal{B}$), we obtain a linear mixed model in the wavelet coefficient domain given by,
\begin{eqnarray}
Y_{sk} &=& H_{sk} + \epsilon_{sk}, \quad \quad s=1,\ldots, S, \quad k=1,\ldots,T \nn
&=& h_{k} + U_{sk} + \epsilon_{sk} \label{eq:4}
\end{eqnarray}
where,
\begin{enumerate}
\item $\bs{h} = (h_1,\ldots, h_T)' = \bs{W}_\mathcal{B}\bs{h}^f \in \ell^2$ with $\bs{h}^f = (h^f(\omega_0),\ldots, h^f(\omega_{T-1}))' \in \mathbb{R}^T$. This is a deterministic sequence of fixed effect wavelet coefficients shared by all replicates in the population.
\item $\bs{U}_{s \cdot} = (U_{s1},\ldots, U_{sT})' = \bs{W}_\mathcal{B} \bs{U}_s^f$ with $\bs{U}_s^f = (U^f_s(\omega_0),\ldots, U^f_s(\omega_{T-1}))'$ for $s=1,\ldots, S$. In particular, we assume that the sequences $\bs{U}_{\cdot k} = (U_{1k},\ldots, U_{Sk})'$ are Gaussian random vectors for each $k=1,\ldots,T$. The assumptions on the variance-covariance structure of the vectors $\bs{U}_{\cdot 1}, \ldots, \bs{U}_{\cdot T}$ are detailed below.
\item $\bs{\epsilon}_{s \cdot} = (\epsilon_{s1},\ldots, \epsilon_{sT})' = \bs{W}_\mathcal{B}\bs{E}_s^f$ with $\bs{E}_s^f = (E_{s}^f(\omega_0),\ldots, E_{s}^f(\omega_{T-1}))'$ for $s = 1,\ldots, S$. The random vectors $\bs{\epsilon}_{s \cdot}$ are sequences of asymptotically independent wavelet noise coefficients with $\mathbb{E}[\epsilon_{sk}] = o_T(1)$ and $\var(\epsilon_{sk}) = \sigma_e^2/T + o_T(T^{-1})$. The noise coefficients are independent between different replicates and independent of the random effects sequences for all $s,k$.
\vspace{-1.5mm}
\end{enumerate}

\subsection{Covariance matrix assumptions} \label{sec:2.2}
Let $\bs{U} = (\bs{U}_{\cdot 1},\ldots, \bs{U}_{\cdot T})$ be the $S \times T$-dimensional random matrix of stacked random effects sequences $\bs{U}_{s \cdot}$. One of our main interests is in allowing for explicit correlation between the random effects sequences of different replicates, therefore we will not assume the covariance matrix of $\te{vec}(\bs{U})$ to be diagonal as is the case in \cite{DA97}, \cite{KHG11}, and \cite{FOvS10}. However, some structure on the covariance matrix $\cov(\te{vec}(\bs{U}))$ is necessary, since consistent estimation in a totally unstructured matrix is impossible (using only $ST$ observations). We consider structural assumptions on the variance-covariance matrix of $\te{vec}(\bs{U})$ as proposed in \cite{MC06} and \cite{M08} in a general functional data analysis context. Since the frequency domain model and the wavelet coefficient domain model are equivalent representations, the structural assumptions in the wavelet domain automatically transfer to assumptions on the variance-covariance structure of the functional random effects in the frequency domain. We assume that the covariance matrix $\bs{G} := \cov(\te{vec}(\bs{U}))$ consists of the Kronecker product of a $T \times T$ within-replicate diagonal covariance matrix and an $S \times S$ between-replicate correlation matrix:
\begin{eqnarray}
\bs{G} &=& \left( \begin{matrix}
\sigma_{u1}^2 & 0 & \ldots & 0 \\
0 & \sigma_{u2}^2 &  \ldots & 0 \\
\vdots & \vdots & \ddots & 0\\
0 & 0 & 0 & \sigma_{uT}^2
\end{matrix} \right) \otimes \left( \begin{matrix}
1 & \rho_{12} & \ldots & \rho_{1S} \\
\rho_{21} & 1 & \ldots & \rho_{2S} \\
\vdots &  \vdots & \ddots & \vdots \\
\rho_{S1} & \rho_{S2} & \ldots & 1
\end{matrix} \right) \nn
& := & \bs{G}_T \otimes \bs{G}_S \nonumber
\end{eqnarray}
where $\bs{G}_S$ is symmetric and positive-semidefinite. By considering a diagonal \emph{within-replicate} covariance matrix $\bs{G}_T$, the random effects coefficients are assumed to be uncorrelated between scale-locations $k=1,\ldots,T$. Note that a diagonal within-replicate covariance matrix $\bs{G}_T$ in the wavelet domain does not mean that the within-replicate covariance matrix in the frequency domain also has to be diagonal. To illustrate, a single non-zero variance component $\sigma_{u1}^2$ corresponding to the variance of the random scaling coefficient at scale-location $(0,0)$ translates to a random shift in the mean of the replicate-specific log-spectra in the frequency domain, thus resulting in highly correlated behavior of the random log-spectra over frequency. Furthermore, the variance components are heterogeneous across coefficients, therefore allowing for very general spatially inhomogeneous behavior of the random log-spectra in the frequency domain across replicates. The unstructured \emph{between-replicate} correlation matrix $\bs{G}_S$ allows for correlation between the random effects coefficients of different replicates at matching scale-locations $k$. We observe that the correlation $\rho_{ss'}$ between two different replicates remains the same across all locations. This is in order to keep the dimensions of the working covariance matrices small, but also to allow for consistent estimation of $\rho_{ss'}$ as the length of the time series increases. Note that $\bs{G}$ is symmetric and positive-semidefinite, since it is the Kronecker product of two symmetric positive-semidefinite matrices. Also, there is no identification issue between the two matrices $\bs{G}_T$ and $\bs{G}_S$, as $\bs{G}_S$ is restricted to have unit diagonal. 
\subsection{Functional space assumptions}\label{sec:2.3}
In order to develop the necessary estimation theory, we impose some regularity (smoothness) conditions on the realized replicate-specific sequences in the wavelet coefficient domain, or equivalently, on the realized discretely sampled log-spectra in the frequency domain. In particular, we assume that the fixed and random effects sequences are \emph{asymptotically sparse} elements of the $\ell_0$-sequence space with respect to the wavelet basis $\mathcal{B}$, defined as:
\begin{equation}
\ell_{0,T}(C) = \{ \bs{x} \in \mathbb{R}^T : \Vert \bs{x} \Vert_0 = C \} \nonumber
\end{equation}
such that $\Vert \bs{x} \Vert_0 = \# \{ k : x_k \neq 0 \}$.
\vspace{1mm}
\begin{ass}\label{ass:1}
Let $\bs{h} \in \ell_{0,T}(k_{h,T})$, with set of indices of non-zero coefficients $K_{h,T}$. We assume that $k_{h,T} = |K_{h,T}| \to \infty$ as $T \to \infty$, but $k_{h,T} = o(T)$. Let $\bs{\sigma}_u^2 = (\sigma_{u1}^2, \ldots, \sigma_{uT}^2)' \in \ell_{0,T}(k_{u,T})$ with set of indices of non-zero coefficients $K_{u,T}$, such that $\sup_k \sigma_{uk}^2 < \infty$. We assume that $K_{u,T} \subseteq K_{h,T}$ and $k_{u,T} = |K_{u,T}| \to \infty$ as $T \to \infty$, but $k_{u,T} = o(T)$.
\end{ass}
\vspace{1mm}
These regularity conditions assert that the fixed and realized random effects sequences or curves increase in complexity with $T$ (almost surely for the random effects), but at a slower rate than $T$. Furthermore, we make the assumption $K_{u,T} \subseteq K_{h,T}$, this is a convenient way to ensure that the population-mean log-spectrum $h^f$ and the realized replicate-specific log-spectra $h_s^f|u_s^f = h^f + u_s^f$ share the same smoothness properties. This complexity constraint allows to disentangle the different parts in the variance components coming from the random effects and the noise terms, but is also important for the sake of interpretation in a functional mixed effects model, as discussed in \cite{G02}, \cite{QG06}, and \cite{AS07}. \\[3mm]
Before presenting the estimation procedure, we recall some useful results on nonlinear thresholding methods in a classical Gaussian sequence model under $\ell_0$-sparsity constraints. Consider the $\ell_0$-Gaussian sequence model:
\begin{equation} \label{eq:5}
y_i = \theta_i + \epsilon_n z_i, \quad \quad i=1,\ldots,n 
\end{equation}
with $\bs{\theta} = (\theta_1,\ldots, \theta_n)' \in \ell_{0,n}(k_n)$ and $z_1,\ldots,z_n \overset{\te{iid}}{\sim} N(0,1)$. Under the $\ell_0$-sparsity constraint $k_n \to \infty$ as $n \to \infty$, but $k_n = o(n)$, the minimax $\ell_2$-risk of estimation for $\bs{\theta}$ satisfies,
\begin{equation} 
R(\ell_{0,n}(k_n), \epsilon_n)\, :=\, \inf_{\hat{\bs{\theta}}} \sup_{\bs{\theta} \in \ell_{0,n}(k_n)} \mathbb{E}\Vert \hat{\bs{\theta}} - \bs{\theta} \Vert^2\, \sim\, 2\epsilon_n^2 k_n \log(n/k_n) \nonumber
\end{equation}
where $\Vert \cdot \Vert$ denotes the Euclidian norm. It is well-known that hard (or soft) nonlinear thresholding of the coefficients asymptotically achieves the minimax risk. In particular, the hard nonlinear thresholding estimator $\hat{\theta}_i = y_i \bs{1}\{ |y_i| \geq \lambda_n \}$ for $i =1,\ldots, n$, with $\lambda_n = \epsilon_n \sqrt{2\log(n/k_n)}$ is an asymptotic minimax estimator, see \cite{J11} for a detailed proof. We note that the nonlinear thresholding estimators $\{\hat{\theta_i} \}_{i=1,\ldots, n}$ are nonadaptive in the sense that the threshold $\lambda_n$ depends on the (typically unknown) smoothness space parameter $k_n$. \cite{ABDJ06} show that in the context of $\ell_0$-Gaussian sequence models, using False Discovery Rate (FDR) nonlinear thresholding, it is possible to asymptotically achieve the minimax risk without requiring knowledge of the smoothness space parameter $k_n$. For details on this FDR-based procedure, and the appropriate choice of its tuning parameter $q_n$, we refer to \cite{ABDJ06}.

\section{Estimation procedure} \label{sec:3}
\subsection{Population-mean log-spectrum} \label{sec:3.1}
We estimate the population-mean log-spectrum $h^f(\omega_\ell)$ at frequencies $\omega_\ell \in [0, 1/2)$ by the projection estimator,
\begin{equation}
\hat{h}^f(\omega_\ell) = \sum_{k=1}^T \hat{h}_k(\bs{Y}_{\cdot k}) \psi_k(\omega_\ell), \quad \quad \ell = 0, \ldots, T-1 \nonumber
\end{equation}
the inverse discrete wavelet transform with respect to the basis $\mathcal{B} = \{ \psi_k \}_{k=1}^\infty$ of the estimated fixed effects sequence of coefficients $\hat{\bs{h}}(\bs{Y}) := \{ \hat{h}_k(\bs{Y}_{\cdot k}) \}_{k=1}^T$, with $\bs{Y}_{\cdot k} = (Y_{1k}, \ldots, Y_{Sk})'$. The sequence $\hat{\bs{h}}(\bs{Y})$ is based on component-wise thresholded generalized least squares estimators,
\begin{equation} 
\hat{h}_k(\bs{Y}_{\cdot k}) = ( \bs{w}_{k}' \bs{Y}_{\cdot k}) \bs{1}\{ k \in \widehat{K}_h(\bs{Y}) \},\quad \quad k=1,\ldots,T \nonumber
\end{equation}
where,
\begin{enumerate}
\item $\bs{w}_{k} = (w_{1k},\ldots, w_{Sk})'$ are generalized least squares weights depending on the between-replicate correlation structure through $w_{sk} = \left( \sum_{i=1}^S \bs{V}_{k[i,s]}^{-1} \right) \cdot \left( \sum_{i,j=1}^S \bs{V}_{k[i,j]}^{-1} \right)^{-1}$. Here, $\bs{V}_k$ denotes the asymptotic covariance matrix of $\bs{Y}_{\cdot k}$ given by $\bs{V}_k = \sigma_{uk}^2 \bs{G}_S + \frac{\sigma_e^2}{T} \te{I}_S$, with $\te{I}_S$ the $S \times S$-identity matrix.\\
\item $\widehat{K}_h(\bs{Y}) = \{ k : | \frac{1}{S} \sum_{s=1}^S Y_{sk} | \geq \lambda_{h,T} \}$ is the estimated set of indices of non-zero coefficients with universal threshold $\lambda_{h,T} = \sqrt{\sigma_e^2/(ST)}\sqrt{2 \log(T/k_{h,T})}$.  The motivation for this thresholded set comes from the observation that the replicate-specific sequences --conditional on the random effects-- are \emph{independent} between replicates and, individually for each replicate, follow an $\ell_{0,T}(k_{h,T})$-sequence model with the same set of non-zero coefficients $K_{h,T}$ for each replicate and noise variance approximately $\sigma_e^2/T$. In the unconditional case, the distributional behavior of the sequences at scale-locations of zero coefficients ($k \notin K_{h,T}$) remains unchanged, allowing for the same control on the number of false positives as in the conditional case. Moreover, under some regularity conditions, the empirical wavelet noise coefficients are asymptotically normal for increasing $T$, this asymptotically justifies the threshold choice $\lambda_{h,T}$ based on a Gaussian sequence model (see Section \ref{sec:4.1}).
\end{enumerate}
\subsection{Random effects covariance matrices}\label{sec:3.2}
\subsubsection{Estimation of the within-replicate covariance matrix}
The within-replicate random effects covariance matrix $\bs{G}_T$ is assumed to be diagonal, with vector of variance components $\bs{\sigma}_u^2 = (\sigma_{u1}^2,\ldots, \sigma_{uT}^2)'\in \ell_{0,T}(k_{u,T})$ on the diagonal. This vector is estimated by the thresholded sample-variances,
\begin{equation}
\hat{\sigma}_{uk}^2(\bs{Y}_{\cdot k}) = \left\{ \frac{1}{S}\sum_{s=1}^S \left( Y_{sk} - \hat{h}_k(\bs{Y}_{\cdot k})\right)^2 - \frac{\sigma_e^2}{T} \right\}_+ \bs{1}\left\{ k \in \widehat{K}_u(\bs{Y}) \right\} \nonumber
\end{equation}
with estimated set of indices of non-zero variance components $\widehat{K}_u(\bs{Y}) = \{ k\, :\, |T_k(\bs{Y}_{\cdot k})| \geq \lambda_{u,T} \}$. Here the statistics $T_k(\bs{Y}_{\cdot k})$ and the threshold $\lambda_{u,T}$  are given by,
\begin{eqnarray}
T_k(\bs{Y}_{\cdot k}) &=& \log\left( \frac{1}{S}\sum_{s=1}^S \left(Y_{sk} - \hat{h}_k(\bs{Y}_{\cdot k})\right)^2 \right) - \log\left( \frac{2\sigma_e^2}{ST}\right) - \psi^{(0)}\left(\frac{S}{2}\right) \label{eq:5b} \\
\lambda_{u,T} &=& \sqrt{\psi^{(1)}(S/2)}\sqrt{2\log(T)} \nonumber
\end{eqnarray}
where $\psi^{(0)}(\cdot)$ and $\psi^{(1)}(\cdot)$ denote the digamma and trigamma function respectively. The motivation for this thresholded set, which has similar structure as the thresholded set $\widehat{K}_h(\bs{Y})$, comes from the observation that --for uncorrelated replicates-- the vector $\{ T_k(\bs{Y}_{\cdot k}) \}_{k=1}^T$ is variance stabilizing and behaves approximately as an $\ell_{0,T}(k_{u,T})$-Gaussian sequence model with noise variance $\psi^{(1)}(S/2)$. This justifies the threshold choice $\lambda_{u,T}$ based on a Gaussian sequence model. For a general between-replicate correlation matrix $\bs{G}_S$, it follows that the  distributional behavior of the zero variance components with indices $k \notin K_{u,T}$ remains the same, thus allowing for the same control on the number of false positives as in the uncorrelated case. 
We note that the threshold $\lambda_{u,T}$ is slightly more conservative than the asymptotic minimax universal threshold as in Section \ref{sec:2.3}. The reason for this is that in the context of a Gaussian sequence model, under the more conservative threshold, \emph{both} the number of false positives and the number of false negatives in the estimated set of indices of non-zero variance components tend to zero almost surely (Corollary \ref{corr:1}). Under the asymptotic minimax threshold, this only holds true for the number of false negatives.

\subsubsection{Estimation of the between-replicate correlation matrix}
The between-replicate correlation matrix $\bs{G}_S$ is estimated elementwise by considering the following sample-correlation based estimators,
\begin{equation}
\hat{\rho}_{ij}(\bs{Y}) = \frac{1}{\hat{k}_u} \sum_{k \in \widehat{K}_u} \frac{(Y_{ik} - \hat{h}_k(\bs{Y}_{\cdot k}))(Y_{jk}-\hat{h}_k(\bs{Y}_{\cdot k}))}{\hat{\sigma}_{uk}^2(\bs{Y}_{\cdot k}) \vee \delta}, \quad \quad i, j = 1,\ldots,S, \quad i \neq j \label{eq:6a}
\end{equation} 
where $\delta > 0$ is a small constant that ensures that the denominator is bounded away from zero, and $\widehat{K}_u = \widehat{K}_u(\bs{Y})$ is the estimated set of indices of non-zero variance components with cardinality $\widehat{k}_u = |\widehat{K}_u(\bs{Y})|$. The intuition behind this estimator comes from the fact that $\cov(Y_{ik}, Y_{jk}) = \sigma_{uk}^2 \rho_{ij}$ for each $k \in K_{u,T}$, whereas $\cov(Y_{ik}, Y_{jk}) = 0$ for each $k \notin K_{u,T}$ as $\sigma_{uk}^2 = 0$ by definition of $K_{u,T}$.\\[3mm]
The estimated matrix $\widehat{\bs{G}}_S(\bs{Y})$ with off-diagonal elements $\hat{\rho}_{ij}(\bs{Y})$ is only an approximate correlation matrix. It is symmetric and has unit diagonal, but is not guaranteed to be positive-semidefinite in a finite sample situation. By \cite{H02}, we compute the correlation matrix $\widetilde{\bs{G}}_S(\bs{Y})$ with minimum distance in Frobenius-norm to the originally estimated matrix $\widehat{\bs{G}}_S(\bs{Y})$,
\begin{equation}
\widetilde{\bs{G}}_S(\bs{Y}) = \te{arg min}_{\bs{X} = \bs{X}^T} \{ \lVert \widehat{\bs{G}}_S(\bs{Y}) - \bs{X}\rVert_F : \bs{X} \te{ is a correlation matrix} \} \nonumber
\end{equation} 
However, replacing the matrix $\widehat{\bs{G}}_S(\bs{Y})$ by the new matrix $\widetilde{\bs{G}}_S(\bs{Y})$, the estimated variance components $\hat{\sigma}_{uk}^2(\bs{Y}_{\cdot k})$ are no longer properly scaled (i.e. $\hat{\sigma}_{uk}^2\widehat{\bs{G}}_S \neq \hat{\sigma}_{uk}^2 \widetilde{\bs{G}}_S$). Instead, we consider the rescaled estimators $\tilde{\sigma}_{uk}^2(\bs{Y})$, such that $\Vert \hat{\sigma}_{uk}^2 \widehat{\bs{G}}_S \Vert_F = \Vert \tilde{\sigma}_{uk}^2 \widetilde{\bs{G}}_S \Vert_F$, which are easily obtained through,
\begin{equation}
\tilde{\sigma}_{uk}^2(\bs{Y}) = \hat{\sigma}_{uk}^2(\bs{Y}_{\cdot k}) \frac{\Vert \widehat{\bs{G}}_S(\bs{Y}) \Vert_F}{\Vert \widetilde{\bs{G}}_S(\bs{Y}) \Vert_F} \nonumber
\end{equation}
Note that the rescaling does not affect the estimated zero variance components corresponding to $k \notin \widehat{K}_u(\bs{Y})$, i.e. if $\hat{\sigma}_{uk}^2(\bs{Y}_{\cdot k}) = 0$, then also $\tilde{\sigma}_{uk}^2(\bs{Y}) = 0$.

\subsection{Iterative estimation scheme} \label{sec:3.3}
In order to estimate the population-mean log-spectrum $h^f(\omega_\ell)$, we consider a generalized least squares estimator with weights depending on the between-replicate correlation structure. On the other hand, estimation of the random effects covariance and correlation matrices $\bs{G}_T$ and $\bs{G}_S$  depends on the population-mean sequence $\bs{h}$, since the sample-variances and sample-correlations need to be centered about their respective means. This is typically the case in linear mixed mdoel estimation where one allows for between-replicate correlation, and in this context, one easy-to-implement approach is to consider an iterative-generalized least squares scheme (see e.g. \cite{J07}). First, we compute the thresholded ordinary least squares estimator of $\bs{h}$ by equally weighting each of the observations across replicates. This does not require any information on the between-replicate correlation structure. Next, we iterate between estimation of $\bs{G}_T$ and $\bs{G}_S$ given the estimate of $\bs{h}$, and estimation of $\bs{h}$ given estimates of $\bs{G}_T$ and $\bs{G}_S$, and we continue iterating until some convergence criterion is satisfied. We note that under a similar random effects variance-covariance structure, considering only the linear part of the estimators (without the thresholding), \cite{J07a} show that an iterative-generalized least squares scheme converges exponentially with probability tending to one as the number of replicates $S$ increases.

\subsection{Replicate-specific log-spectra} \label{sec:3.4}
The replicate-specific log-spectra $H_s^f(\omega_\ell)$ at frequencies $\omega_\ell \in [0, 1/2)$ are predicted by the projection estimators,
\begin{equation}
\widehat{H}_s^f(\omega_\ell) = \sum_{k=1}^T \widehat{H}_{sk}(\bs{Y}) \psi_k(\omega_\ell), \quad \quad s=1,\ldots,S, \quad \ell=0,\ldots, T-1 \nonumber
\end{equation}
the inverse discrete wavelet transform with respect to the basis $\mathcal{B}$ of the predicted replicate-specific sequence of coefficients $\widehat{\bs{H}}_s(\bs{Y}) := \{ \widehat{H}_{sk}(\bs{Y}) \}_{k=1}^T$. Prediction in the wavelet coefficient domain reduces to prediction in a linear mixed model, and we can find estimated predictors of the random effects sequences $\bs{U}_{\cdot k} = (U_{1k},\ldots, U_{Sk})'$ through,
\begin{equation}
\widehat{\bs{U}}_k(\bs{Y}) = \widehat{\bs{G}}_k \widehat{\bs{V}}_k^{-1}(\bs{Y}_{\cdot k} - \hat{h}_k \bs{1}_S) \nonumber
\end{equation}
Here, $\widehat{\bs{G}}_k = \tilde{\sigma}^2_{uk}\widetilde{\bs{G}}_S$ and $\widehat{\bs{V}}_k = \tilde{\sigma}^2_{uk}\widetilde{\bs{G}}_S + \frac{\sigma_e^2}{T}\te{I}_S$ are plug-in estimators, and $\bs{1}_S$ denotes an $S$-dimensional vector of ones. Note that if $\widehat{\bs{G}}_k$ and $\widehat{\bs{V}}_k^{-1}$ are replaced by the true matrices $\bs{G}_k$ and $\bs{V}_k^{-1}$, and $\hat{\bs{h}}$ is replaced by the best linear unbiased estimator, then the $\widehat{\bs{U}}_{\cdot k}$ are the best linear unbiased predictors of the random effects sequences $\bs{U}_{\cdot k}$ in a linear mixed model,  see \cite{S92}. In general, due to the nonlinear thresholding of coefficients, $\hat{\bs{h}}$ ceases to be an unbiased estimator of $\bs{h}$. By combining the estimator for the fixed effects sequence $\bs{h}$ and the predictors for the random effects sequences $\bs{U}_{\cdot k}$, the replicate-specific sequences of coefficients $\bs{H}_s = \{ h_k + U_{sk} \}_{k=1}^T$ are predicted through,
\begin{equation}
\widehat{\bs{H}}_s(\bs{Y}) = \left\{ \hat{h}_k(\bs{Y}_{\cdot k}) + \widehat{U}_{sk}(\bs{Y}) \right\}_{k=1}^T \nonumber
\end{equation}
\section{Estimation theoretical results} \label{sec:4}
\subsection{Risk bounds for the population-mean log-spectrum} \label{sec:4.1}
In this section we derive finite-sample upper bounds for the $\ell_2$-risk of the estimated fixed effects sequence of coefficients $\hat{\bs{h}}(\bs{Y}) = \{ \hat{h}_k(\bs{Y}_{\cdot k} \}_{k=1}^T$ with respect to $\bs{h}$. By Parseval's relation, for any given number of replicates, the $\ell_2$-risk in the wavelet coefficient domain is asymptotically equal to the $L_2$-risk of the projected estimators in the frequency domain. It then follows that the same expression derived for the $\ell_2$-risk in the wavelet coefficient domain also gives an upper bound for the $L_2$-risk of the estimated population-mean log-spectrum $\hat{h}^f$. \\
The derivation of the $\ell_2$-risk bounds is based on the observation that, under some regularity conditions, the non-Gaussian linear mixed model in the wavelet coefficient domain (eq.(\ref{eq:4})) is asymptotically equivalent to a Gaussian linear mixed model ($T \to \infty$) as the empirical wavelet noise coefficients are essentially local averages of the log-periodogram ordinates, and the random effects wavelet coefficients are assumed to be normally distributed. This allows us to calculate the $\ell_2$-risk of the sequence $\hat{\bs{h}}(\bs{Y})$ first under an accompanying Gaussian sequence model, and relate this to the $\ell_2$-risk under the non-Gaussian sequence model. The asymptotic equivalence between the two models is based on a uniform asymptotic normality result for the empirical wavelet noise coefficients. 
\cite{N96} already establishes uniform asymptotic normality for the empirical wavelet noise coefficients of periodogram ordinates for a general non-Gaussian process $X(t)$ in the context of a single long time series. Since the technical considerations in \cite{N96} are not the main focus of this paper, here we derive uniform asymptotic normality of the empirical wavelet noise coefficients of log-periodogram ordinates only for a weakly dependent Gaussian process $X_s(t)$ as in \cite[Chapter 5]{B81} with a given (log-)spectrum, i.e. conditional on the functional random effects in the frequency domain. By the (conditional) Gaussianity assumption for the time series replicates, we can derive cumulant bounds for the realized replicate-specific log-periodogram ordinates in the frequency domain using results from \cite{T79}. These cumulant bounds are used to derive uniform asymptotic normality of the empirical wavelet noise coefficients for an increasing number of coefficients, along the same lines as \cite{N96} for a single time series replicate. Note that this result is conditional on the functional random effects in the frequency domain, however since the random effects wavelet coefficients in eq.(\ref{eq:4}) are assumed to be normally distributed, unconditional uniform asymptotic normality of the wavelet coefficients of the random replicate-specific log-periodograms follows as well. We point out that asymptotic normality of empirical wavelet noise coefficients of the log-periodogram has already been suggested without proof by \cite{G97}, \cite{M94}, and \cite{FOvS10} under the approximate additive noise model with $\epsilon_k \sim (0, \pi^2/6)$. In order for a certain summation effect to work, we make the additional assumption that, for increasing $T$, the set of non-zero fixed effects coefficients is bounded away from the finest wavelet scale, intuitively this means that the finest wavelet scale contains virtually only noise and no signal as $T$ increases. This is a typical assumption in the wavelet literature, and in an ordinary signal plus noise model this is commonly used for estimation of the noise variance through the empirical coefficients located only at the finest wavelet scale (see \cite{V99}).
\vspace{1mm}
\begin{ass}\label{ass:2} Define the set,
\begin{equation}
J_{T,\alpha} := \{1\} \cup \{k \geq 2 \,|\, 2^{\lfloor \log_2(k-1) \rfloor} \leq CT^{1-\alpha} \} \label{eq:7}
\end{equation} 
for some constant $C > 0$. We assume that there exist some $T^*> 0$ and $0 < \alpha^* < 1$, such that for $T \geq T^*$, $K_{h,T} \subseteq J_{T,\alpha^*}$.
\end{ass}
\vspace{1mm}
\begin{ass}\label{ass:3} Conditional on $U^f_s(\omega) = u^f_s(\omega)$,  $\{X_s(t)\}_{t>0}$ is a Gaussian process satisfying $\sum_{h=-\infty}^\infty |h||\cov(X_s(t), X_s(t+h)| < \infty$ for each $s =1,\ldots, S$.
\end{ass}
\vspace{1mm}
\begin{theorem}\label{thm:1} Under assumptions \ref{ass:1}-\ref{ass:3}, let $\alpha$ such that $0 < \alpha \leq \alpha^*$. Consider the estimators,
\begin{equation}
\hat{h}_{k}(\bs{Y}_{\cdot k}) = \left( \bs{w}_{k}'\bs{Y}_{\cdot k} \right) \bs{1}\{ | \bar{\bs{Y}}_{k} | \geq \lambda_{h,T}, k \in J_{T, \alpha} \}, \quad \quad k=1,\ldots, T \nonumber
\end{equation}
where $\bar{\bs{Y}}_k = \frac{1}{S}\sum_{s=1}^S Y_{sk}$ and $\lambda_{h,T} = \sqrt{\sigma_e^2/(ST)}\sqrt{2 \log( T/k_{h,T})}$. For $T$ sufficiently large, the $\ell_2$-risk of $\hat{\bs{h}}(\bs{Y})$ satisfies,
\begin{equation}
\sup_{\bs{h} \in \ell_{0,T}(k_{h,T})} \mathbb{E}\Vert \hat{\bs{h}}(\bs{Y}) - \bs{h} \Vert^2 \lesssim \frac{k_{h,T}}{ST} \log\left( \frac{T}{k_{h,T}} \right) + k_{u,T} \left( \sup_k \bs{w}_{k}'\bs{V}_k\bs{w}_{k} - \frac{\sigma_e^2}{TS} \right) \label{eq:8}
\end{equation}
where $\lesssim$ denotes the inequality $\leq$ up to a multiplicative positive constant.
\end{theorem}
\vspace{1mm}
The first term on the right-hand side in eq.(\ref{eq:8}) is equivalent to the minimax rate of estimation in an $\ell_{0,T}(k_{h,T})$-Gaussian sequence model with noise variance of order $(ST)^{-1}$. The second term arises from introducing the random effects and is an upper bound of the integrated error made in estimating $\bs{h}$ by taking a weighted sample average over a finite number of replicates. We observe that $\bs{w}_{k}'\bs{V}_k \bs{w}_{k} = \var(\bs{w}_{k}'\bs{Y}_{\cdot k})$ for $\bs{Y}_{\cdot k} \sim (h_k\bs{1}_S, \bs{V}_k)$, and this term is minimized by the generalized least squares weights $\bs{w}_{k}$ as in Section \ref{sec:3.1}. In the case of sub-optimal ordinary least squares weights $\bs{w}_{k} = (\frac{1}{S}, \ldots, \frac{1}{S})'$ the second term becomes:
\begin{eqnarray}
k_{u,T} \left( \sup_k \bs{w}_{k}'\bs{V}_k\bs{w}_{k} - \frac{\sigma_e^2}{ST} \right) &=& k_{u,T} \sup_k \frac{\sigma_{uk}^2 \bs{1}'_S \bs{G}_S\bs{1}_S}{S^2} \nonumber
\end{eqnarray}
This implies that if $\frac{k_{u,T}}{S^2}\bs{1}_S'\bs{G}_S \bs{1}_S \to 0$ as $S, T \to \infty$ the thresholded ordinary least squares estimator remains a consistent estimator of $\bs{h}$. For uncorrelated replicates this term is for instance of the order $k_{u,T}/S$. The expression $\bs{1}_S'\bs{G}_S \bs{1}_S$ is always nonnegative, and is decreasing as replicates become more negatively correlated. This might seem surprising, but can be illustrated by the following simple bi-replicate example: suppose one observes two random replicate-curves that are highly negatively correlated, with high probability the true population mean-curve lies in between the two replicate-curves and the error term due the random effects should therefore be smaller than in the independent curve situation; if the curves are perfectly negatively correlated, the true population mean-curves lies exactly in between the two replicate-curves, and the error term due to random effects should disappear completely. 
\subsection{Consistent estimation of random effects covariance matrices} \label{sec:4.2}
In this section, we derive some asymptotic results for the estimators $\hat{\sigma}_{uk}^2(\bs{Y}_{\cdot k})$ of the variance components in the within-replicate covariance matrix $\bs{G}_T$, and the estimators $\hat{\rho}_{ij}(\bs{Y})$ of the correlation coefficients in the between-replicate correlation matrix $\bs{G}_S$. The derived results crucially rely on the condition $\Vert \bs{G}_S \Vert_F = o(S)$, which controls the level of correlation between different replicates. Essentially it requires that the effective number of uncorrelated replicates increases with the total number of replicates $S$. To illustrate, for uncorrelated replicates $\Vert \bs{G}_S \Vert_F/S = 1/\sqrt{S}$, whereas for perfectly correlated replicates $\Vert \bs{G}_S \Vert_F/S = 1$. In order to simplify the proofs, as in \cite{KHG11}, \cite{G97}, \cite{M94}, and \cite{FOvS10}, we work under the approximate model where the empirical wavelet noise coefficients are mean zero with variance $\sigma_e^2/T$, which is the asymptotic version of the model as $T \to \infty$. Note that we do not assume normality of the empirical wavelet noise coefficients, nor independence between different scale-locations $k$ within a single replicate.
\vspace{1mm}
\begin{theorem} \label{thm:2}
Suppose that $\epsilon_{sk} \sim (0, \sigma_e^2/T)$ with $\mathbb{E}[\epsilon_{sk}^4] = O(T^{-2})$ for each $s=1,\ldots,S$ and $k = 1,\ldots,T$, and that there exist uniform consistent estimators $\sup_k |\hat{h}_k - h_k| = o^{S,T}_{p}(1)$, with $|\hat{h}_k - h_k| = o^{S,T}_{p}(T^{-1/2})$ for $k \notin K_{u,T}$. If $\Vert \bs{G}_S \Vert_F = o(S)$, and $C \leq \lambda_{u,T} = o(\log(T))$  for some constant $C > 0$, then
\begin{eqnarray}\label{eq:9}
P(\widehat{K}_u(\bs{Y}) = K_{u,T}) &\to & 1, \quad \quad \te{as } S,T \to \infty \\ 
\sup_k |\hat{\sigma}_{uk}^2(\bs{Y}_{\cdot k}) - \sigma_{uk}^2| &\overset{P}{\to}& 0, \quad \quad \te{as } S,T \to \infty \label{eq:10}
\end{eqnarray}
If $0 < \delta \leq \inf_{k \in K_{u,T}} \sigma_{uk}^2$ for $\hat{\rho}_{ij}(\bs{Y})$ in eq.(\ref{eq:6a}), then for each $i,j$ with $i\neq j$,
\begin{eqnarray}
\hat{\rho}_{ij}(\bs{Y}) &\overset{P}{\to}& \rho_{ij}, \quad \quad \te{as } S,T \to \infty \nonumber
\end{eqnarray}
\end{theorem}
\vspace{1mm}
The conditions $\mathbb{E}[\epsilon_{sk}^4] = O(T^{-2})$ and $|\hat{h}_k - h_k| = o_p^{S,T}(T^{-1/2})$ for $k \notin K_{u,T}$ are needed in order to control the number of false positives in the set of estimated non-zero variance components in eq.(\ref{eq:9}). Under assumptions \ref{ass:1}-\ref{ass:3}, by the cumulant bounds derived in the proof of Theorem \ref{thm:1} (see Appendix), it follows that $\mathbb{E}[\epsilon_{sk}^4] = O(T^{-2})$ for each $s,k$. Furthermore, if $\log(T)/S \to 0$ as $S,T \to \infty$, it can be verified that $|\hat{h}_k - h_k| = o_p^{S,T}(T^{-1/2})$ for $k \notin K_{u,T}$ for the nonlinear estimators $\hat{h}_k(\bs{Y}_{\cdot k})$ in Theorem \ref{thm:1} with ordinary least squares weights $\bs{w}_{k} = (\frac{1}{S},\ldots, \frac{1}{S})'$. \\
The following corollary gives the theoretical justification for the form of the statistics $T_k(\bs{Y}_{\cdot k})$ as given in eq.(\ref{eq:5b}), which are based on the accompanying Gaussian sequence model where the empirical wavelet noise coefficients are exactly normally distributed. Under the Gaussian sequence model, it is possible to  adopt the threshold $\lambda_{u,T} = \sqrt{\psi^{(1)}(S/2)}\sqrt{2\log(T-k_{u,T})}$, which converges to zero if $\log(T)/S \to 0$ as $S,T \to \infty$, while preserving the consistency result that both the number of false positives and false negatives in the estimated set of non-zero variance components is zero with probability tending to one.
\vspace{1mm}
\begin{corr}\label{corr:1} Suppose that $\epsilon_{sk} \overset{\textnormal{iid}}{\sim} N(0,\sigma_e^2/T)$ for each $s=1,\ldots,S$ and $k=1,\ldots,T$, such that $\bs{\xi}_k \sim N(h_k\bs{1}_S, \bs{V}_k)$, and consider the statistics $T_k(\bs{\xi}_k)$ as in eq.(\ref{eq:5b}) with $\hat{h}_k(\bs{\xi}_k)$ replaced by the true coefficients $h_k$.
For uncorrelated replicates ($\bs{G}_S = \te{I}_S$), the vector $\{ T_k(\bs{\xi}_k) \}_{k=1}^T$ converges to an $\ell_{0,T}(k_{u,T})$-Gaussian sequence model as $S \to \infty$, 
\begin{eqnarray} \left\{ \begin{array}{lll} 
\dfrac{1}{\sqrt{\psi^{(1)}(S/2)}}\left( T_k(\bs{\xi}_k) - \log\left( \frac{\sigma_{uk}^2 T + \sigma_e^2}{\sigma_e^2} \right)\right) & \overset{d}{\to} & N(0,1), \quad  \te{if } k \in K_{u,T} \\
\dfrac{1}{\sqrt{\psi^{(1)}(S/2)}} T_k(\bs{\xi}_k) & \overset{d}{\to} & N(0,1), \quad  \te{if } k \notin K_{u,T} \label{eq:11} 
\end{array}\right.
\end{eqnarray}
For correlated replicates, with general correlation matrix $\bs{G}_S$, it remains true that for $S \to \infty$, 
\begin{eqnarray}
\frac{1}{\sqrt{\psi^{(1)}(S/2)}} T_k(\bs{\xi}_k) & \overset{d}{\to} &  N(0,1), \quad  \te{if } k \notin K_{u,T} \nonumber
\end{eqnarray}
If $\Vert \bs{G}_S \Vert_F = o(S)$ and $\sqrt{\psi^{(1)}(S/2)}\sqrt{2 \log(T - k_{u,T})} \leq \lambda_{u,T} = o(\log(T))$, then as in eq.(\ref{eq:9})
\begin{equation}
P(\widehat{K}_u(\bs{\xi}) = K_{u,T}) \to 1 \quad \quad \te{as } S,T \to \infty. \nonumber
\end{equation}
\end{corr}
\section{Confidence regions} \label{sec:5}
In this section, we develop asymptotic confidence regions for the discretely sampled population mean log-spectrum, where it is important to take into account the possible correlation between different replicate-specific curves to avoid the use of erroneous confidence sets. In the wavelet coefficient domain, the estimated sequence of fixed effect coefficients is a nonlinear biased estimator of $\bs{h}$, and for this reason it is generally difficult to derive asymptotic confidence bounds directly from the asymptotic distribution of $\hat{\bs{h}}(\bs{Y})$, even under Gaussian model assumptions. As proposed in \cite{RV06} and \cite{GW05} among others, instead we consider an estimator of the squared norm $\Vert \bs{h} - \hat{\bs{h}} \Vert^2$ (conditional on $\hat{\bs{h}}$), and we derive the asymptotic distribution of this estimator instead of the asymptotic distribution of the original estimator $\hat{\bs{h}}$. Asymptotic $\ell_2$-confidence regions for $\bs{h}$ can then be constructed by restricting the norm of $\bs{h}$ with respect to the estimated sequence $\hat{\bs{h}}$. Moreover, by the norm equivalence between the functional (i.e. frequency) domain and wavelet coefficient domain, we can easily transfer the confidence regions for $\bs{h}$ in the wavelet domain to confidence regions for $\bs{h}^f$ in the frequency domain. For convenience, we work under the approximate model assumption that the wavelet noise coefficients are exactly normally distributed, with mean zero and variance $\sigma_e^2/T$. The derived confidence regions are therefore approximate in the sense that they are based on asymptotic distributional behavior of the estimator of the pivot quantity ($S \to \infty$), but also on the fact that the empirical wavelet noise coefficients are only asymptotically normally distributed ($T \to \infty$) under appropriate model conditions as discussed in Section \ref{sec:4.1}.\\
The method is based on the assumption that we can split the sample $\bs{\xi} = (\bs{\xi}_1, \ldots, \bs{\xi} _T) \in \mathbb{R}^{S \times T}$, with $\bs{\xi}_k \sim N(h_k \bs{1}_S, \bs{V}_k)$, into two sets of independent observations $\bs{\xi}^{(1)}, \bs{\xi}^{(2)} \in \mathbb{R}^{S \times T}$. Suppose that the covariance matrices $\bs{V}_k$ are known, one simple approach to split the sample into two independent samples at the cost of making the variance twice as large is to consider,
\begin{eqnarray}
\bs{\xi}_k^{(1)} := \bs{\xi}_k - \bs{X}_k  & \quad & \bs{\xi}_k^{(2)} := \bs{\xi}_k + \bs{X}_k \quad \quad \te{for all } k=1,\ldots, T \nonumber
\end{eqnarray}
where the vectors $\bs{X}_k \sim N(\bs{0}, \bs{V}_k)$ are independent of $\bs{\xi}_k$. We estimate the sequence $\bs{h}$ using only the observations in $\bs{\xi}^{(1)}$, and construct the confidence regions from the additional independent set of observations $\bs{\xi}^{(2)}$ conditional on $\hat{\bs{h}}(\bs{\xi}^{(1)})$. The nature of the estimator $\hat{\bs{h}}$ is irrelevant for the construction of the confidence region, however, since the radius of the confidence region is proportional to $\Vert \bs{h} - \hat{\bs{h}} \Vert$, better estimators $\hat{\bs{h}}$ (in terms of $\ell_2$-risk) will lead to smaller confidence regions.\\
Suppose that we have split the sample into two independent parts $\bs{\xi}^{(1)}, \bs{\xi}^{(2)}$, with $\bs{\xi}^{(1)}_k, \bs{\xi}^{(2)}_k \sim N(h_k\bs{1}_S, 2\bs{V}_k)$ for $k =1,\ldots,T$ and we have computed $\hat{\bs{h}} := \hat{\bs{h}}(\bs{\xi}^{(1)})$. The next step is to find an estimator of the pivot quantity $\Vert \bs{h} - \hat{\bs{h}} \Vert^2$, conditional on $\hat{\bs{h}}$, using only the set of observations $\bs{\xi}^{(2)}$. We consider the unbiased estimator,
\begin{equation}
\widehat{R}(\bs{\xi}^{(2)}, \hat{\bs{h}}) = \sum_{k=1}^T \left[\sum_{s=1}^S \left[w_{sk}(\xi_{sk}^{(2)} - \hat{h}_k)^2\right] - 2\sigma_{k,T}^2 \right] \nonumber
\end{equation}
where $\bs{w}_{k} = (w_{1k},\ldots, w_{Sk})'$ is a vector of (generalized least squares) weights such that $\sum_s w_{sk} = 1$.
Straightforward calculus shows that, conditional on $\hat{\bs{h}}$, $\widehat{R}(\bs{\xi}^{(2)}, \hat{\bs{h}})$ is an unbiased estimator of $\Vert \bs{h} - \hat{\bs{h}} \Vert^2$ and,
\begin{equation}
\tau^2(\bs{h}, \hat{\bs{h}}) := \var\left(\widehat{R}(\bs{\xi}^{(2)}, \hat{\bs{h}})\right) = \sum_{k=1}^T \left[ 8 \Vert \te{diag}(\bs{w}_{k}) \bs{V}_k \Vert_F^2 + 8(h_k - \hat{h}_k)^2 \bs{w}_{k}' \bs{V}_k \bs{w}_{k} \right] \nonumber
\end{equation}
where $\te{diag}(\bs{w}_{k})$ is a diagonal matrix with the vector $\bs{w}_{k}$ on the diagonal.
\vspace{1mm}
\begin{theorem}\label{thm:3}  Suppose that $\Vert \bs{\Gamma}_k \Vert_1/ \Vert \bs{\Gamma}_k \Vert_F \to 0$ as $S \to \infty$, where $\bs{\Gamma}_k := \bs{V}_k^{1/2} \textnormal{diag}(\bs{w}_{k}) \bs{V}_k^{1/2}$ with $\bs{V}_k^{1/2}$ a symmetric matrix square root of $\bs{V}_k$. For a given confidence level $1-\alpha$, consider the confidence set
\begin{equation}
\widehat{C}_\alpha(\bs{\xi}) = \left\{ \bs{h} \in \ell_2\ :\ \Vert \bs{h} - \hat{\bs{h}} \Vert \leq \sqrt{z_\alpha \tau(\bs{h}, \hat{\bs{h}}) + \widehat{R}(\bs{\xi}^{(2)}, \hat{\bs{h}})} \right\} \nonumber
\end{equation}
with standard normal quantile $z_\alpha$, and $\hat{\bs{h}} = \hat{\bs{h}}(\bs{\xi}^{(1)})$ such that $\bs{\xi}_k^{(1)}, \bs{\xi}_k^{(2)} \sim N(h_k \bs{1}_S, 2\bs{V}_k)$ with $\bs{\xi}^{(1)}_k, \bs{\xi}^{(2)}_k$ independent for all $k=1,\ldots,T$. Then,
\begin{equation}
\liminf_{S\to \infty} \inf_{\bs{h} \in \ell_2} P( \bs{h} \in \widehat{C}_\alpha(\bs{\xi})) \geq 1-\alpha \nonumber
\end{equation}
\end{theorem}
\vspace{1mm}
The validity of the asymptotic confidence regions relies on the condition $\Vert \bs{\Gamma}_k \Vert_1/\Vert \bs{\Gamma}_k \Vert_F \to 0$, which  requires the maximum absolute row sum (or column sum by symmetry) of $\bs{\Gamma}_k$ to be dominated by its Frobenius-norm for increasing $S$. This condition implies that the number of relevant principal components of the matrix $\bs{\Gamma}_k$ is increasing, or in other words, the vector of eigenvalues of $\bs{\Gamma}_k$ should not be dominated by one or a few large values as $S$ increases. Although in a somewhat different spirit than the condition $\Vert \bs{G}_S \Vert_F = o(S)$, this condition also implies that the effective number of independent replicates should increase with the total number of replicates $S$. Note that considering a vector of equal weights $\bs{w}_{k} = (\frac{1}{S},\ldots, \frac{1}{S})'$, this condition can be restated in terms of the between-replicate correlation matrix as $\Vert \bs{G}_S \Vert_1/ \Vert \bs{G}_S \Vert_F \to 0$ for $S \to \infty$. This ratio has an optimal rate $1/\sqrt{S}$ when $\bs{G}_S$ is equal to the identity matrix, and it can be verified that this condition implies $\Vert \bs{G}_S \Vert_F = o(S)$, (the other direction does not hold).\\[3mm]
In Theorem \ref{thm:3} we have constructed asymptotic confidence regions only for the sequence of fixed effect coefficients in the wavelet coefficient domain, however by the $\ell_2$-normalization of the wavelet basis we have that $\frac{1}{\sqrt{T}}\Vert \bs{h}^f - \hat{\bs{h}}^f \Vert = \Vert \bs{h} - \hat{\bs{h}} \Vert$, thus we can consider the scaled confidence regions in the frequency domain given by,
\begin{equation}
\widehat{C}^f_{\alpha}(\bs{\xi}) = \left\{ \bs{h}^f \in \ell_2 : \Vert \bs{h}^f - \hat{\bs{h}}^f \Vert \leq \sqrt{T}\sqrt{z_\alpha \tau(\bs{h}, \hat{\bs{h}}) + \widehat{R}(\bs{\xi}^{(2)}, \hat{\bs{h}}}) \right\} \nonumber
\end{equation}
and by Theorem \ref{thm:3}, the asymptotic coverage probability also satisfies,
\begin{equation}
\liminf_{S \to \infty} \inf_{\bs{h}^f \in \ell_2} P(\bs{h}^f \in \widehat{C}_{\alpha}^f(\bs{\xi})) \geq 1 - \alpha \nonumber
\end{equation}
\begin{remark} 
Note that the confidence regions are constructed under the assumption that the covariance matrices $\bs{V}_k$ are known. In practice, these covariance matrices are unknown, and we therefore replace them by plug-in estimators $\widehat{\bs{V}}_k$. The generalized least squares weights $\bs{w}_{k}$, which typically also depend on the covariance matrices $\bs{V}_k$, can be replaced for instance by the sub-optimal ordinary least squares weights $\bs{w}_{k} = (\frac{1}{S}, \ldots, \frac{1}{S})'$. This does not change the asymptotic normality result of the estimator $\widehat{R}(\bs{\xi}^{(2)}, \hat{\bs{h}})$ in the proof of Theorem \ref{thm:3}, but it comes at the cost of increasing its variance, thereby increasing the radius of the confidence regions. 
\end{remark}
\section{Simulated data examples} \label{sec:6}
In this section, we assess the finite-sample performance of the developed estimators by some simulated data examples. In Algorithm 1 below, we describe a procedure to simulate replicated time series $\{ X_s(t), s=1,\ldots,S \}$ with random log-spectra $H_s^f$ by means of their discrete Cram\'er representations. In short, given a wavelet basis $\mathcal{B}$, population-mean transfer function $a^f$ (with population-mean log-spectrum $h^f(\omega) = \log(|a^f(\omega)|^2)$), within-replicate covariance matrix $\bs{G}_T$, and between-replicate correlation matrix $\bs{G}_S$, we generate replicate-specific random transfer functions $A^f_s(\omega)$ which are inserted into discrete Cram\'er representations to generate the replicated time series. 
\begin{algorithm}[H] 
\begin{small}
\caption{(Generating replicated time series)}
\begin{algorithmic}[1]
\State \quad $\bs{h}^f \gets \log(\te{Mod}(\bs{a}^f)^2)$
\State \quad $\bs{h} \gets \te{DWT}_{\mathcal{B}}(\bs{h}^f)$,\ \  the discrete wavelet transform w.r.t. the basis $\mathcal{B}$.
\State \quad $\bs{U} \gets \bs{0}_{S \times T}$,\ \ an $(S \times T)$-matrix of zeroes.
\State \quad \textbf{For} $k = 1,\ldots, T$,
\State \quad \quad \textbf{if} $\sigma_{uk}^2 > 0$ \textbf{then} put $\bs{U}_{[,k]} \sim N(\bs{0}, \sigma_{uk}^2 \bs{G}_S)$
\State \quad \textbf{For} $s=1,\ldots, S$,
\State \quad \quad $U_s^f \gets \te{I-DWT}_{\mathcal{B}}(\bs{U}_{[s,]})$,\ \ the inverse discrete wavelet transform w.r.t. the basis $\mathcal{B}$.
\State \quad \quad $A_s^f \gets \bs{a}^f \sqrt{\exp(U_s^f)}$
\State \quad \quad $X_{s}(t) \gets \frac{1}{\sqrt{2T}} \sum_{\ell = -(T-1)}^T A_s^f(\omega_\ell) \exp(i 2\pi \omega_{\ell} t) \xi_{\ell}^s$
\end{algorithmic}
\end{small}
\end{algorithm}
\noindent In Algorithm 1, $\xi_\ell^s$ denotes a complex-valued normal random variable with independent real and imaginary parts, such that $\xi^s_\ell = \xi^{s*}_{-\ell}$.
\subsection{Population-mean log-spectrum}\label{sec:6.1}
\begin{figure}
\centering
\includegraphics[width=1\linewidth]{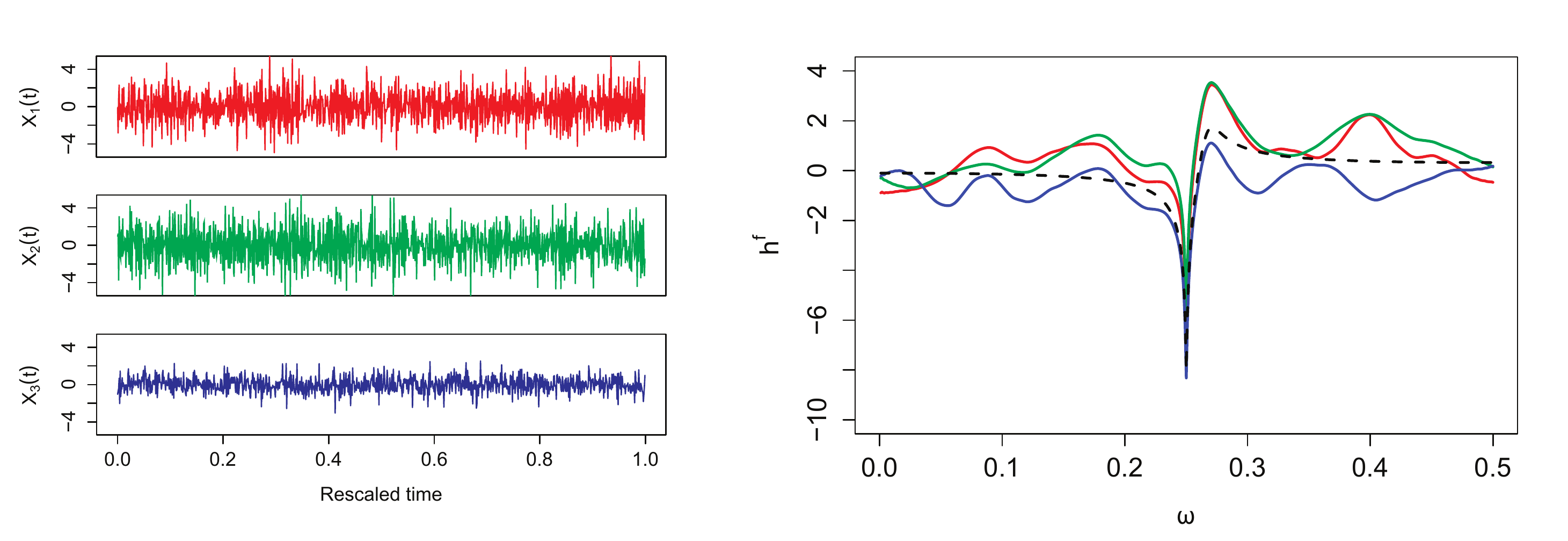}
\caption{Simulated replicated time series (left), and corresponding replicate-specific log-spectra, with underlying population-mean log-spectrum (right).} \label{fig:1}
\end{figure}

We consider data generated under a single population-mean log-spectrum $h^f$ coming from an $\te{ARMA}(2,2)$ process with parameters $\bs{\phi} = (-0.2,-0.9)$, $\bs{\theta} = (0,1)$ and white noise variance $\sigma_w^2=1$. The log-spectrum of this $\te{ARMA}(2,2)$ process is particularly difficult to estimate due to some sharp local features. 
In the right image of Figure \ref{fig:1} (dashed line), the considered population-mean log-spectrum $\bs{h}^f = (h^f(\omega_0), \ldots, h^f(\omega_{T-1}))'$ is shown for $\omega_\ell \in [0,1/2)$ with $T = 1024$. In fact, the displayed curve is a relatively sparse $\ell_0$-approximation of $\bs{h}^f$ under a Daubechies extremal-phase wavelet basis $\mathcal{B}$ with $N = 6$ vanishing moments, where we have thresholded all coefficients with $|h_k| < T^{-1}$. Here, we have used the \texttt{WaveThresh} package in \texttt{R}, see \cite[Chapter 2]{N10} for more details.
\subsection{Random effects covariance matrices}\label{sec:6.2}
\begin{figure}
\centering
\includegraphics[width=1\linewidth]{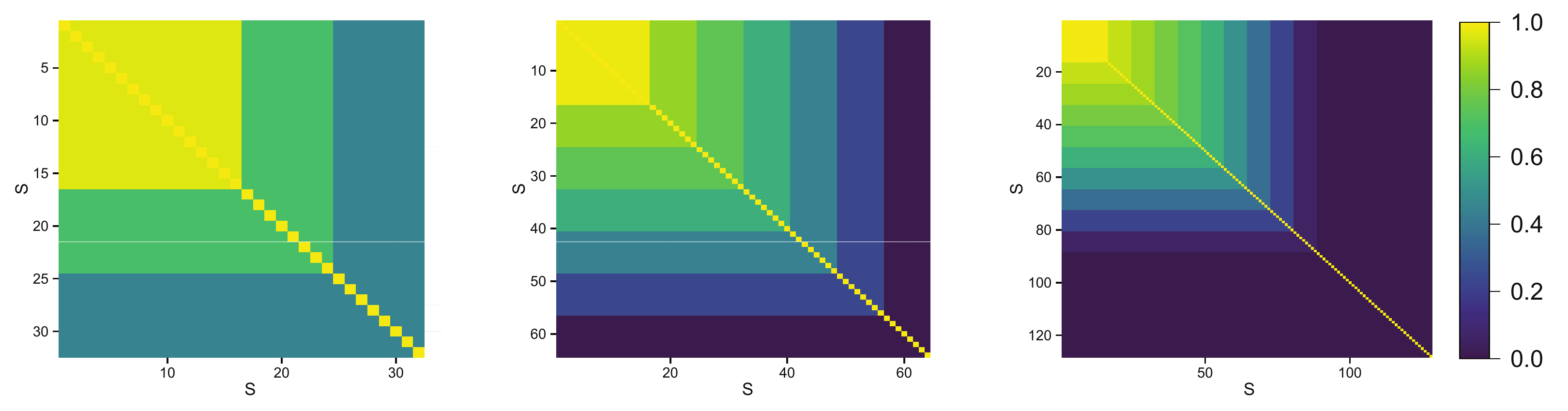}
\caption{Contour correlation matrices $\bs{G}_S$ for $S = \{ 32, 64, 128 \}$.}
\label{fig:2}
\end{figure}

For the $T \times T$-dimensional covariance matrix $\bs{G}_T$, we consider the diagonal matrix $\te{diag}(\bs{\sigma}_u^2)$ with set of indices of non-zero variance components given by,
\begin{equation}
K_{u,T} = \{ k \in K_{h,T}\, :\, \lfloor \log_2(k-1) \rfloor < J \} \nonumber
\end{equation}
which are simply all the indices contained in the wavelet scales $\{ j\, :\, 0\leq j < J \}$ with the additional constraint that $K_{u,T} \subseteq K_{h,T}$. For the magnitudes of the variance components, we consider $\sigma_{uk}^2$ decaying with a factor 2 per increasing wavelet scale, i.e. for some constant $C > 0$, let $\sigma_{u1}^2 = C$ and define for $k > 1$,
\begin{equation}
\sigma_{uk}^2 = \left( C \cdot 2^{-\lfloor \log_2(k-1) \rfloor - 2} \right) \bs{1}\left\{ k \in K_{u,T} \right\} \nonumber
\end{equation}
Under this specific model, Figure \ref{fig:1} shows generated random log-spectra for three replicates (two of which are highly correlated) and corresponding simulated replicate-specific time series, using a Daubechies extremal-phase wavelet basis with $N=6$ vanishing moments and parameters $C=0.5$, $J=4$, which are also the values used in the subsequent simulation studies. \\[3mm]
For the between-replicate $S\times S$-dimensional correlation matrix $\bs{G}_S$ we consider two different scenarios:
\begin{enumerate}
\item A symmetric block-diagonal matrix containing a single $(S/2 \times S/2)$ dimensional block of highly correlated replicates with $\rho_{ij} = 0.9$ for $1 \leq i,j \leq S/2$. The constructed correlation matrix satisfies $\Vert \bs{G}_S \Vert_F = o(S)$ and is positive-semidefinite.
\item A symmetric contour-matrix that consists of layers of block matrices with decaying levels of correlation, see Figure \ref{fig:2}. The layers are chosen such that again $\Vert \bs{G}_S \Vert_F = o(S)$ and $\bs{G}_S$ is positive-semidefinite. The correlation matrix $\bs{G}_S$, for $S \geq 16$ dyadic, is constructed as follows. Divide an $S \times S$-identity matrix into blocks $B_{ij}$ of size $8 \times 8$ with $1 \leq i,j \leq S/8$. Fix $j \geq 2$, for $i < j$, set all elements of $B_{ij}$ equal to $ \{ 1 - \frac{j^2}{S} \}_+$, and do the same for $B_{jj}$ only for its off-diagonal elements. Similarly, fixing $i \geq 2$, for $j < i$, set all elements of $B_{ij}$ equal to $\{ 1 - \frac{i^2}{S} \}_+$, and finally put $B_{11} = B_{22}$.
\end{enumerate}
\subsection{Simulation study}\label{sec:6.3}
\begin{table}
\caption{Average squared errors (and standard errors $\hat{\sigma}/M$) of the estimates of $\bs{h}^f$ ($\times 10^{-1})$, $\bs{G}_T$ ($\times 10^{-4}$), and $\bs{G}_S$ ($\times 10^{-1}$). \label{tab:1}}
\centering 
\resizebox{\textwidth}{!}{\fbox{%
\begin{tabular}{llccccccc}
&&&& &&&& \\[-1mm]
& & \multicolumn{3}{c}{$T=512$} & \ & \multicolumn{3}{c}{$T=1024$} \\[0mm]
\cmidrule(lr{0.9em}){3-5} \cmidrule(lr{0.9em}){7-9} \\[-2mm]
$S$ & Approach & $\te{ASE}(\hat{\bs{h}}^f)$ & $\te{ASE}(\widehat{\bs{G}}_T)$ & $\te{ASE}(\widehat{\bs{G}}_S)$ & \ & $\te{ASE}(\hat{\bs{h}}^f)$ & $\te{ASE}(\widehat{\bs{G}}_T)$ & $\te{ASE}(\widehat{\bs{G}}_S)$ \\[1mm]
\hline \\[2mm]
\multicolumn{5}{l}{\textbf{Block-diagonal correlation matrix $\bs{G}_S$ (Case (a))}} &&&&\\[2mm]
32 & OLS & 2.74 (0.05) & - & - & \ & 2.53 (0.05) & - & -  \\
& Non-adaptive & 2.15  (0.04) & 0.65 (0.02) & 1.29 (0.01) & \ & 2.12 (0.04) & 0.31 (0.01) & 1.24 (0.01) \\
& Adapt. ($q=0.1$) & 2.10 (0.04) & 0.66 (0.02) & 1.30 (0.01) & \ & 2.08 (0.04) & 0.31 (0.01) & 1.24 (0.01) \\
& Adapt. ($q=0.001$) & 2.15 (0.05) & 0.64 (0.02) & 1.31 (0.01) & \ & 2.12 (0.04) & 0.31 (0.01) & 1.25 (0.01) \\
& Oracle ($q=0.001$) & 1.18 (0.02) & 0.42 (0.01) & 0.78 (0.01) & \ & 1.05 (0.02) & 0.20 (0.01) & 0.69 (0.01) \\[3mm]
64 & OLS & 2.61 (0.05) & - & - & \ & 2.42 (0.05) & - & -  \\
& Non-adaptive & 1.70 (0.03) & 0.52 (0.02) & 1.38 (0.01) & \ & 1.76 (0.04) & 0.27 (0.01) & 1.36 (0.01) \\
& Adapt. ($q=0.1$) & 1.69 (0.04) & 0.52 (0.02) & 1.39 (0.01) & \ & 1.73 (0.04) & 0.28 (0.01) & 1.35 (0.01) \\
& Adapt. ($q=0.001$) & 1.75 (0.04) & 0.51 (0.02) & 1.41 (0.01) & \ & 1.74 (0.04) & 0.27 (0.01) & 1.36 (0.01) \\
& Oracle ($q=0.001$) & 0.86 (0.01) & 0.24 (0.01) & 0.80 (0.01) & \ & 0.74 (0.01) & 0.12 (0.005) & 0.72 (0.01) \\[3mm]
128 & OLS & 2.56 (0.05) & - & - & \ & 2.38 (0.05) & - & -  \\
& Non-adaptive & 1.57 (0.03) & 0.41 (0.01) & 1.56 (0.01) & \ & 1.61 (0.03) & 0.21 (0.01) & 1.55 (0.01)  \\
& Adapt. ($q=0.1$) & 1.56 (0.03) & 0.41 (0.01) & 1.57 (0.01) & \ & 1.61 (0.04) & 0.21 (0.01) & 1.56 (0.01) \\
& Adapt. ($q=0.001$) & 1.58 (0.03) & 0.40 (0.01) & 1.58 (0.01) & \ & 1.62 (0.04) & 0.21 (0.01) & 1.57 (0.01) \\
& Oracle ($q=0.001$) & 0.69 (0.01) & 0.23 (0.01) & 0.96 (0.01) & \ & 0.58 (0.01) & 0.09 (0.01) & 0.90 (0.01) \\[5mm] 
\multicolumn{5}{l}{\textbf{Contour correlation matrix $\bs{G}_S$ (Case (b))}} &&&&\\[2mm]
32 & OLS & 7.21 (0.16) & - & - & \ & 7.24 (0.16) & - & -  \\
& Non-adaptive & 6.64  (0.15) & 1.95 (0.03) & 4.34 (0.03) & \ & 6.45 (0.15) & 1.04 (0.02) & 4.01 (0.03) \\
& Adapt. ($q=0.1$) & 6.60 (0.15) & 1.93 (0.03) & 4.35 (0.03) & \ & 6.41 (0.15) & 1.04 (0.02) & 4.03 (0.03) \\
& Adapt. ($q=0.001$) & 6.56 (0.15) & 1.92 (0.03) & 4.31 (0.03) & \ & 6.35 (0.15) & 1.03 (0.02) & 4.05 (0.03) \\
& Oracle ($q=0.001$) & 5.95 (0.13) & 1.26 (0.03) & 3.40 (0.02) & \ & 5.85 (0.13) & 0.76 (0.02) & 2.98 (0.01) \\[3mm]
64 & OLS & 4.61 (0.10) & - & - & \ & 4.60 (0.10) & - & -  \\
& Non-adaptive & 2.66 (0.06) & 1.07 (0.06) & 1.95 (0.02) & \ & 2.65 (0.06) & 0.54 (0.02) & 1.93 (0.02) \\
& Adapt. ($q=0.1$) & 2.64 (0.06) & 1.05 (0.05) & 1.95 (0.02) & \ & 2.63 (0.06) & 0.54 (0.02) & 1.94 (0.02) \\
& Adapt. ($q=0.001$) & 2.67 (0.06) & 1.05 (0.05) & 1.98 (0.02) & \ & 2.66 (0.06) & 0.53 (0.02) & 1.95 (0.02) \\
& Oracle ($q=0.001$) & 1.48 (0.03) & 0.53 (0.02) & 0.99 (0.01) & \ & 1.35 (0.03) & 0.24 (0.01) & 0.85 (0.01) \\[3mm]
128 & OLS & 2.70 (0.05) & - & - & \ & 2.64 (0.05) & - & -  \\
& Non-adaptive & 1.72 (0.03) & 0.84 (0.07) & 1.34 (0.01) & \ & 1.65 (0.03) & 0.37 (0.03) & 1.30 (0.01)  \\
& Adapt. ($q=0.1$) & 1.71 (0.03) & 0.84 (0.07) & 1.35 (0.01) & \ & 1.64 (0.03) & 0.37 (0.03) & 1.30 (0.01) \\
& Adapt. ($q=0.001$) & 1.72 (0.03) & 0.85 (0.07) & 1.36 (0.01) & \ & 1.67 (0.04) & 0.36 (0.03) & 1.31 (0.01) \\
& Oracle ($q=0.001$) & 0.75 (0.01) & 0.28 (0.01) & 0.91 (0.01) & \ & 0.64 (0.01) & 0.11 (0.005) & 0.85 (0.01)\\[2mm] 
\end{tabular}}}
\end{table} 
\begin{table}
\caption{Empirical coverage ($\times 10^2$) of confidence regions for $\bs{h}^f$. \label{tab:2}}
\centering
\resizebox{\textwidth}{!}{\fbox{%
\begin{tabular}{lccccccccc}
&&&&& &&&&\\[-1mm]
& \multicolumn{4}{c}{$\alpha=0.05$} & \ & \multicolumn{4}{c}{$\alpha=0.1$}\\[0mm]
\cmidrule(lr{0.9em}){2-5} \cmidrule(lr{0.9em}){7-10} \\[-2mm]
& \multicolumn{2}{c}{$T=512$} & \multicolumn{2}{c}{$T=1024$} & \ & \multicolumn{2}{c}{$T=512$} & \multicolumn{2}{c}{$T=1024$} \\[0mm]
\cmidrule(lr{0.9em}){2-3} \cmidrule(lr{0.9em}){4-5} \cmidrule(lr{0.9em}){7-8} \cmidrule(lr{0.9em}){9-10} \\[-2mm]
 & $S=64$ & $S=128$ &  $S=64$ & $S=128$ & \ & $S=64$ & $S=128$ & $S=64$ & $S=128$ \\[1mm]
\hline\\[2mm]
\multicolumn{5}{l}{\textbf{Block-diagonal correlation matrix $\bs{G}_S$ (Case (a))}} &&&&&\\[2mm]
Asymptotic (Scen. 1) & 95.1 & 93.5 & 95.6 & 93.7 & \ & 88.7 & 86.0 & 88.5 & 87.2\\
Asymptotic (Scen. 2) & 96.7 & 96.5 & 98.3 & 98.2 & \ & 86.1 & 84.4 & 89.8 & 89.7\\
Bootstrap & 99.1 & 99.8 & 96.7 & 97.5 & \ & 98.2 & 99.2 & 92.7 & 95.1 \\[5mm]
\multicolumn{5}{l}{\textbf{Contour correlation matrix $\bs{G}_S$ (Case (b))}} &&&&&\\[2mm]
Asymptotic (Scen. 1) & 99.5 & 96.4 & 99.6 & 96.7 & \ & 96.9 & 90.3 & 97.2 & 91.1\\
Asymptotic (Scen. 2) & 97.2 & 97.2 & 98.4 & 98.5 & \ & 86.4 & 85.6 & 90.5 & 90.1\\
Bootstrap & 97.3 & 99.5 & 95.5 & 96.2 & \ & 94.4 & 99.0 & 89.6 & 93.5\\[2mm]
\end{tabular}}} 
\end{table}
We assess the performance of the proposed estimation procedure and compare this to the performance of several related alternatives. First, we consider a naive ordinary least squares approach (OLS), where we smooth the replicate-specific log-periodograms using FDR thresholding with tuning parameter $q = 0.001$ and estimate $\bs{h}^f$ by averaging the smoothed curves over replicates, thereby not taking into account the between-replicate dependence structure. Second, we consider the non-adaptive iterative-generalized least squares approach, where the smoothness space parameter $k_{h,T}$ is assumed to be known and the set $K_{h,T}$ is estimated using the universal threshold $\lambda_{h,T}$ as in Section \ref{sec:3.1}. Third, we consider an adaptive iterative-generalized least squares approach, in which $k_{h,T}$ is not known. In order to estimate the set $K_{h,T}$ we use FDR thresholding with tuning parameters $q=\{ 0.1, 0.001 \}$. Finally, in order to assess the increase in estimation error due to the iteration scheme, we also compute oracle estimators $\hat{\bs{h}}^f$ and $\widehat{\bs{G}}_S(\hat{\bs{h}})$, where we assume the true generalized least squares-weights (depending on $\bs{G}_S$) to be known in estimating $\bs{h}^f$, so that no iteration of the estimators is required. Note that in this final scenario we do not assume knowledge of $k_{h,T}$ nor $k_{u,T}$, and the set $K_{h,T}$ is estimated as in the third scenario with FDR tuning parameter $q = 0.001$. The performance of the estimator $\hat{\bs{h}}^f = (\hat{h}^f(\omega_0), \ldots, \hat{h}^f(\omega_{T-1}))'$ is assessed through the squared error averaged over Fourier frequencies. Similarly, the performance estimators $\widehat{\bs{G}}_T$ and $\widehat{\bs{G}}_S$ is evaluated through the squared error averaged over matrix elements.\\
In Table \ref{tab:1} we show average squared errors with in parenthesis corresponding standard errors ($\hat{\sigma}/\sqrt{M}$) for $M=1000$ replicated simulation experiments. It should not come as a surprise that knowledge of the true generalized least squares-weights significantly improves the estimation error. This is seen by comparing the estimation error for $\bs{h}^f$ of the naive ordinary least squares approach, which does not take into account the between-replicate correlation structure, with that of the oracle estimator. The iterative-generalized least squares scheme then inflates the estimation error for $\bs{h}^f$ relative to the oracle estimator, but still outperforms the ordinary least squares approach under all of the considered scenarios. Another important observation is that the performance of the adaptive estimators --regardless of the choice of the FDR tuning parameter-- is similar to the performance of the nonadaptive estimators in essentially all of the considered scenarios. The adaptive estimators slightly outperform the nonadaptive estimators in some cases, which is most likely due to the fact that the asymptotic minimax threshold $\lambda_{h,T}$ is somewhat conservative in a finite-sample situation. \\
\begin{figure}
\centering
\includegraphics[width=1\linewidth]{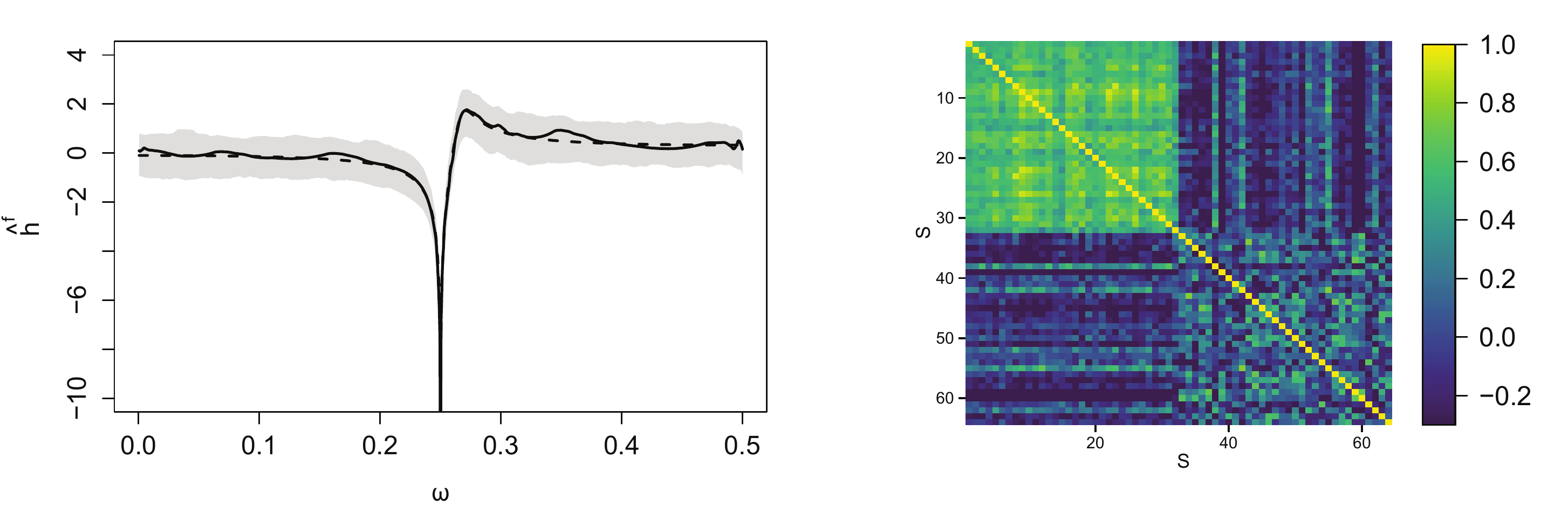}
\vspace{-4mm}
\caption{Estimates of $\bs{h}^f$ (solid black line) and $\bs{G}_S$ for a single simulation experiment, with $(\hat{\bs{h}}^f_{0.01}, \hat{\bs{h}}^f_{0.99})$-empirical pointwise quantiles for 1000 repititions of the experiment.} \label{fig:3}
\end{figure}
\begin{figure}
\centering
\includegraphics[width=1\linewidth]{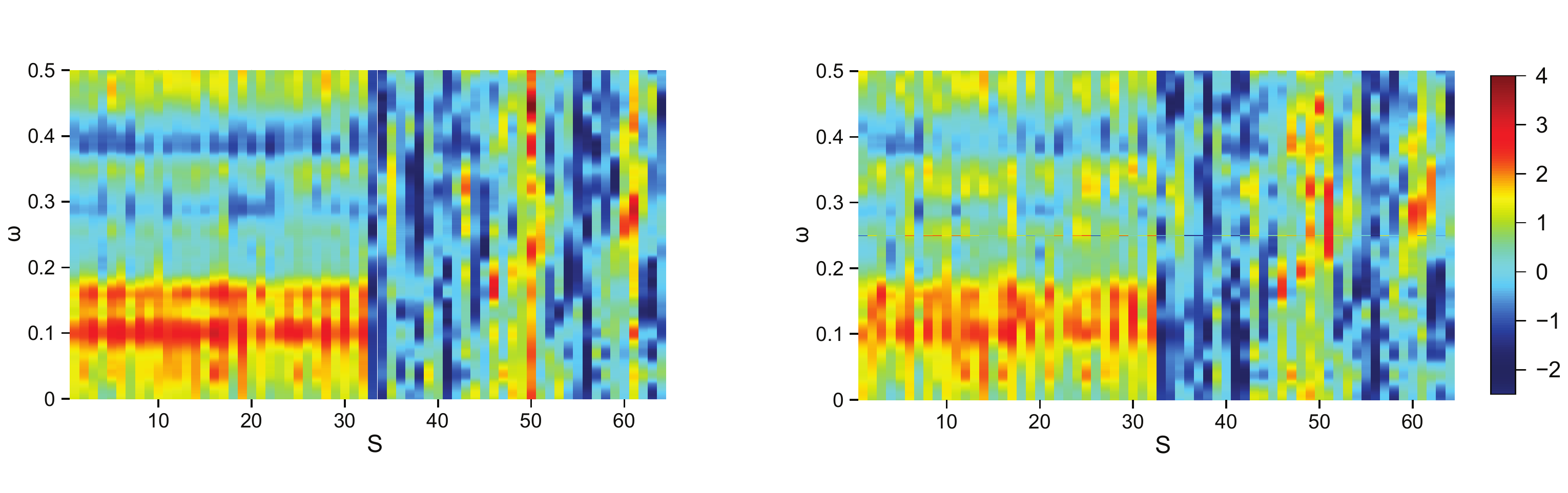}
\vspace{-6mm}
\caption{True random effects curves $U_s^f(\omega)$ (left) and predicted random effects curves $\widehat{U}_s^f(\omega)$ (right) for a single simulation experiment.} \label{fig:4}
\end{figure}
In Figure \ref{fig:3} and \ref{fig:4}, we give some visual representations of the estimates under a block-diagonal between-replicate correlation structure ($T = 512$, $S = 64$). Figure \ref{fig:3} shows the estimated population-mean log-spectrum and between-replicate correlation matrix for a single simulation experiment using the adaptive approach with FDR tuning parameter $q=0.001$, and Figure \ref{fig:4} shows true and predicted random effects curves for a single simulation experiment under the same scenario.

\subsection{Confidence region coverage}\label{sec:6.4}
We also assess the validity of the constructed confidence regions for $\bs{h}^f$ by computing their empirical coverage under some of the simulated models considered before. Since we are interested in the (negative) impact on the empirical coverage caused by replacing the true covariance matrices $\bs{V}_k$ by estimates $\widehat{\bs{V}}_k$ we consider two different scenarios. In the first scenario, we assume the true matrices $\bs{V}_k$ to be known, both in performing the sample splitting procedure and in constructing the confidence regions. In the second scenario, we consider the matrices $\bs{V}_k$ to be unknown, therefore replacing them by plug-in estimators $\widehat{\bs{V}}_k$. For the weight vectors $\bs{w}_k$, we consider ordinary least squares weights,  equally weighting each replicate. As a benchmark procedure, we also compute the empirical coverage of parametric bootstrap confidence regions for $\bs{h}^f$ with $B=1000$ bootstrap samples as proposed in \cite{KHG11} and \cite{FO16}. The bootstrap confidence regions are constructed using the true covariance matrices $\bs{V}_k$, and in this sense are oracle confidence regions, that should be compared to the asymptotic confidence regions under the first scenario. In Table \ref{tab:2}, the empirical coverage for the different approaches is shown for $M = 5000$ replicated simulation experiments ($M =1000$ for the bootstrap confidence regions),  the simulation results are shown only for the adaptive approach (i.e. $k_{h,T}$ unknown) using FDR tuning parameter $q =0.001$, the results for the other approaches considered in Table \ref{tab:1} are similar. 
\section{Analysis of brain signals: replicated LFP time series} \label{sec:7}
\begin{figure}
\centering
\includegraphics[width=1\linewidth]{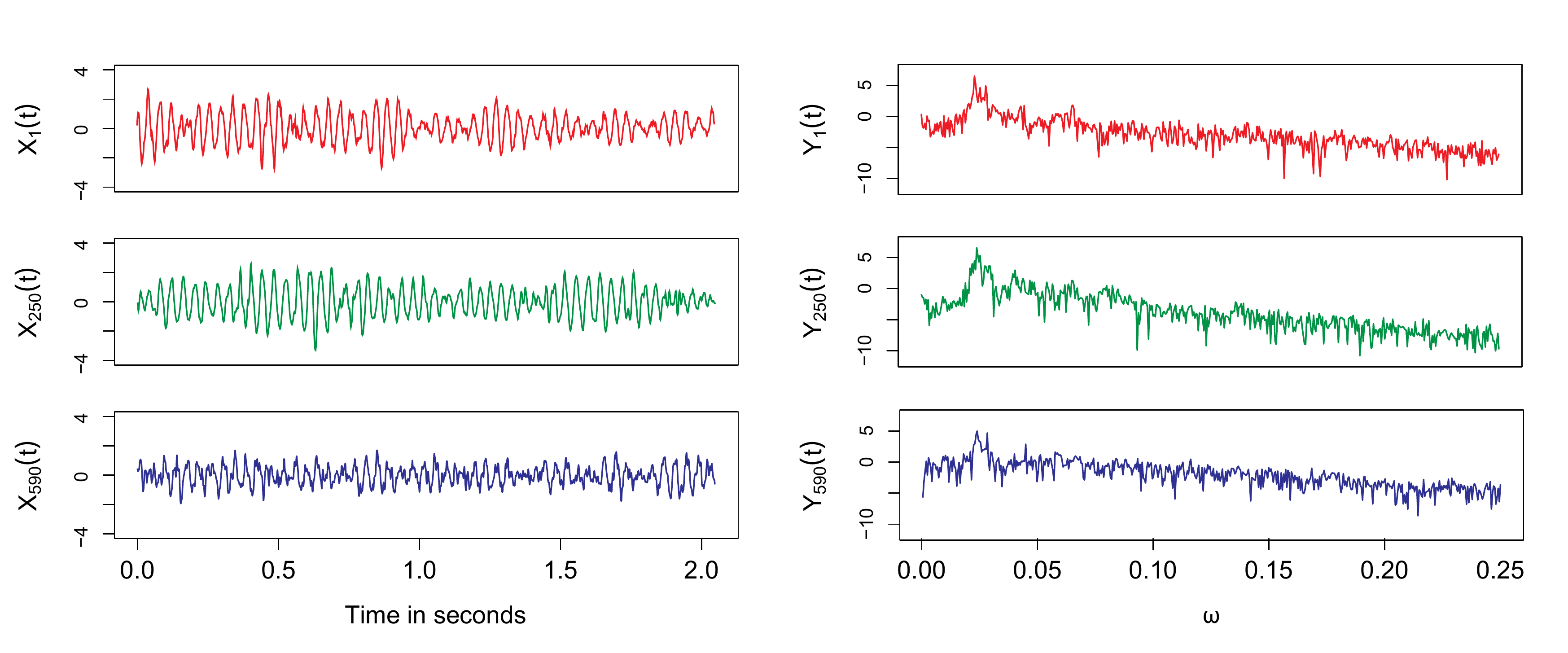}
\caption{LFP time series (left) and corresponding raw log-periodograms (right) for trials 1, 250, and 590.} \label{fig:5}
\end{figure}
\begin{figure}
\centering
\includegraphics[width=1\linewidth]{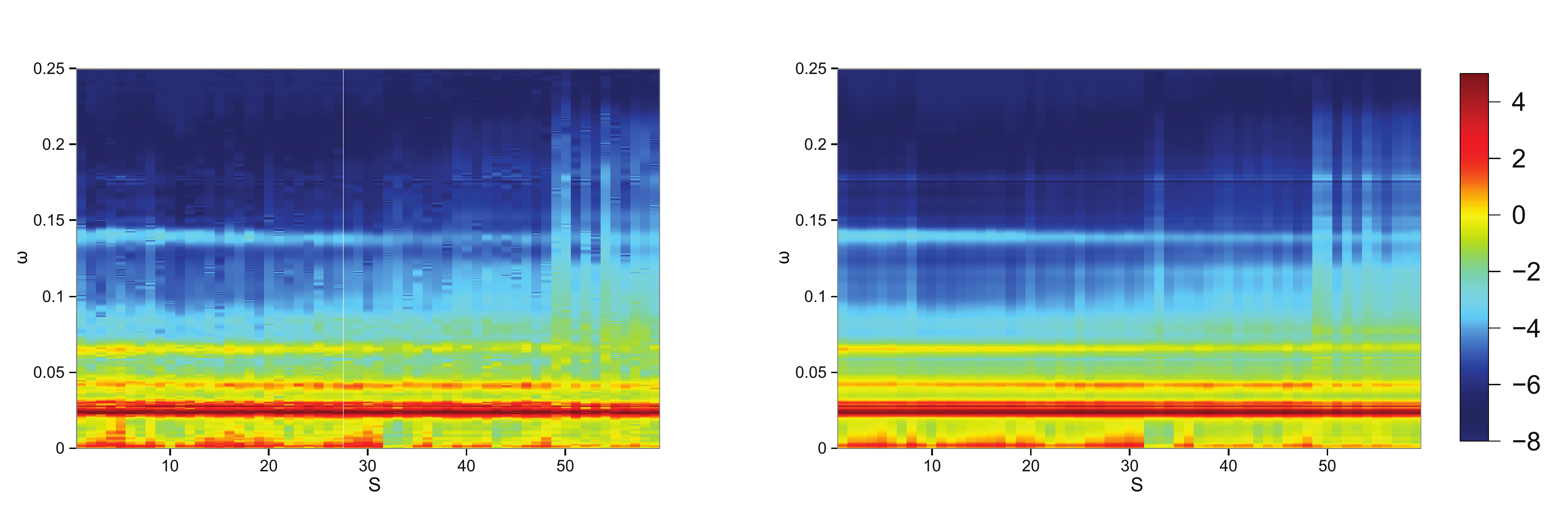}
\vspace{-5mm}
\caption{Individually smoothed log-periodograms (left) and predicted trial-specific log-spectra $\hat{\bs{h}}_s^f$ (right).} \label{fig:7}
\end{figure}

To conclude, we analyze brain signal data recorded during an associative learning experiment. The dataset consists of local field potential (LFP) time series traces, measuring electrical activity in the brain over the course of the experiment, see \cite{G12} and \cite{FO16} for a more detailed description. During the experiment, a male macaque learns the association between one set of objects (four pictures) and another set of objects (doors located in four different quadrants of the visual field) by means of trial-and-error. In each trial, the macaque was first presented with a picture and then required to select one of the four doors. Each time the macaque made the correct association it was given a reward in the form of a small quantity of juice. Over the course of the associative learning experiment, the electrical activity in the brain of the macaque was measured using local field potentials. Local field potentials measure electrical activity in the brain directly via chronically implanted probes, in contrast to other commonly-used non-invasive recording techniques, such as electroencephalograms (EEGs) and functional magnetic resonance imaging (fMRI). In this analysis, we consider univariate local field potential time series data recorded in the nucleus accumbens (NAc) region, which is a region in the brain that has been demonstrated to be highly implicated in cognitive processes involving memory and reward. After preprocessing of the LFP time series data, there remains a total of 590 (univariate) time series traces of length 2048 sampled at 1000 Hz, thus roughly corresponding to 2 seconds of data. \\
The goal of the analysis is to study trial-population spectral characteristics, and in particular the evolving spectral properties of the time series over the course of the experiment. In Figure \ref{fig:5}, we show the recorded LFP time series for three trials (start, middle, and end of the experiment) and their corresponding raw log-periodograms, which show clear common frequency behavior across the three different trials. The left image of Figure \ref{fig:7} shows initial smoothed log-periodograms across all the trials in the experiment, using FDR thresholding with tuning parameter $q=0.001$, without taking into account the between-trial dependence structure. Note that instead of the individually smoothed log-periodograms, we show blockwise-average log-periodograms over 10 adjacent individual trials in order to improve visibility of the image. \\
In general, the log-periodograms in Figure \ref{fig:7} (left) display very common frequency behavior across trials, however towards the end of the experiment the overall power in the middle- and high-frequency range ($\omega > 0.1$) seems to increase (except for the frequency-band around $\omega \approx 0.14$), and we also note that power in the very low-frequency range ($\omega$ close to zero) fades out after approximately half of the trials. Simply averaging the spectral estimates across all trials will not take into account, nor give us any information, about the dependence structure of the underlying brain dynamics over the course of the experiment. Therefore, we need a model that allows for explicit correlation between trial-replicates in the population.
\begin{figure}[H]
\centering
\includegraphics[width=1\linewidth]{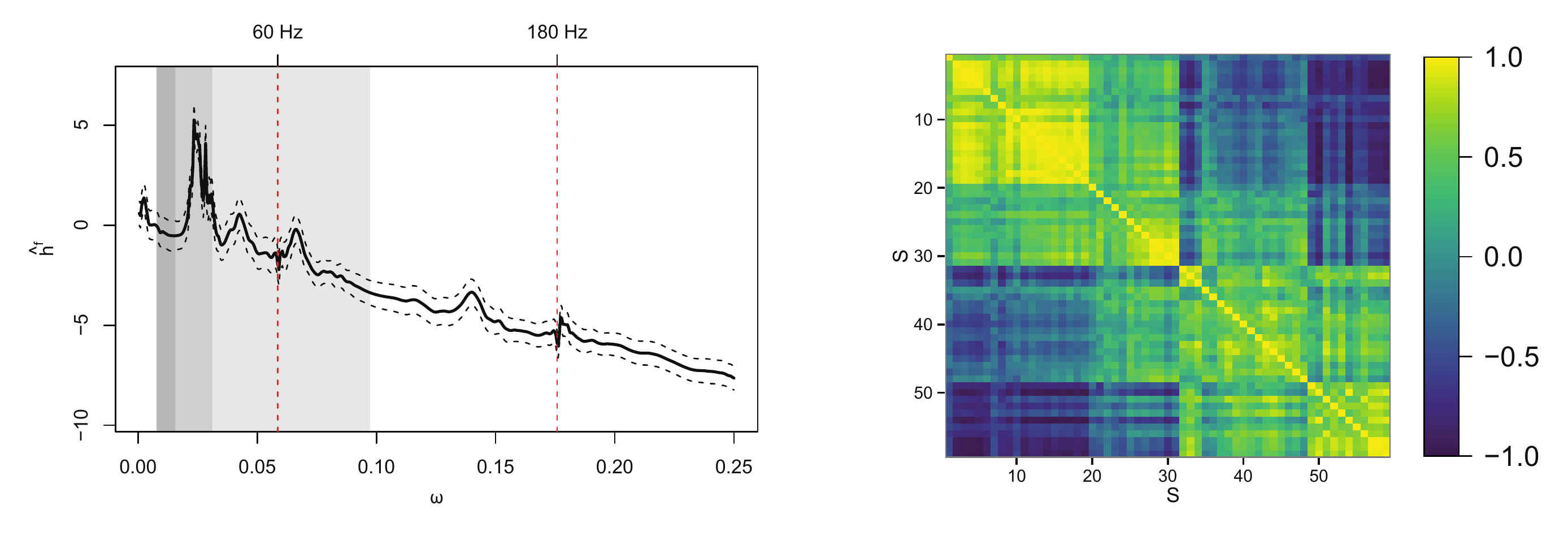}
\vspace{-6mm}
\caption{Estimates of the population-mean log-spectrum $\bs{h}^f$ (left) and between-trial correlation matrix $\bs{G}_S$ (right).} \label{fig:6}
\end{figure}
In the data analysis we consider frequency content up to 256 Hz ($\omega = 0.25$), since higher frequency behavior is typically attributed to noise and not physiological behavior in the brain, and we do not want the estimated between-trial correlation matrix to be dominated by this very high-frequency content ($\omega > 0.25$). In the left image of Figure \ref{fig:6}, we show the estimated population-mean log-spectrum $\bs{h}^f = (h^f(\omega_0), \ldots, h^f(\omega_{511}))'$, with in grey several regions of interest for the neuroscientist: the $\alpha$-band (8-16 Hz), the $\beta$-band (16-32 Hz), $\gamma$-band (32-100 Hz). The black dotted lines correspond to the variability (square root of the diagonal of the estimate of $\bs{G}_T^f = \bs{W}_\mathcal{B}'\bs{G}_T\bs{W}_\mathcal{B}$) of the random effects curves in the frequency domain. Note that there are two small dips at 60 Hz and 180 Hz, these are artifacts remaining after the application of two Butterworth band-stop filters in order to filter out the power line frequency around 60 Hz (in North America) and one of its harmonics at 180 Hz. The right image in Figure \ref{fig:6} shows the estimate of the between-trial correlation matrix $\bs{G}_S$ for blocks of 10 adjacent individual trials. Note that the estimated correlation matrix clearly demonstrates the correlation structure between the trials that we observed for the initial smoothed log-periodograms in Figure \ref{fig:7} (left). Trials at the beginning of the experiment are highly correlated and trials at the end of the experiment are highly correlated, and the correlation between trials decays as the lag between trials increases. This suggests that the trial-specific log-spectra evolve over the course of the learning experiment, as  already discussed in \cite{FO16}, and it is important to take this behavior into account in the data analysis in order to improve estimation of the population-mean curve and especially prediction of the replicate-specific curves, but also to avoid misleading results in any subsequent inference procedures. The right image of Figure \ref{fig:7} shows the predicted trial-specific log-spectra (again for blocks of 10 adjacent trials). Comparing the two images in Figure \ref{fig:7}, we observe that the predicted trial-specific log-spectra perform better in suppressing the overall noise than the individual smoothed log-periodograms, since the functional mixed-effects model pools information across different trials, on the other hand we are still able to capture most of the relevant features present for the individual smoothed log-periodograms. 
\section{Conclusion} \label{sec:8}
In the context of spectral analysis for replicated time series, where the focus is on population spectral characteristics rather than the behavior of individual time series, we propose to model replicate-specific log-spectra as random curves based on a nonparametric functional mixed effects approach. We address the specific problem of analyzing spectra that are characterized by localized peaks or troughs by successfully using projection estimators, and in particular nonlinear wavelet thresholding. Here we benefit both from a convenient linear mixed effects structure in the wavelet coefficient domain, and the possibility to constrain the complexity of this nonparametric estimation problem by natural $\ell_0$-sparsity constraints. The paradigm of $\ell_0$-driven sparsity leads to simple constraints ensuring that the replicate-specific curves and the population-mean curve share the same level of complexity (as in \cite{G02}), but also allows to come up with practical near-optimal threshold choices for both fixed- and random effects estimation. The good performance of this smoothing device is also confirmed by our empirical results. \\
As an additional important ingredient, we introduce a generic correlation model for the population of random effects curves, where both intra- and inter-subject correlation are modeled in a convenient nonparametric way in the wavelet coefficient domain. The importance of including correlated functions in the general setting of functional response regression has already been underlined by \cite{M14}. The author clearly expresses that the overwhelming majority of existing work cannot model correlation between curves and is hence only suitable for independently sampled functions. This is not realistic for many functional data problems, and can lead to estimators that are statistically inefficient, or even give misleading inferences. As an example of replicated time series data with explicit correlation between different replicates in the population, we analyze empirical brain signal data over the course of an associative learning experiment. There is a clear indication that the spectral behavior of trial-replicated time series evolves over the course of the experiment, and we are able to reproduce a meaningful correlation structure over time series replicates that demonstrates this evolutionary behavior over the course of the learning experiment.
We note that our fully nonparametric approach, although developed in the framework of spectral analysis for time series data, can equally well be applied in a general functional data analysis context in the presence of correlated random curves, where we benefit from the twofold adaptation properties of wavelets towards sparse and localized structures. \\
To conclude, we discuss two important directions in which to generalize the proposed model. First, replicated time series spectral analysis of brain data eventually calls for a \emph{multivariate} treatment in order to reveal dependence structures between different regions in the brain through cross-spectral analysis of different components of the multivariate time series (e.g. multivariate EEG time series data from different regions in the brain). Recent unpublished work by \cite{K16} treats the problem of analyzing associations between power spectra of multivariate time series and cross-sectional outcomes by an approach based on a tensor-product spline model, in frequency and outcome, of Cholesky components of outcome-dependent power spectra. However, to the  best of our knowledge, no quantitative analysis that embeds replicate-specific spectral matrices into a multivariate \emph{functional mixed effects model} exists so far, not even for the case of independent replicates. We are currently generalizing our functional mixed effects approach developed for replicated univariate time series to this more challenging setting. Second, there is considerable evidence (see e.g. \cite{OvSG05}) that for long EEG recordings, the second-order stationary assumption for the time series is too strong. It is preferable to weaken this assumption and to consider for instance a variance-covariance structure that slowly changes over time. In the context of an individual time series, time-varying spectral analysis is a challenging task since it can lead to estimators with extremely high variance (see e.g. \cite{NvS97}). We expect that the methodology presented here will become very efficient in this context, since it allows for pooling of information across the different time series replicates.
\section*{Acknowledgments}
We thank the UC Irvine Space-Time Modeling Group, and in particular Hernando Ombao for useful discussions regarding this work, and Dr. Emad Eskandar (Massachussetts General Hospital) for the local field potential data that was used to illustrate the methodology. This research is supported by IAP research network P7/06 of the Belgian government (Belgian Science Policy), and by the contract `Projet d'Actions de Recherche Concert\'ees' (ARC) No. 12/17-045 of the `Communaut\'e fran\c{c}aise de Belgique' granted by the Acad\'emie universitaire Louvain. The first author gratefully acknowledges funding from the Belgian Fund for Scientific Research (FRIA/FRS-FNRS).

\section*{Appendix: Proofs}

\section{Key components in the proof of Theorem \ref{thm:1}}
We outline the proof of Theorem \ref{thm:1} through several key lemmas. 
The first lemma gives the uniform asymptotic normality result for empirical wavelet coefficients of the log-periodogram ordinates and relates the $\ell_2$-risk of $\hat{\bs{h}}_\lambda(\bs{Y})$ under the sequence model in the wavelet coefficient domain in Section \ref{sec:2.1.2} to the $\ell_2$-risk of $\hat{\bs{h}}_\lambda(\bs{\xi})$ under an accompanying Gaussian sequence model. Here, $\lambda \geq 0$ is an arbitrary nonlinear threshold.
\begin{lemma}\label{II-lem:1} Under assumptions \ref{ass:1} and \ref{ass:3}, uniformly in $k \in J_{T,\alpha}$, for arbitrary $0 < \alpha < 1$,
\begin{equation}
P\left( (Y_{sk} - h_k)/\sigma_{k,T} \geq x_s \right) = (1 + o_T(1))(1 - \Phi(x_s)) \quad \quad \te{for all } s = 1,\ldots, S \nonumber
\end{equation}
with $-\infty < x_s \leq \Delta_T \sim T^\nu$ for some $\nu > 0$, where $\sigma_{k,T}^2 = \sigma_{uk}^2 + \sigma_e^2/T$. Furthermore,
\begin{equation}
\sum_{k \in J_{T,\alpha}} \mathbb{E}[(\hat{h}_{k,\lambda}(\bs{Y}_{\cdot k}) - h_k)^2] = (1 + o_T(1)) \sum_{k \in J_{T,\alpha}} \mathbb{E}[(\hat{h}_{k,\lambda}(\bs{\xi}_k) - h_k)^2] + O(S^{-\mu}T^{-\mu+1}) \nonumber
\end{equation}
for arbitrary $0 < \mu < \infty$, and with $\bs{\xi}_k \sim N(h_k\bs{1}_S, \bs{V}_k)$ Gaussian random vectors.
\end{lemma}
This lemma implies that it suffices to derive an $\ell_2$-risk upper bound of $\hat{\bs{h}}(\bs{Y})$ under the accompanying Gaussian sequence model.  In the lemma below, we derive exact expressions of the mean-squared errors of $\hat{h}_k(\bs{\xi}_k)$ with respect to $h_k$ under the Gaussian model, which is equivalent to exact normality of the empirical wavelet noise coefficients. 
\begin{lemma}\label{II-lem:2} Suppose that $\epsilon_{1k},\ldots, \epsilon_{Sk} \overset{\textnormal{iid}}{\sim} N(0, \sigma_e^2/T)$, such that $\bs{\xi}_k \sim N(h_k \bs{1}_S, \bs{V}_k)$ for each $k=1,\ldots,T$. An exact expression of the mean squared error of $\hat{h}_{k,\lambda}(\bs{\xi}_k)$ with threshold $\lambda \geq 0$ is given by,
\begin{eqnarray*}
\lefteqn{\mathbb{E}[(\hat{h}_{k,\lambda}(\bs{\xi}_k)- h_k)^2] =} \\
&& \left(2\bs{w}_{k}'\bs{V}_k\bs{w}_{k} - h_k^2 \right) + \left(h_k^2 - \bs{w}_{k}'\bs{V}_k\bs{w}_{k} \right) \Bigg( \Phi\left(\frac{\sqrt{S}(\lambda-h_k)}{\tilde{\sigma}_{k,T}}\right) \\
&&+ \Phi\left(\frac{\sqrt{S}(\lambda+h_k)}{\tilde{\sigma}_{k,T}} \right) \Bigg) + \frac{ (\bs{w}_{k}'\bs{V}_k\bs{1}_S)^2}{S \tilde{\sigma}_{k,T}^2} \left( \frac{\sqrt{S}(\lambda-h_k)}{\tilde{\sigma}_{k,T}} \right) \phi\left( \frac{\sqrt{S}(\lambda-h_k)}{\tilde{\sigma}_{k,T}} \right) \\
&&+ \frac{ (\bs{w}_{k}'\bs{V}_k\bs{1}_S)^2}{S \tilde{\sigma}_{k,T}^2} \left( \frac{\sqrt{S}(\lambda+h_k)}{\tilde{\sigma}_{k,T}} \right) \phi\left( \frac{\sqrt{S}(\lambda+h_k)}{\tilde{\sigma}_{k,T}} \right) 
\end{eqnarray*}
where $\tilde{\sigma}_{k,T}^2 = \frac{\sigma_{uk}^2}{S}\bs{1}'_S \bs{G}_S \bs{1}_S + \sigma_e^2/T$, and $\phi(\cdot)$ denotes the standard normal probability density function. Furthermore, for each $k$ 
\begin{equation}
\mathbb{E}[(\hat{h}_{k,\lambda}(\bs{\xi}_k) - h_k)^2] \leq \bs{w}_{k}'\bs{V}_k\bs{w}_{k} + \lambda^2 \nonumber
\end{equation}
These upper bounds are sharp when either $\lambda/\tilde{\sigma}_{k,T} \to \infty$, in which case $\mathbb{E}[(\hat{h}_{k,\lambda}(\bs{\xi}_k) - h_k)^2] \uparrow \lambda^2$, or when $\lambda/\tilde{\sigma}_{k,T} \to 0$, in which case $\mathbb{E}[(\hat{h}_{k,\lambda}(\bs{\xi}_k) - h_k)^2] \uparrow \bs{w}_{k}'\bs{V}_k\bs{w}_{k}$.
\end{lemma}
The derivations follow from straightforward calculus and the proofs are therefore omitted. The expression for the mean squared error generalizes an expression for the mean squared error of a nonlinear hard threshold estimator in a classical Gaussian sequence model found in \cite{BG96}. Note that we find back the expression in \cite{BG96} when $S=1$ and $\tilde{\sigma}_{k,T}^2 = 1$. The main result in Theorem \ref{thm:1} on the $\ell_2$-risk of $\hat{\bs{h}}(\bs{Y})$ now follows from combining Lemma \ref{II-lem:1}, the upper bounds in Lemma \ref{II-lem:2}, and plugging in the threshold $\lambda_{h,T}$ (see Section \ref{II-sec:2}).

\subsection{Proof of Lemma \ref{II-lem:1}}
\begin{proof}
Let us write $\bs{Y}_{\cdot k} = \bs{h}_k^u + \bs{\epsilon}_{\cdot k}$, where $\bs{h}_k^u = \bs{h}_k + \bs{U}_{\cdot k}$, with $\bs{h}_k = h_k\bs{1}_S$ and $\bs{\epsilon}_{\cdot k} = (\epsilon_{1k},\ldots, \epsilon_{Sk})'$ independent noise terms. Then, conditional on the random effects coefficients, we will show that, uniformly in $k \in J_{T,\alpha}$ with $\alpha > 0$, 
\begin{equation}
\frac{P\left( \frac{\sqrt{T}}{\sigma_e}(Y_{sk} - h_{sk}^u) \geq x_s\, |\, U_{sk} = u_{sk} \right)}{1-\Phi(x_s)} \overset{T \to \infty}{\to} 1 \quad \quad \te{for all } s=1,\ldots, S \nonumber
\end{equation}
where $-\infty < x_s \leq T^\nu$ for some $\nu > 0$. In order to prove this asymptotic normality result, we derive upper bounds on the $n$-th order cumulants ($n \geq 2$) of $\epsilon_{sk} = \langle \bs{E}_s^f, \psi_k \rangle_T$, the empirical wavelet coefficients of the log-periodogram error terms $\bs{E}_s^f$ with respect to the wavelet basis function $\psi_k$. Since $\var(\epsilon_{sk}) = O(T^{-1})$, the $n$-th order cumulant of $\epsilon_{sk}$ (for all $s,k$) can be written in terms of joint cumulants as,
\begin{eqnarray}
\lefteqn{\cum_n(\epsilon_{sk}/\sqrt{\var(\epsilon_{sk})}) \lesssim }\nn
&& \cum\left(\frac{1}{\sqrt{T}}\sum_{\ell_1 = 1}^T E_s^f(\omega_{\ell_1})\psi_k(\omega_{\ell_1}), \ldots, \frac{1}{\sqrt{T}}\sum_{\ell_n = 1}^T E_s^f(\omega_{\ell_n})\psi_k(\omega_{\ell_n})\right) \nn
&=& T^{-n/2} \sum_{\ell_1} \cdots \sum_{\ell_n} \psi_k(\omega_{\ell_1})\cdots \psi(\omega_{\ell_n}) \cum( E_s^f(\omega_{\ell_1}), \ldots, E_s^f(\omega_{\ell_n})) \nn
&=& T^{-n/2} \sum_{\ell_1} \psi_k(\omega_{\ell_1})^n \cum(E_s^f(\omega_{\ell_1}),\ldots, E_s^f(\omega_{\ell_1})) \nn
&& +\ T^{-n/2} \sum_{j = 1}^n \sum_{\ell_1 \neq \ell_2} \psi_k(\omega_{\ell_1})^j \psi_k(\omega_{\ell_2})^{n-j} \cum(E_s^f(\omega_{\ell_1}),\ldots, E_s^f(\omega_{\ell_1}),\nn
&& \quad E_s^f(\omega_{\ell_2}),\ldots, E_s^f(\omega_{\ell_2})) \nn
&& \quad \vdots \nn
&& +\ T^{-n/2} \sum_{\ell_1 \neq \ldots \neq \ell_n} \psi_k(\omega_{\ell_1}) \ldots \psi_k(\omega_{\ell_n}) \cum(E_s^f(\omega_{\ell_1}), \ldots, E_s^f(\omega_{\ell_n}))\quad \quad \label{eq:12}
\end{eqnarray} 
Under \ref{ass:3}, conditional on the random effects coefficients, for each $s=1,\ldots,S$, $X_s(t)$ is a stationary Gaussian process such that $\sum_{h=-\infty}^\infty |h||\cov(X_s(t), X_s(t+h))| < \infty$.  By Lemma 2 in \cite{T79} it then follows that, conditional on the random effects coefficients, for each $s=1,\ldots,S$ and $n\geq 2$: 
\begin{eqnarray}
\lefteqn{\te{cum}(E_s^f(\omega_1), \ldots, E_s^f(\omega_{n_1}), E_s^f(\omega_{n_1+1}),\ldots, E_s^f(\omega_{n_1+n_2}), \ldots, E_s^f(\omega_{n})) =} \nn
&& \hspace{8cm} \left\{ \begin{array}{ll}
O(T^{-m}) & \te{if } m \geq 2 \\
O(1) & \te{if } m = 1 
\end{array}\right. \nonumber
\end{eqnarray}
where, $n_1+\ldots + n_m = n$, and 
\begin{equation}
0 < \omega_1 = \ldots, = \omega_{n_1}< \omega_{n_1+1} = \ldots = \omega_{n_1 + n_2} < \omega_{n - n_m + 1 } = \ldots = \omega_n \nonumber
\end{equation}
Furthermore, since wavelet basis functions at wavelet scale $j$ are of the order $2^{j/2}$, for all $k \in J_{T,\alpha}$, $| \psi_k(\omega_\ell)| \leq  CT^{(1-\alpha)/2}$ uniformly in $\omega_\ell$ for some $C > 0$. From eq.(\ref{eq:12}) we obtain,
\begin{eqnarray}
\lefteqn{\cum_n(\epsilon_{sk}/\sqrt{\var(\epsilon_{sk})}) \lesssim} \nn
&&  T^{-n/2+1} C^{n-2}(T^{(1-\alpha)/2})^{n-2} \int \psi_k(\omega)^2 d\omega + T^{-n/2} C^n B_n (T^{(1-\alpha)/2})^n \nn
&\leq &  T^{-n/2+1}C^{n-2}(T^{(1-\alpha)/2})^{n-2} + T^{-n/2}C^n n! (T^{(1-\alpha)/2})^n \nn
&\lesssim & C^n n! (T^{-\alpha/2})^{n-2} \label{eq:13} 
\end{eqnarray}
where by orthonomality of the wavelet basis functions $\int \psi_k(\omega)^2 d\omega = 1$, and where $B_n$ denotes the $n$-th Bell number satisfying $B_n \leq n!$ for all $n \in \mathbb{N}$. \\[3mm]
By the same arguments as in \cite{N96}, due to the cumulant bounds in eq.(\ref{eq:13}) and Lemma 1 in \cite{RSS78},
\begin{equation} \label{eq:14}
P\left(\frac{\epsilon_{sk} - \mathbb{E}[\epsilon_{sk}]}{\sqrt{\var(\epsilon_{sk})}} \geq x_s\right) = (1 + o_T(1))(1- \Phi(x_s)) \quad \quad \te{for all } s =1,\ldots, S
\end{equation}
for $-\infty < x_s \leq \Delta_T \sim T^{\alpha/6}$. \\[3mm]
Furthermore, under \ref{ass:3}, $\mathbb{E}[E_s^f(\omega_\ell)] = O(T^{-1})$ uniformly in $\omega_\ell$ by \cite{T79}, thus for $k \in J_{T,\alpha}$,
\begin{equation} \label{eq:15}
\mathbb{E}[\epsilon_{sk}] = \frac{1}{T}\sum_{\ell=1}^T \mathbb{E}[E_s^f(\omega_\ell)] \psi_k(\omega_\ell) \leq \sup_{\ell} | \psi_k(\omega_\ell) | O(T^{-1}) = O(T^{-(1+\alpha)/2})
\end{equation}
and  since $\var(\epsilon_{sk}) = O(T^{-1})$, the standardized bias satisfies $b := \mathbb{E}[\epsilon_{sk}]/\sqrt{\var(\epsilon_{sk})} = O(T^{-\alpha/2})$. Rewriting eq.(\ref{eq:14}) gives,
\begin{equation} \label{eq:16}
\frac{P\left( \frac{\sqrt{T}}{\sigma_e} \epsilon_{sk} \geq x_s \right)}{1-\Phi(x_s)} = (1 + o_T(1))\frac{1-\Phi(x_s + b))}{1-\Phi(x_s)}
\end{equation}
Let w.l.o.g. $b \geq 0$ and fix some $c > 1$, then for $x_s \leq c$
\begin{equation}
\left| \frac{1-\Phi(x_s + b)}{1-\Phi(x_s)} - 1 \right| = \frac{|\Phi(x_s + b) -\Phi(x_s)|}{1-\Phi(x_s)} \to 0 \quad \quad \te{as } T \to \infty \nonumber
\end{equation}
On the other hand, for $c < x_s \leq \Delta_T$ by a formula for Mill's ratio (see \cite{N96}),
\begin{equation}
\left| \frac{1-\Phi(x_s + b)}{1-\Phi(x_s)} - 1 \right|  \leq \frac{b \phi(x_s)}{1-\Phi(x_s)} \leq \frac{bx_s}{1-1/x_s^2} \to 0 \quad \quad \te{as } T \to \infty \nonumber
\end{equation}
Since $\bs{Y}_k - \bs{h}_k^u = \bs{\epsilon}_k$, we conclude from eq.(\ref{eq:16}) that conditional on the random effects coefficients,
\begin{equation}
\frac{P\left( \frac{\sqrt{T}}{\sigma_e}(Y_{sk} - h_{sk}^u) \geq x_s\, |\, U_{sk} = u_{sk} \right)}{1-\Phi(x_s)} \overset{T \to \infty}{\to} 1 \quad \quad \te{for all } s = 1,\ldots, S \nonumber
\end{equation}
for $-\infty < x_s \leq \Delta_T$ and uniformly in $k \in J_{T,\alpha}$. Moreover, for any given $S$,
\begin{equation}
\frac{\prod_{s=1}^S P\left( \frac{\sqrt{T}}{\sigma_e}(Y_{sk} - h_{sk}^u) \geq x_s\, |\, U_{sk} = u_{sk} \right)}{\prod_{s=1}^S (1-\Phi(x_s))} \overset{T \to \infty}{\to} 1, \quad \quad \te{for } x_s \leq \Delta_T \nonumber
\end{equation}
and since conditional on the random effects coefficients the terms $Y_{sk} - h_{sk}^u$ are all independent across replicates, this is equivalent to:
\begin{equation}  
P\left(\frac{\sqrt{T}}{\sigma_e}\te{I}_S \cdot (\bs{Y}_{\cdot k} -\bs{h}_k^u) \geq \bs{x}\, |\, \bs{U}_{\cdot k} = \bs{u}_{\cdot k} \right) = (1 + o_T(1)) P(\bs{Z}_k \geq \bs{x}), \quad \te{for } \bs{x} \leq \bs{\Delta}_T \nonumber
\end{equation}
where $\bs{x} = (x_1,\ldots, x_S)' \in \mathbb{R}^S$, $\bs{\Delta}_T \sim (T^\nu, \ldots, T^\nu)' \in \mathbb{R}^S$ with $\nu > 0$, and $\bs{Z}_k \in \mathbb{R}^S$ a vector of independent standard normal random variables. Let us write $\widetilde{\bs{Y}}_k = \bs{V}_k^{-1/2}(\bs{Y}_{\cdot k} - \bs{h}_k)$, where $\bs{V}_k^{-1/2}$ is the square root matrix of $\bs{V}_k^{-1}$, (recall that $\bs{V}_k = \sigma_{uk}^2 \bs{G}_S + \frac{\sigma_e^2}{T}\te{I}_S$). The random effects coefficients are assumed to be jointly multivariate normal, therefore the unconditional version follows as well,
\begin{equation}
P( \widetilde{\bs{Y}}_k \geq \bs{x} ) = (1+o_T(1)) P(\bs{Z}_k \geq \bs{x}), \quad \quad \te{for } \bs{x} \leq \bs{\Delta}_T \label{eq:17}
\end{equation}
In the following part, we relate the mean squared error of $\hat{h}_{k,\lambda}(\bs{Y}_{\cdot k})$ w.r.t. $h_k$ to the mean squared error of $\hat{h}_{k,\lambda}(\bs{\xi}_k)$ w.r.t. $h_k$, and show that they are asymptotically equivalent as $T \to \infty$. We split up,
\begin{eqnarray}
\mathbb{E}[(\hat{h}_{k,\lambda}(\bs{Y}_{\cdot k}) - h_k)^2] &=& \mathbb{E}\left[(\hat{h}_{k,\lambda}(\bs{Y}_{\cdot k}) - h_k)^2 \bs{1}_{\{ \bs{h}_k - \bs{V}_k^{1/2}\bs{\Delta}_T \leq \bs{Y}_{\cdot k} \leq \bs{h}_k + \bs{V}_k^{1/2}\bs{\Delta}_T \}} \right] \nn
&& + \mathbb{E}\left[(\hat{h}_{k,\lambda}(\bs{Y}_{\cdot k}) - h_k)^2 \bs{1}_{\{ |\widetilde{\bs{Y}}_k| > \bs{\Delta}_T \}} \right] \nn
&:= & R_1 + R_2 \label{eq:18}
\end{eqnarray}
According to eq.(\ref{eq:17}) above, there exist $C_T^{(\ell)}$ and $C_T^{(u)}$ tending to 1 as $T \to \infty$ (uniformly in $k \in J_{T,\alpha}$), such that
\begin{equation} 
C_T^{(\ell)}P(\bs{Z}_k \geq \bs{x}) \leq P( \widetilde{\bs{Y}}_k \geq \bs{x}) \leq C_T^{(u)} P(\bs{Z}_k \geq \bs{x}) \quad \quad \te{for } \bs{x} \leq \bs{\Delta}_T \nonumber
\end{equation} 
which is equivalent to,
\begin{equation}
C_T^{(\ell)}P(\bs{\xi}_k \geq \bs{x}) \leq P( \bs{Y}_{\cdot k} \geq \bs{x}) \leq C_T^{(u)} P(\bs{\xi}_k \geq \bs{x}) \quad \quad \te{for } \bs{x} \leq \bs{h}_k + \bs{V}_k^{1/2}\bs{\Delta}_T
\label{eq:19}
\end{equation}
with $\bs{\xi}_k = \bs{h}_k + \bs{V}_k^{1/2} \bs{Z}_k \sim N(\bs{h}_k, \bs{V}_k)$.\\[3mm]
In the argument below we use that for $g : \mathbb{R}^S \to \mathbb{R}$ measurable, if $\bs{a} \in \mathbb{R}^S$ is such that $P(g(\bs{X}) \geq g(\bs{a}))=1$, then
\begin{equation} \label{eq:20}
\mathbb{E}[g(\bs{X})] = \int g(\bs{x})\ dP(\bs{X} \leq \bs{x}) = g(\bs{a}) + \int_{\bs{a}}^{(\infty, \ldots, \infty)} P(\bs{X} \geq \bs{x})\ dg(\bs{x}) 
\end{equation}
Let $g(\bs{x}) = (\hat{h}_{k,\lambda}(\bs{x}) - h_k)^2 \bs{1}_{\{\bs{x} \in \bs{B}_k\}}$, where we recall that $\hat{h}_{k,\lambda}(\cdot)$ is the (deterministic) thresholding rule that defines our estimator, and let $\bs{a} = \inf_{\bs{x} \in \bs{B}_k} (\hat{h}_{k,\lambda}(\bs{x}) - h_k)^2$, with $\bs{B}_k = \{ \bs{x}\, :\, \bs{h}_k - \bs{V}_k^{1/2} \bs{\Delta}_T \leq \bs{x} \leq \bs{h}_k + \bs{V}_k^{1/2} \bs{\Delta}_T \}$. Note that $\bs{a}$ exists and is attained for $\bs{x} \in \bs{B}_k$, since $\bs{B}_k$ is closed and bounded and therefore compact.\\[3mm]
Using eq.(\ref{eq:19}) and eq.(\ref{eq:20}), we can upper bound $R_1$ by,
\begin{eqnarray}
R_1 &=& \int g(\bs{x}) \ dP(\bs{Y}_{\cdot k} \leq \bs{x}) \nn
&=& g(\bs{a}) + \int_{\bs{a}}^{(\infty, \ldots, \infty)} P(\bs{Y}_{\cdot k} \geq \bs{x})\ d\big[ (\hat{h}_{k,\lambda}(\bs{x}) - h_k)^2 \bs{1}_{\{\bs{x} \in \bs{B}_k \}}\big] \nn
&\leq & g(\bs{a}) + C_T^{(u)}\int_{\bs{a}}^{(\infty, \ldots, \infty)} P(\bs{\xi}_k \geq \bs{x})\ d\big[ (\hat{h}_{k,\lambda}(\bs{x}) - h_k)^2 \bs{1}_{\{\bs{x} \in \bs{B}_k \}}\big] \nn
&\leq & (C_T^{(u)} \vee 1) \left\{ g(\bs{a}) + \int_{\bs{a}}^{(\infty, \ldots, \infty)} P(\bs{\xi}_k \geq \bs{x})\ d\big[ (\hat{h}_{k,\lambda}(\bs{x}) - h_k)^2 \bs{1}_{\{\bs{x} \in \bs{B}_k \}}\big] \right\} \nn
&=& (C_T^{(u)} \vee 1) \mathbb{E}\left[ (\hat{h}_{k,\lambda}(\bs{\xi}_k)-h_k)^2 \bs{1}_{\{ \bs{\xi}_k \in \bs{B}_k \}} \right] \label{eq:21}
\end{eqnarray}
Completely analogous, we can lower bound $R_1$ by,
\begin{eqnarray} \label{eq:22}
R_1 &\geq & (C_T^{(\ell)} \wedge 1) \mathbb{E}\left[ (\hat{h}_{k,\lambda}(\bs{\xi}_k)-h_k)^2 \bs{1}_{\{ \bs{\xi}_k \in \bs{B}_k \}} \right] 
\end{eqnarray}
Combining eq.(\ref{eq:21}) and eq.(\ref{eq:22}), and using that $C_T^{(\ell)}, C_T^{(u)} \to 1$ as $T \to \infty$, we conclude that 
\begin{equation}
R_1 = (1 + o_T(1))\mathbb{E}\left[ (\hat{h}_{k,\lambda}(\bs{\xi}_k) - h_k)^2\bs{1}_{\{ \bs{\xi}_k \in \bs{B}_k \}} \right] \label{eq:23}
\end{equation}
uniformly in $k \in J_{T,\alpha}$. For the other term $R_2$ in eq.(\ref{eq:18}), with $\bs{\Delta}_T = (\Delta_{T,1},\ldots, \Delta_{T,S})' = (T^\nu, \ldots, T^\nu)$ for some $\nu > 0$, using an upper bound for multivariate Gaussian tail probabilities we find that,
\begin{eqnarray}
P(|\widetilde{\bs{Y}}_k| \geq \bs{\Delta_T}) & \leq & C_T^{(u)}P(\bs{Z}_k \geq \bs{\Delta}_T)  \nn
&\leq & C_T^{(u)} \left( \prod_{s=1}^S \Delta_{T,s} \right)^{-1} (2\pi)^{-S/2} \exp\left( -\frac{1}{2}\bs{\Delta}_T'\te{I}_S \bs{\Delta}_T \right) \nn
& \lesssim & C_T^{(u)} T^{-S\nu} \exp\left( -\frac{S}{2} T^{2\nu} \right) = O((ST)^{-\mu})
\label{eq:24}
\end{eqnarray}
for arbitrary $0< \mu < \infty$, since the exponential rate above decays faster than any arbitrary polynomial rate. \\[3mm]
Furthermore, it can be verified that
\begin{equation}
\mathbb{E}[(\hat{h}_{k,\lambda}(\bs{Y}_{\cdot k}) - h_k)^4] \leq \mathbb{E}[(|\bs{w}_{k}'\bs{Y}_{\cdot k} - h_k| + |h_k|)^4]  \leq \max_s \mathbb{E}[(|Y_{sk} - h_k| + |h_k|)^4] \label{eq:25}
\end{equation}
By Parseval's relation,
\begin{equation}
\sup_k |h_k| = \Vert \bs{h} \Vert_\infty \leq \Vert \bs{h} \Vert_{2} = \frac{1}{\sqrt{T}}\Vert \bs{h}^f \Vert_2 = \Vert h^f \Vert_{L_2} + o_T(1) = O(1) 
\end{equation}
where the last equality is due to the fact that $h^f \in L_2([0,1/2])$ (here $\Vert \cdot \Vert_2$ denotes the Euclidian norm). By Jensen's inequality,
\begin{equation}
\mathbb{E}[|Y_{sk} - h_k|^n] \leq \sqrt{\mathbb{E}[(Y_{sk} - h_k)^{2n}]} = \sqrt{\mathbb{E}[(U_{sk} + \epsilon_{sk})^{2n}]}\quad \quad \te{for }  1 \leq n \leq 4 
\end{equation}
By eq.(\ref{eq:15}) and the cumulant bounds in eq.(\ref{eq:13}) for all $s =1,\ldots,S$, 
\begin{eqnarray}
\cum_1(\epsilon_{sk}) &=& O(T^{-1/2}T^{-\alpha/2}) = O(T^{-1/2}) \nn
\cum_n(\epsilon_{sk}) &=& O(T^{-n/2}(T^{-\alpha/2})^{n-2}) = O(T^{-n/2}) \quad \quad \te{for } n \geq 2 \nonumber
\end{eqnarray}
Therefore,
\begin{equation}
\mathbb{E}[\epsilon_{sk}^n] = O\left(\sum_{m=1}^n \prod_{i_1,\ldots,i_m\,:\, i_1+ \ldots + i_m = n, i_j \geq 1} |\cum_{i_j}(\epsilon_{sk})| \right) = O(T^{-n/2}) \label{eq:28}
\end{equation}
Also, since the random effects coefficients are assumed to be Gaussian $\mathbb{E}[U_{sk}^n] = O(1)$ for $1 \leq n \leq 8$, we obtain from eq.(\ref{eq:25})-(\ref{eq:28}) that
\begin{equation} \label{eq:29}
\mathbb{E}[(\hat{h}_{k,\lambda}(\bs{Y}_{\cdot k}) - h_k)^4] = O(1)
\end{equation}
Combining the Cauchy-Schwarz inequality and eq.(\ref{eq:24}) and eq.(\ref{eq:29}) we find,
\begin{equation}
R_2 \leq \sqrt{P(|\widetilde{\bs{Y}}_k| \geq \Delta_T)} \sqrt{\mathbb{E}[(\hat{h}_{k,\lambda}(\bs{Y}_{\cdot k}) - h_k)^4]} = O((ST)^{-\mu}) \nonumber
\label{eq:30}
\end{equation}
for arbitrary $0 < \mu < \infty$, (abusing notation for $\mu$). Combining eq.(\ref{eq:18}), eq.(\ref{eq:23}), and eq.(\ref{eq:30}) yields,
\begin{eqnarray}
\mathbb{E}[(\hat{h}_{k,\lambda}(\bs{Y}_{\cdot k}) - h_k)^2] &=& (1 + o_T(1)) \mathbb{E}[(\hat{h}_{k,\lambda}(\bs{\xi}_k) - h_k)^2] + O((ST)^{-\mu}) \quad \quad \quad
\end{eqnarray}
Since the above equation holds uniformly over $k \in J_{T,\alpha}$, with $|J_{T,\alpha}| = O(T^{1-\alpha})$ for $\alpha > 0$, it follows that,
\begin{eqnarray}
\lefteqn{\sum_{k \in J_{T,\alpha}} \mathbb{E}[(\hat{h}_{k,\lambda}(\bs{Y}_{\cdot k}) - h_k)^2]  =} \nn
&& (1 + o_T(1)) \sum_{k \in J_{T,\alpha}} \mathbb{E}[(\hat{h}_{k,\lambda}(\bs{\xi}_k) - h_k)^2] + \sum_{k \in J_{T,\alpha}} O((ST)^{-\mu}) \nn
&=&  (1 + o_T(1)) \sum_{k \in J_{T,\alpha}} \mathbb{E}[(\hat{h}_{k,\lambda}(\bs{\xi}_k) - h_k)^2] + O(S^{-\mu} T^{-\mu + 1}) \nonumber
\end{eqnarray}
for arbitrary $0 < \mu < \infty$, which concludes the proof.
\end{proof}

\section{Proof of Theorem \ref{thm:1}} \label{II-sec:2}
\begin{proof}
We split up, using Lemma \ref{II-lem:1}:
\begin{eqnarray}
\lefteqn{\mathbb{E}\Vert \hat{\bs{h}}(\bs{Y}) - \bs{h} \Vert^2 = \sum_{k \in J_{T,\alpha}} \mathbb{E}[(\hat{h}_{k}(\bs{Y}_{\cdot k}) - h_k)^2] + \sum_{k \notin J_{T, \alpha}} h_k^2} \nn
&=& (1 + o_T(1)) \sum_{k \in J_{T,\alpha}} \mathbb{E}[(\hat{h}_{k}(\bs{\xi}_k) - h_k)^2] + O(S^{-\mu}T^{1-\mu}) + \sum_{k \notin J_{T, \alpha}} h_k^2 \quad \quad
\label{eq:31}
\end{eqnarray}
for arbitrary $0 < \mu < \infty$. For $T$ sufficiently large ($T \geq T^*$) since $0 < \alpha \leq \alpha^*$ with $T^*,\alpha^*$ as in \ref{ass:2}, the last term on the right-hand side disappears by \ref{ass:2}. It remains to show that the remaining part behaves according to the claimed rates.\\[3mm]
By \ref{ass:1}, $\bs{h} \in \ell_{0,T}(k_{h,T})$ and $\bs{\sigma}^2_u \in \ell_{0,T}(k_{u,T})$ with $K_{u,T} \subseteq K_{h,T}$. We decompose the first sum on the right-hand side above into three different regions $\{ k \notin K_{h,T} \}$, $\{ k \in K_{h,T} \setminus K_{u,T} \}$, and $\{ k \in K_{h,T} \cap K_{u,T} \}$, and upper bound by Lemma \ref{II-lem:2},
\begin{eqnarray}
\lefteqn{\sum_{k \in J_{T,\alpha}} \mathbb{E}[\hat{h}_{k}(\bs{\xi}_k) - h_k)^2] \leq} \nn
&& (T-k_{h,T})\left( \frac{2\sigma_e^2}{TS}\left( 1 - \Phi\left( \frac{\sqrt{TS}\lambda_{h,T}}{\sigma_e} \right) \right) + \frac{2\sigma_e\lambda_{h,T}}{\sqrt{TS}}\phi\left( \frac{\sqrt{TS}\lambda_{h,T}}{\sigma_e}\right) \right) \nn
&&+\ (k_{h,T} - k_{u,T}) \left( \frac{\sigma_e^2}{TS} + \lambda_{h,T}^2 \right) + k_{u,T} \sup_k \left[ \bs{w}_{k}'\bs{V}_k\bs{w}_{k} + \lambda_{h,T}^2 \right]
\label{eq:32}
\end{eqnarray} 
We observe that $\lambda_{h,T} = \frac{\sigma_e}{\sqrt{TS}}\sqrt{2\log(T/k_{h,T})} > \frac{\sigma_e}{\sqrt{TS}}$ for $T$ large, since $k_{h,T}/T \to 0$ as $T \to \infty$. Therefore, for $T$ sufficiently large,
\begin{eqnarray}
\frac{2\sigma^2_e}{TS} \left(1-\Phi\left( \frac{\sqrt{TS}\lambda_{h,T}}{\sigma_e}\right)\right) &=& \frac{2\sigma^2_e}{TS} \int_{\sqrt{TS}\lambda_{h,T}/\sigma_e}^\infty \frac{1}{\sqrt{2\pi}} \exp\left( -\frac{1}{2}z^2 \right)\ dz \nn
&\leq & \frac{2\sigma^2_e}{TS} \frac{\sigma_e}{\sqrt{TS}\lambda_{h,T}} \int_{\sqrt{TS}\lambda_{h,T} /\sigma_e}^\infty \frac{z}{\sqrt{2\pi}} \exp\left( -\frac{1}{2}z^2 \right)\ dz \nn
&\leq & \frac{2\sigma_e\lambda_{h,T}}{\sqrt{TS}} \phi\left(\frac{\sqrt{TS}\lambda_{h,T}}{\sigma_e} \right)\nn
&=& \frac{2\sigma_e^2}{TS}\sqrt{2\log\left(\frac{T}{k_{h,T}}\right)} \phi(0) \frac{k_{h,T}}{T}
\label{eq:33}
\end{eqnarray}
where in the second step we use $\frac{z}{\sqrt{TS}\lambda_{h,T} / \sigma_e} \geq 1 $ and in the third step  $\frac{\sigma_e^2}{TS} \leq \lambda_{h,T}^2$ and $\int_{\lambda}^\infty z \phi(z) dz = \phi(\lambda)$. 
By eq.(\ref{eq:33}) and plugging in $\lambda_{h,T}$, the right-hand side in eq.(\ref{eq:32}) is upper bounded by:
\begin{eqnarray}
\sum_{k \in J_{T,\alpha}} \mathbb{E}[\hat{h}_k(\bs{\xi}_k) - h_k)^2] &\leq & (T-k_{h,T})\left( \frac{4\sigma_e^2}{TS}\sqrt{2\log\left(\frac{T}{k_{h,T}}\right)} \phi(0) \frac{k_{h,T}}{T} \right) \nn
&&+ (k_{h,T} - k_{u,T}) \left( \frac{\sigma_e^2}{TS} + \frac{2\sigma_e^2}{TS}\log\left( \frac{T}{k_{h,T}} \right) \right) \nn
&& + k_{u,T} \sup_k \left[ \bs{w}_{k}'\bs{V}_k\bs{w}_{k} + \frac{2\sigma_e^2}{TS} \log\left( \frac{T}{k_{h,T}} \right) \right] \nn
& \leq & \frac{k_{h,T} \sigma_e^2}{TS} \left(1 + 4\phi(0)\sqrt{2\log\left(\frac{T}{k_{h,T}}\right)} + 2\log\left( \frac{T}{k_{h,T}} \right) \right) \nn
&& + k_{u,T} \left( \sup_k \bs{w}_{k}'\bs{V}_k\bs{w}_{k} - \frac{\sigma_e^2}{TS} \right) \nn
& \lesssim & \frac{k_{h,T}}{TS} \log\left( \frac{T}{k_{h,T}} \right) + k_{u,T} \left( \sup_k \bs{w}_{k}'\bs{V}_k\bs{w}_{k} - \frac{\sigma_e^2}{TS} \right)\nn
\label{eq:34}
\end{eqnarray}
By plugging eq.(\ref{eq:34}) into eq.(\ref{eq:31}), we obtain for $T$ sufficiently large
\begin{eqnarray}
\mathbb{E}\Vert \hat{\bs{h}}(\bs{Y}) - \bs{h} \Vert^2 & \lesssim &  \frac{k_{h,T}}{TS} \log\left( \frac{T}{k_{h,T}} \right) + k_{u,T} \left( \sup_k \bs{w}_k'\bs{V}_k\bs{w}_k - \frac{\sigma_e^2}{TS} \right) + S^{-\mu}T^{1-\mu} \nonumber
\end{eqnarray}
Since $0< \mu < \infty$ can be chosen arbitrarily large, for $\mu > 2$ the term $\frac{k_{h,T}}{TS} \log\left( \frac{T}{k_{h,T}} \right)$ dominates $S^{-\mu}T^{1-\mu}$, thus concluding the proof.
\end{proof}

\section{Proof of Theorem \ref{thm:2}}\label{II-sec:3}
\subsection{Almost sure convergence of the estimated set $\widehat{K}_u(\bs{Y})$} \label{II-sec:3.1}
\begin{proof}
First we show that the inclusion $K_{u,T} \subseteq \widehat{K}_u(\bs{Y})$ holds with probability tending to $1$ as $S,T \to \infty$. Since $\widehat{K}_u(\bs{Y}) = \{ k\,:\, |T_k(\bs{Y}_{\cdot k})| \geq \lambda_{u,T} \}$, it suffices to show that
\begin{equation}
P\left( \inf_{k \in K_{u,T}} |T_k(\bs{Y}_{\cdot k})| \geq \lambda_{u,T} \right) \to 1, \quad \quad \te{as } S,T \to \infty \label{eq:35}
\end{equation}
Writing $\sigma_{k,T}^2 = \sigma_{uk}^2 + \sigma_e^2/T$ and $\hat{\sigma}_k^2(\bs{Y}_{\cdot k}) = \frac{1}{S}\sum_{s=1}^S (Y_{sk} - \hat{h}_k)^2$, since $\inf_{k \in K_{u,T}} \sigma_{k,T}^2 \geq \delta > 0$ we can lower bound, 
\begin{eqnarray}
\inf_{k \in K_{u,T}} |T_k(\bs{Y}_{\cdot k})| &=& \inf_{k \in K_{u,T}} \left| \log( \hat{\sigma}_k^2(\bs{Y}_{\cdot k} )) + \log\left( \frac{T}{\sigma_e^2} \right) - \left( \log\left( \frac{2}{S} \right) + \psi^{(0)}\left(\frac{S}{2} \right)\right) \right| \nn
& \geq & \inf_{k \in K_{u,T}} \left| \log(\sigma_{k,T}^2) + \log\left( \frac{T}{\sigma_e^2} \right) - \left( \log\left( \frac{2}{S} \right) + \psi^{(0)}\left(\frac{S}{2} \right)\right) \right| \nn
&& - \sup_{k \in K_{u,T}} \left| \log(\hat{\sigma}_k^2(\bs{Y}_{\cdot k}) - \log(\sigma_{k,T}^2) \right| \nonumber
\end{eqnarray}
It can be verified that $\psi^{(0)}(x) = \log(x) + O(x^{-1})$, therefore $|\log(2/S) + \psi^{(0)}(S/2)| = o_S(1)$, and the first term on the right-hand side is seen to grow at the rate $\sim \log(T)$ for $S,T$ increasing. For the second term, we show that $\sup_{k \in K_{u,T}} | \log(\hat{\sigma}_k^2(\bs{Y}_{\cdot k}) - \log(\sigma_{k,T}^2)| = o_{p}^{S,T}(1)$. First, decompose
\begin{eqnarray}
\hat{\sigma}_k^2(\bs{Y}_{\cdot k}) &=& \frac{1}{S}\sum_{s=1}^S ( (Y_{sk} - h_k) + (h_k-\hat{h}_k))^2 \nn
&=& \underbrace{\frac{1}{S}\sum_{s=1}^S (Y_{sk}-h_k)^2}_{\te{(i)}} + \underbrace{\frac{2}{S}\sum_{s=1}^S (Y_{sk} - h_k)(h_k-\hat{h}_k)}_{\te{(ii)}} + \underbrace{\frac{1}{S}\sum_{s=1}^S (h_k-\hat{h}_k)^2}_{\te{(iii)}} \nn \label{eq:35a}
\end{eqnarray}
For $k \in K_{u,T}$, term (ii) and term (iii) uniformly converge to zero in probability as $S,T \to \infty$, which follows from the fact that $\sup_k|h_k - \hat{h}_k| = o^{S,T}_{p}(1)$ and $\frac{2}{S}\sum_{s=1}^S (Y_{sk}-h_k) = O_p(1)$ since $\var(Y_{sk}) < \infty$. \\[3mm]
For term (i), by linearity of the expectation and recalling that $Y_{sk} = h_k + U_{sk} + \epsilon_{sk}$,
\begin{equation}
\mathbb{E}\left[ \frac{1}{S}\sum_{s=1}^S (Y_{sk} - h_k)^2 \right] = \frac{1}{S}\sum_{s=1}^S \var(U_{sk} + \epsilon_{sk}) = \sigma_{uk}^2 + \sigma_e^2/T = \sigma_{k,T}^2 \nonumber
\end{equation}
where we use that by assumption $\epsilon_{sk} \sim (0, \sigma_e^2/T)$. Also,
\begin{eqnarray}
\lefteqn{\var\left( \frac{1}{S} \sum_{s=1}^S (Y_{sk} - h_k)^2 \right) =}\nn
&& \frac{1}{S^2}\sum_{s=1}^S \var((U_{sk} + \epsilon_{sk})^2 ) + \frac{1}{S^2}\sum_{s \neq s'} \cov((U_{sk} + \epsilon_{sk})^2, (U_{s'k} + \epsilon_{s'k})^2) \nn
&=& \frac{1}{S^2}\sum_{s=1}^S \var((U_{sk} + \epsilon_{sk})^2 ) + \frac{1}{S^2}\sum_{s \neq s'} \cov(U^2_{sk}, U^2_{s'k}) \nonumber
\end{eqnarray}
where in the second step we use that $\epsilon_{sk} \perp U_{s'k}$ for all $s,s'$, $\epsilon_{sk} \perp \epsilon_{s'k}$ for $s \neq s'$, and the fact that $\mathbb{E}[\epsilon_{sk}] = 0$ for all $s,k$. We observe that $\var((U_{sk} + \epsilon_{sk})^2) = O(1)$ uniformly over $k=1,\ldots,T$, since $U_{sk} \sim N(0, \sigma_{uk}^2)$ with $\sup_k \sigma_{uk}^2 < \infty$, and $\epsilon_{sk} \sim (0, \sigma_e^2/T)$ with $\mathbb{E}[\epsilon_{sk}^4] = O(T^{-2})$ is independent of $U_{sk}$ for all $s,k$. Moreover, by Gaussianity of the random effects coefficients, it can be verified that
\begin{equation}
\cov(U_{sk}^2, U_{s'k}^2) = 2\rho_{ss'}^2 \sigma_{uk}^4 \nonumber
\end{equation}
Therefore, using that $\sup_k \sigma_{uk}^2 < \infty$, 
\begin{eqnarray}
\var\left(\frac{1}{S}\sum_{s=1}^S (Y_{sk} - h_k)^2 \right) = O(S^{-1}) + 2\sigma_{uk}^4 \frac{\Vert \bs{G}_S \Vert^2_F}{S^2} = O(S^{-1}) + O\left( \frac{\Vert \bs{G}_S \Vert_F^2}{S^2} \right) \nonumber
\end{eqnarray}
which converges to zero uniformly over $k \in \{1,\ldots,T\}$, since $\Vert \bs{G}_S \Vert_F/S \to 0$ by assumption. By Chebychev's inequality $\frac{1}{S}\sum_{s=1}^S (Y_{sk} - h_k)^2 \overset{P}{\to} \sigma_{k,T}^2$ uniformly over $k\in K_{u,T}$ as $S \to \infty$, and by Slutsky's lemma it follows from eq.(\ref{eq:35a}) that $\hat{\sigma}_k^2(\bs{Y}_k) \overset{P}{\to} \sigma_{k,T}^2$ uniformly over $k \in K_{u,T}$ as $S, T \to \infty$. \\[3mm]
For the uniform convergence in probability of $\log(\hat{\sigma}_k^2(\bs{Y}_{\cdot k}))$ over $k \in K_{u,T}$, we show that for any $\epsilon, \gamma > 0$, there exist $S_0, T_0$ sufficiently large (depending on only $\epsilon, \gamma$) such that
\begin{equation}
P(|\log(\hat{\sigma}_k^2(\bs{Y}_{\cdot k})) - \log(\sigma_{k,T}^2)| > \epsilon) < \gamma \label{eq:36}
\end{equation}
for $S > S_0, T > T_0$ and all $k \in K_{u,T}$. For arbitrary $\epsilon > 0$, by the law of total probability,
\begin{eqnarray}
\lefteqn{P(|\log(\hat{\sigma}_k^2) - \log(\sigma_{k,T}^2)| > \epsilon) =} \nn
&& P(|\log(\hat{\sigma}_k^2) - \log(\sigma_{k,T}^2)| > \epsilon : \hat{\sigma}_k^2 \geq \delta/2) P(\hat{\sigma}_k^2 \geq \delta/2) + P(\hat{\sigma}_k^2 < \delta/2) \nonumber 
\end{eqnarray}
Since $\log(x)$ has bounded derivative on $x \in [\delta/2, \infty)$ for $\delta$ bounded away from zero, it is uniformly continuous on the domain $[\delta/2, \infty)$. Thus, there exists $\epsilon_1 > 0$ such that $|x - x_0| \leq \epsilon_1$ implies $|\log(x) - \log(x_0)| \leq \epsilon$ for all $x,x_0 \in [\delta/2, \infty)$. Using that $\inf_{k \in K_{u,T}} \sigma_{k}^2 \geq \delta$, for this choice of $\epsilon_1$ we get,
\begin{eqnarray}
\lefteqn{P(|\log(\hat{\sigma}_k^2) - \log(\sigma_{k,T}^2)| > \epsilon : \hat{\sigma}_k^2 \geq \delta/2)P(\hat{\sigma}_k^2 \geq \delta/2) \leq} \nn
&\hspace{3cm}& P(|\hat{\sigma}_k^2 - \sigma_{k,T}^2| > \epsilon_1 : \hat{\sigma}_k^2 \geq \delta/2)P(\hat{\sigma}_k^2 \geq \delta/2) \nn
&\hspace{3cm} \leq & P(|\hat{\sigma}_k^2-\sigma_{k,T}^2| > \epsilon_1) \nonumber
\end{eqnarray}
By the uniform convergence in probability of $\hat{\sigma}_k^2(\bs{Y}_{\cdot k})$, there exist $S_0^{(1)}, T_0^{(1)}$ such that
\begin{equation}
P(|\hat{\sigma}_k^2 - \sigma_{k,T}^2| > \epsilon_1) < \gamma/2 \nonumber
\end{equation}
for $S > S_0^{(1)}, T > T_0^{(1)}$ and all $k \in K_{u,T}$. Similarly, using again that $\inf_{k \in K_{u,T}} \sigma_{k,T}^2 \geq \delta$, there exist $S_0^{(2)}, T_0^{(2)}$ such that 
\begin{equation}
P(\hat{\sigma}^2_k < \delta/2) < P(|\hat{\sigma}_k^2 - \sigma_{k,T}^2| > \delta/2) < \gamma/2 \nonumber
\end{equation}
for $S > S_0^{(2)}, T > T_0^{(2)}$ and all $k \in K_{u,T}$. The uniform convergence in probability in eq.(\ref{eq:36}) now follows from the above arguments with $S_0 = S_0^{(1)} \vee S_0^{(2)}$ and $T_0 = T_0^{(1)} \vee T_0^{(2)}$.\\[3mm]
We conclude that the left-hand side inside the probability in eq.(\ref{eq:35}) grows at the rate $\sim \log(T)$ in probability for increasing $S,T$. On the other hand, $\lambda_{u,T} = o(\log(T))$ by assumption. Combining these two results implies that $K_{u,T} \subseteq \widehat{K}_u(\bs{Y})$ with probability tending to 1 as $S,T \to \infty$.\\[3mm]
Next, we show that the other inclusion $\widehat{K}_u(\bs{Y}) \subseteq K_{u,T}$ also holds with probability tending to 1 as $S,T \to \infty$, which is equivalent to showing 
\begin{equation}
P\left( \sup_{k \notin K_{u,T}} |T_k(\bs{Y}_{\cdot k})| \geq \lambda_{u,T} \right) \to 0, \quad \quad \te{as } S,T \to \infty \nonumber
\end{equation}
For $k \notin K_{u,T}$,  eq.(\ref{eq:35a}) term (ii) and term (iii) are $o_p^{S,T}(T^{-1})$, which is obtained by combining $\sup_{k \notin K_{u,T}} |h_k - \hat{h}_k| = o_p^{S,T}(T^{-1/2})$ and $(Y_{sk} - h_k) = O_p(T^{-1/2})$ since $\var(Y_{sk}) = \sigma_e^2/T$. Term (i) in eq.(\ref{eq:35a}) satisfies uniformly over $k \notin K_{u,T}$,
\begin{eqnarray}
\frac{1}{S}\sum_{s=1}^S (Y_{sk} - h_k)^2 &=& \mathbb{E}[(Y_{sk}-h_k)^2] + O_p\left(\sqrt{\var\left(\frac{1}{S}\sum_{s=1}^S (Y_{sk} - h_k)^2\right)}\right) \nn
&=& \frac{\sigma_e^2}{T} + o^{S,T}_p(T^{-1})
\end{eqnarray}
since $\var((Y_{sk}-h_k)^2) = \var(\epsilon_{sk}^2) = O(T^{-2})$ for $k \notin K_{u,T}$. Combining the three terms it follows that,
\begin{equation}
\sup_{k \notin K_{u,T}} \left| \frac{T}{\sigma_e^2} \hat{\sigma}_k^2(\bs{Y}_{\cdot k}) - 1 \right| \overset{P}{\to} 0, \quad \quad \te{as } S,T \to \infty
\end{equation} 
By continuity of the logarithm, for arbitrary $\epsilon > 0$ and fixing $x_0 = 1$, there exists $\epsilon_1 = \epsilon_1(x_0)$ independent of $k$, such that $|x - x_0| \leq \epsilon_1$ implies $|\log(x) - \log(x_0)| \leq \epsilon$. For this choice of $\epsilon_1$,
\begin{equation}
P(|\log( T\hat{\sigma}_k^2(\bs{Y}_{\cdot k})/\sigma_e^2) - \log(1)| > \epsilon) \leq P(|T\hat{\sigma}_k^2(\bs{Y}_{\cdot k})/\sigma_e^2 - 1| > \epsilon_1) \nonumber
\end{equation}
Since $\epsilon_1$ only depends on $x_0 = 1$, and $T\hat{\sigma}_k^2(\bs{Y}_k)/\sigma_e^2 \overset{P}{\to} 1$ uniformly over $k \notin K_{u,T}$, we obtain
\begin{equation}
\sup_{k \notin K_{u,T}} |\log(\hat{\sigma}_k^2(\bs{Y}_{\cdot k})) + \log(T/\sigma_e^2)| \overset{P}{\to} 0, \quad \quad \te{as } S,T \to \infty \nonumber
\end{equation}
Therefore, by the triangle inequality, 
\begin{eqnarray}
\sup_{k \notin K_{u,T}} |T_k(\bs{Y}_{\cdot k})| & \leq & \sup_{k \notin K_{u,T}} |\log(\hat{\sigma}_k^2(\bs{Y}_{\cdot k})) + \log(T/\sigma_e^2) | + | \log( 2/S) + \psi^{(0)}( S/2 ) | \nn
&\overset{P}{\to}& 0, \quad \te{as } S,T \to \infty \nonumber
\end{eqnarray}
since $|\log(2/S) + \psi^{(0)}(S/2)| = o_S(1)$. On the other hand, $\lambda_{u,T} \geq C$ for some constant $C > 0$ by assumption. Combining these two results implies that $\widehat{K}_u(\bs{Y}) \subseteq K_{u,T}$ with probability tending to 1 as $S,T \to \infty$. To conclude, since both $P(K_{u,T} \subseteq \widehat{K}_u(\bs{Y})) \to 1$ and $P(\widehat{K}_u(\bs{Y}) \subseteq K_{u,T}) \to 1$ as $S,T \to \infty$, this also implies $P(\widehat{K}_u(\bs{Y}) = K_{u,T}) \to 1$ as $S,T \to \infty$.
\end{proof}

\subsection{Uniform consistency of the estimators of $\sigma_{uk}^2$}\label{II-sec:3.2}
\begin{proof}
From the proof in Section \ref{II-sec:3.1} we already know that $\sup_{k} |\hat{\sigma}_k^2(\bs{Y}_{\cdot k}) - \sigma_{k,T}^2| \overset{P}{\to} 0$ as $S,T \to \infty$ for $k \in \{1,\ldots, T\}$. It remains to show that $\sup_k | \hat{\sigma}_{uk}^2(\bs{Y}_{\cdot k}) - \sigma_{uk}^2 | = o_{p}^{S,T}(1)$ with $k \in \{1,\ldots, T\}$. For the linear part of the estimator, consider the function $g(x) = \{ x - \frac{\sigma_e^2}{T} \}_+$ which is uniformly continuous on $x \in [0,\infty)$ since $|g(x) - g(x_0)| \leq |x - x_0|$ for all $x,x_0 \in [0,\infty)$. Therefore, by a similar argument as in Section \ref{II-sec:3.1},
\begin{equation}
\sup_{k} \left| \left\{ \hat{\sigma}_k^2(\bs{Y}_{\cdot k}) - \frac{\sigma_e^2}{T} \right\}_+ - \sigma_{uk}^2 \right| \overset{P}{\to} 0, \quad \quad \te{as } S,T \to \infty \label{eq:37}
\end{equation}
In order to show the uniform convergence in probability of the nonlinear estimators $\hat{\sigma}_{uk}^2(\bs{Y}_{\cdot k})$ over $k \in \{1,\ldots,T\}$, it suffices to show that for any $\epsilon, \gamma > 0$, there exist $S_0,T_0$ sufficiently large (depending only on $\epsilon, \gamma$) such that
\begin{equation}
P(|\hat{\sigma}_{uk}^2(\bs{Y}_{\cdot k}) - \sigma_{uk}^2| > \epsilon ) < \gamma \nonumber
\end{equation}
for $S > S_0$, $T > T_0$ and all $k \in \{1,\ldots, T\}$.\\[3mm]
First, consider the case $k \in K_{u,T}$, by the law of total probability for arbitrary $\epsilon > 0$, 
\begin{eqnarray}
\lefteqn{P(|\hat{\sigma}_{uk}^2 - \sigma_{uk}^2| > \epsilon \,|\, k \in K_{u,T}) =} \nn
&\hspace{2cm}& P(| \hat{\sigma}_{uk}^2 - \sigma_{uk}^2| > \epsilon\, |\, k \in \widehat{K}_u)P(k \in \widehat{K}_u \, |\, k \in K_{u,T}) \nn
&\hspace{2cm}&+\ P(|\hat{\sigma}_{uk}^2 - \sigma_{uk}^2 | > \epsilon\, |\, k \notin \widehat{K}_u)P(k \notin \widehat{K}_u\, |\, k \in K_{u,T}) \nn
&\hspace{2cm} \leq & P( | \{ \hat{\sigma}_k^2 - \sigma_e^2/T \}_+ - \sigma_{uk}^2 | > \epsilon) + P(k \notin \widehat{K}_u\, |\, k \in K_{u,T}) \nonumber
\end{eqnarray}
For any $\gamma > 0$, there exist $S_0^{(1)}, T_0^{(1)}$ such that for $S > S^{(1)}_0$, $T > T^{(1)}_0$ we have $P( | \{ \hat{\sigma}_k^2 - \sigma_e^2/T \}_+ - \sigma_{uk}^2 | > \epsilon) < \gamma/2$ for all $k \in K_{u,T}$ due to eq.(\ref{eq:37}), and $P(k \notin \widehat{K}_u\, |\, k \in K_{u,T}) < \gamma/2$ for all $k \in K_{u,T}$, since $P(K_{u,T} \subseteq \widehat{K}_u(\bs{Y})) \to 1$. Thus for any $\epsilon, \gamma > 0$, 
\begin{equation}
P(|\hat{\sigma}_{uk}^2(\bs{Y}_{\cdot k}) - \sigma_{uk}^2| > \epsilon) < \gamma \nonumber
\end{equation}
for $S > S_0^{(1)}, T > T_0^{(1)}$ and all $k \in K_{u,T}$.\\[3mm]
Second, consider the case $k \notin K_{u,T}$, again by the law of total probability for arbitrary $\epsilon > 0$, 
\begin{eqnarray}
\lefteqn{P(|\hat{\sigma}_{uk}^2 - \sigma_{uk}^2| > \epsilon \, |\, k \notin K_{u,T}) =} \nn
&\hspace{2cm}& P(| \hat{\sigma}_{uk}^2 - \sigma_{uk}^2| > \epsilon\, |\, k \in \widehat{K}_u)P(k \in \widehat{K}_u\, |\, k \notin K_{u,T}) \nn
&\hspace{2cm}&+\ P(|\hat{\sigma}_{uk}^2 - \sigma_{uk}^2 | > \epsilon\, |\, k \notin \widehat{K}_u)P(k \notin \widehat{K}_u\, |\, k \notin K_{u,T}) \nn
&\hspace{2cm} \leq & P( | \{ \hat{\sigma}_k^2 - \sigma_e^2/T \}_+ - \sigma_{uk}^2 | > \epsilon) + P(\sigma_{uk}^2 > \epsilon\, |\, k \notin K_{u,T}) \nonumber
\end{eqnarray}
We note that $P(\sigma_{uk}^2 > \epsilon\, |\, k \notin K_{u,T}) = 0$ for all $\epsilon > 0$ since $\sigma_{uk}^2 = 0$ for $k \notin K_{u,T}$, and again by eq.(\ref{eq:35a}) for any $\gamma > 0$ there exist $S_0^{(2)}, T_0^{(2)}$ such that $P( | \{ \hat{\sigma}_k^2 - \sigma_e^2/T \}_+ - \sigma_{uk}^2 | > \epsilon) < \gamma$ for all $k \notin K_{u,T}$. Thus for any $\epsilon,\gamma > 0$, 
\begin{equation}
P(|\hat{\sigma}_{uk}^2(\bs{Y}_{\cdot k}) - \sigma_{uk}^2| > \epsilon) < \gamma \nonumber
\end{equation}
for $S > S_0^{(2)}$, $T > T_0^{(2)}$ and all $k \notin K_{u,T}$. Let $T_0 = T_0^{(1)} \vee T_0^{(2)}$ and $S_0 = S_0^{(1)} \vee S_0^{(2)}$, then the uniform convergence in probability with $k \in \{1,\ldots, T\}$ in accordance with Theorem 2 follows from the above arguments.
\end{proof}

\subsection{Consistency of the estimators of $\rho_{ij}$}\label{II-sec:3.3}
\begin{proof}
We show consistency of the estimators $\hat{\rho}_{ij}(\bs{Y})$ marginally for each $i,j$ with $i \neq j$. Writing $\widehat{K}_u = \widehat{K}_u(\bs{Y})$ and $\hat{k}_u = |\widehat{K}_u(\bs{Y})|$, we decompose:
\begin{eqnarray}
\lefteqn{\hat{\rho}_{ij}(\bs{Y}) = \frac{1}{\hat{k}_u} \sum_{k \in \widehat{K}_u} \frac{(Y_{ik} - \hat{h}_k)(Y_{jk}-\hat{h}_k)}{\hat{\sigma}_{uk}^2(\bs{Y}_{\cdot k}) \vee \delta}} \nn
&=& \frac{k_{u,T}}{\hat{k}_u} \Bigg( \underbrace{\frac{1}{k_{u,T}} \sum_{k \in K_{u,T}} \frac{(Y_{ik}-\hat{h}_k)(Y_{jk}-\hat{h}_k)}{\hat{\sigma}_{uk}^2(\bs{Y}_{\cdot k}) \vee \delta}}_{\te{(i)}} + \underbrace{\frac{1}{k_{u,T}}\sum_{k \in \widehat{K}_u \setminus K_{u,T}} \frac{(Y_{ik}-\hat{h}_k)(Y_{jk}-\hat{h}_k)}{\hat{\sigma}_{uk}^2(\bs{Y}_{\cdot k}) \vee \delta}}_{\te{(ii)}} \nn
&& -\ \underbrace{\frac{1}{k_{u,T}} \sum_{k \in K_{u,T} \setminus \widehat{K}_u} \frac{(Y_{ik}-\hat{h}_k)(Y_{jk}-\hat{h}_k)}{\hat{\sigma}_{uk}^2(\bs{Y}_{\cdot k}) \vee \delta} \Bigg)}_{\te{(iii)}} \label{eq:38}
\end{eqnarray}
In part (i) below it is shown that term (i) converges to $\rho_{ij}$ in probability as $S,T \to \infty$. In part (ii) it is shown that the terms (ii) and (iii) converge to zero in probability as $S,T \to \infty$. Then, combining these results, and observing that $k_{u,T} / \hat{k}_u \overset{P}{\to} 1$ as $S,T \to \infty$ since $P(\widehat{K}_u = K_{u,T}) \to 1$, an application of Slutsky's lemma yields the claimed result.\\[3mm]
\textbf{\emph{Part (i)}} \quad From the proof in Section \ref{II-sec:3.2}, we know that $\sup_{k \in K_{u,T}} |\hat{\sigma}_{uk}^2(\bs{Y}_{\cdot k}) - \sigma_{uk}^2| = o_p^{S,T}(1)$. Therefore, term (i) in eq.(\ref{eq:38}) can be rewritten as,
\begin{eqnarray}
\lefteqn{\frac{1}{k_{u,T}} \sum_{k \in K_{u,T}} \frac{(Y_{ik} - \hat{h}_k)(Y_{jk}-\hat{h}_k)}{\hat{\sigma}_{uk}^2(\bs{Y}_{\cdot k}) \vee \delta} =} \nn 
&&\frac{1}{k_{u,T}} \sum_{k \in K_{u,T}} \frac{(Y_{ik}-\hat{h}_k)(Y_{jk}-\hat{h}_k)}{\sigma_{uk}^2} \cdot \frac{\sigma_{uk}^2}{(\sigma_{uk}^2 + o_p^{S,T}(1)) \vee \delta} \nn
&=& (1 + o_p^{S,T}(1)) \cdot \left(\frac{1}{k_{u,T}} \sum_{k \in K_{u,T}} \frac{(Y_{ik}-\hat{h}_k)(Y_{jk}-\hat{h}_k)}{\sigma_{uk}^2}\right) \label{eq:39}
\end{eqnarray}
where we also use that $\inf_{k \in K_{u,T}} \sigma_{uk}^2 \geq \delta$. We show that $\frac{1}{k_{u,T}}\sum_{k \in K_{u,T}} \frac{(Y_{ik}-\hat{h}_k)(Y_{jk}-\hat{h}_k)}{\sigma_{uk}^2} \overset{P}{\to} \rho_{ij}$ as $S,T \to \infty$. Writing out further,
\begin{eqnarray}
\lefteqn{\frac{1}{k_{u,T}}\sum_{k\in K_{u,T}} \frac{(Y_{ik}-\hat{h}_k)(Y_{jk}-\hat{h}_k)}{\sigma_{uk}^2} =} \nn
&& \underbrace{\frac{1}{k_{u,T}}\sum_{k \in K_{u,T}} \frac{(Y_{ik}-h_k)(Y_{jk}-h_k)}{\sigma_{uk}^2}}_{\te{(i)}} + \underbrace{\frac{1}{k_{u,T}}\sum_{k\in K_{u,T}} \frac{(Y_{jk}-h_k)(\hat{h}_k-h_k)}{\sigma_{uk}^2}}_{\te{(ii)}} \nn
&& +\ \underbrace{\frac{1}{k_{u,T}}\sum_{k \in K_{u,T}} \frac{(Y_{ik}-h_k)(\hat{h}_k-h_k)}{\sigma_{uk}^2}}_{\te{(iii)}} + \underbrace{\frac{1}{k_{u,T}}\sum_{k \in K_{u,T}}\frac{(\hat{h}_k - h_k)^2}{\sigma_{uk}^2}}_{\te{(iv)}} \label{eq:40}
\end{eqnarray}
Term (ii), (iii), and (iv) converge to zero in probability as $S,T \to \infty$. This is observed from combining $\sup_k (\hat{h}_k - h_k) = o_{p}^{S,T}(1)$, $\inf_{k \in K_{u,T}} \sigma_{uk}^2 > 0$, and respectively $\frac{1}{k_{u,T}}\sum_{k \in K_{u,T}} (Y_{jk} - h_k) = O_p(1)$ for term (ii) and $\frac{1}{k_{u,T}}\sum_{k \in K_{u,T}} (Y_{ik} - h_k) = O_p(1)$ for term (iii). \\[3mm]
It remains to show that term (i) in eq.(\ref{eq:40}) converges to $\rho_{ij}$ in probability. By linearity of the expectation,
\begin{eqnarray}
\mathbb{E}\left[ \frac{1}{k_{u,T}}\sum_{k \in K_{u,T}} \frac{(Y_{ik}-h_k)(Y_{jk}-h_k)}{\sigma_{uk}^2} \right] = \frac{1}{k_{u,T}}\sum_{k \in K_{u,T}} \frac{\sigma_{uk}^2\rho_{ij}}{\sigma_{uk}^2} = \rho_{ij} \nonumber
\end{eqnarray}
Thus, by Chebychev's inequality:
\begin{eqnarray}
\lefteqn{P\left( \left| \frac{1}{k_{u,T}}\sum_{k \in K_{u,T}} \frac{(Y_{ik}-h_k)(Y_{jk}-h_k)}{\sigma_{uk}^2} - \rho_{ij} \right| > \epsilon \right) \leq} \nn
&\hspace{3cm}& \frac{1}{\epsilon^2}\var\left( \frac{1}{k_{u,T}}\sum_{k \in K_{u,T}} \frac{(Y_{ik}-h_k)(Y_{jk}-h_k)}{\sigma_{uk}^2}\right) \quad \quad \quad \label{eq:41}
\end{eqnarray}
Since $E[Y_{ik}^4] = O(1)$ for all $i =1,\ldots,S$, by Cauchy-Schwarz's inequality $\sup_{k \in K_{u,T}} \var((Y_{ik} - h_k)(Y_{jk}-h_k))/\sigma_{uk}^4 = O(1)$ for all $i \neq j$. Recalling that $Y_{ik} = h_k + U_{ik} + \epsilon_{ik}$, it can be verified that,
\begin{eqnarray}
\lefteqn{\cov\left(\frac{(Y_{ik}-h_k)(Y_{jk}-h_k)}{\sigma_{uk}^2}, \frac{(Y_{ik'}-h_k')(Y_{jk'}-h_k')}{\sigma_{uk'}^2} \right) =} \nn
&\hspace{4cm}& \frac{1}{\sigma_{uk}^2\sigma_{uk'}^2}\cov(\epsilon_{ik},\epsilon_{ik'})\cov(\epsilon_{jk},\epsilon_{jk'}) \nonumber
\end{eqnarray}
where we use that, $\cov(U_{ik},U_{jk'})=0$ for all $i,j$ and $k \neq k'$, 
$U_{ik} \perp \epsilon_{jk'}$ for all $i,j$ and $k, k'$, and $\epsilon_{ik} \perp \epsilon_{jk'}$ for all $i \neq j$ and $k, k'$. We argue that $\epsilon_{ik}$ and $\epsilon_{ik'}$ are asymptotically uncorrelated for all $i$ and $k \neq k'$. In the frequency domain, the noise terms at different frequencies $E^f_{i}(\omega_\ell)$ and $E^f_{i}(\omega_{\ell'})$ for $\ell \neq \ell'$ are asymptotically independent (see \cite{B81}), therefore
\begin{eqnarray}
\cov(\bs{E}^f_i) = \sigma_e^2\te{I}_T + \bs{\Delta}_i^f, \quad \quad \quad \te{s.t. } \Vert \bs{\Delta}_i^f \Vert_F = o^T(1) \nonumber 
\end{eqnarray}
where $\bs{E}^f_i = (E^f_{i}(\omega_0),\ldots,E^f_{i}(\omega_{T-1}))'$. Projecting to the coefficient domain with orthogonal discrete wavelet transform-matrix $\bs{W}$ and $\bs{\epsilon}_i = (\epsilon_{i1},\ldots,\epsilon_{iT})'$, this yields
\begin{eqnarray}
\cov(\bs{\epsilon}_i) \ =\ \cov( \bs{W} \bs{E}_i^f) &=& \sigma_e^2\bs{W}\te{I}_T \bs{W}' + \bs{W}\bs{\Delta}_i^f\bs{W}' \nn
&=& \frac{\sigma_e^2}{T} \te{I}_T + \bs{W}\bs{\Delta}_i^f\bs{W}' \nonumber
\end{eqnarray}
and by the rotational invariance of the Frobenius-norm, 
\begin{equation}
\Vert \bs{W} \bs{\Delta}_i^f \bs{W}'\Vert_F = \Vert \bs{\Delta}_i^f \bs{W}'\bs{W} \Vert_F = \frac{1}{T} \Vert \bs{\Delta}_i^f \Vert_F = o^T(T^{-1}) \nonumber
\end{equation}
from which we conclude that $\epsilon_{ik}$ and $\epsilon_{ik'}$ are asymptotically uncorrelated for all $k \neq k$.
Combining the arguments above, it follows that,
\begin{eqnarray}
\var\left( \frac{1}{k_{u,T}}\sum_{k \in K_{u,T}} \frac{(Y_{ik}-h_k)(Y_{jk}-h_k)}{\sigma_{uk}^2}\right) &\lesssim &  \frac{1}{k_{u,T}} + \frac{(k_{u,T}^2 - k_{u,T})}{k_{u,T}^2} \cdot o^T(T^{-1}) \nn
&& \to 0, \quad \te{as } T \to \infty \nonumber
\end{eqnarray}
using that $k_{u,T} \to \infty$ as $T \to \infty$. From eq.(\ref{eq:40}), eq.(\ref{eq:41}) and Slutsky's lemma, we conclude that,
\begin{eqnarray}
\frac{1}{k_{u,T}}\sum_{k \in K_{u,T}} \frac{(Y_{ik}-\hat{h}_k)(Y_{jk}-\hat{h}_k)}{\sigma_{uk}^2} \overset{P}{\to} \rho_{ij} \quad \quad \te{as } S,T \to \infty \nonumber
\end{eqnarray}
Returning to eq.(\ref{eq:39}), 
\begin{eqnarray}
\frac{1}{k_{u,T}} \sum_{k \in K_{u,T}} \frac{(Y_{ik} - \hat{h}_k)(Y_{jk}-\hat{h}_k)}{\hat{\sigma}_{uk}^2(\bs{Y}_{\cdot k}) \vee \delta} &=& (1+o_p^{S,T}(1)) \cdot (\rho_{ij} + o_p^{S,T}(1))\nn
& \overset{P}{\to} & \rho_{ij} \quad \quad \te{as } S,T \to \infty \nonumber
\end{eqnarray}
\textbf{\emph{Part (ii)}} \quad First we show that term (ii) in eq.(\ref{eq:38}) converges to zero in probability. Since $\hat{\sigma}_{uk}^2(\bs{Y}_{\cdot k}) \vee \delta \geq \delta$,
\begin{eqnarray}
\lefteqn{P\left( \Bigg| \frac{1}{k_{u,T}} \sum_{k \in \widehat{K}_u \setminus K_{u,T}} \frac{(Y_{ik}-\hat{h}_k)(Y_{jk}-\hat{h}_k)}{\hat{\sigma}_{uk}^2(\bs{Y}_{\cdot k}) \vee \delta} \Bigg| > \epsilon \right) \leq} \nn
&\hspace{4cm}& P\left( \Bigg| \frac{1}{k_{u,T}} \sum_{k \in \widehat{K}_u \setminus K_{u,T}} \frac{(Y_{ik}-\hat{h}_k)(Y_{jk}-\hat{h}_k)}{\delta} \Bigg| > \epsilon \right) \nonumber
\end{eqnarray}
Therefore it suffices to show that the probability on the right-hand side converges to zero for all $\epsilon > 0$ as $S,T \to \infty$. For ease of notation write $\zeta_k := \frac{(Y_{ik}-\hat{h}_k)(Y_{jk}-\hat{h}_k)}{\delta}$. By the law of total probability, for all $\epsilon > 0$,
\begin{eqnarray}
\lefteqn{P\left( \Bigg| \frac{1}{k_{u,T}} \sum_{k \in \widehat{K}_u \setminus K_{u,T}} \zeta_k \Bigg| > \epsilon \right) =} \nn
&& P \left(  \Bigg| \frac{1}{k_{u,T}} \sum_{k \in \widehat{K}_u \setminus K_{u,T}} \zeta_k \Bigg| > \epsilon\ :\ \#\{\widehat{K}_u \setminus K_{u,T} \} = 0 \right) P\left( \#\{ \widehat{K}_u \setminus K_{u,T} \} = 0 \right) \nn
&& +\ P \left(  \Bigg| \frac{1}{k_{u,T}} \sum_{k \in \widehat{K}_u \setminus K_{u,T}} \zeta_k \Bigg| > \epsilon\ :\ \#\{\widehat{K}_u \setminus K_{u,T} \} > 0 \right) P\left( \#\{ \widehat{K}_u \setminus K_{u,T} \} > 0 \right) \nn
& \leq & 0\cdot P\left( \# \{ \widehat{K}_u \setminus K_{u,T} = 0 \right) + 1 \cdot P\left( \#\{ \widehat{K}_u \setminus K_{u,T} \} > 0 \right) \nonumber
\end{eqnarray}
Since $P(\widehat{K}_u = K_{u,T}) \to 1$ as $S, T \to \infty$, also $P( \#\{ \widehat{K}_u \setminus K_{u,T}\} = 0) \to 1$, thus the right-hand side above converges to zero as $S,T \to \infty$. \\[3mm]
Completely analogous, using that $P(\#\{ K_{u,T} \setminus \widehat{K}_u \} = 0) \to 1$ as $S,T \to \infty$, we find that term (iii) in eq.(\ref{eq:38}) converges to zero in probability as $S,T \to \infty$.
\end{proof}

\section{Proof of Corollary \ref{corr:1}}\label{II-sec:4}
\begin{proof}
The proof consists of two parts. In the first part the convergence in distribution in the first part of the Corollary is shown. In the second part we show that $\widehat{K}_u(\bs{\xi}) = K_{u,T}$ with probability tending to 1 as $S, T \to \infty$.\\[3mm]
\textbf{\emph{Part (i)}} \quad For uncorrelated replicates ($\bs{G}_S = \te{I}_S$), we can write $\bs{\xi}_1, \ldots, \bs{\xi}_T \overset{\te{iid}}{\sim} N(\bs{0}, (\sigma_{uk}^2 + \sigma_e^2/T)\te{I}_S)$, and thus
\begin{equation}
\log\left\{ \frac{1}{S} \sum_{s=1}^S(\xi_{sk} - h_k)^2 \right\} \overset{d}{=} \log( \sigma_{uk}^2 + \sigma_e^2/T) + \log( A_S^2/S ), \quad \quad \te{for } k =1,\ldots, T \nonumber
\end{equation} 
for some $A_S^2 \sim \chi_S^2$. The term $\log( A_S^2/S )$ tends to a normal distribution as $S \to \infty$, and by \cite{P15},
\begin{eqnarray}
\mathbb{E}\left[ \log( A_S^2/S) \right] &=& \log( 2/S )+ \psi^{(0)}(S/2) \nn
\var(\log(A_S^2/S)) &=& \psi^{(1)}(S/2) \nonumber
\end{eqnarray}
where $\psi^{(0)}(\cdot)$ and $\psi^{(1)}(\cdot)$ denote the digamma and trigamma function. Combining the above results with the definition of $T_k(\bs{\xi}_k)$ implies the weak convergence in the first part of the Corollary. For correlated replicates, the noise coefficients $\epsilon_{sk}$ remain independent across replicates, and since $\sigma_{uk}^2 = 0$ for $k \notin K_{u,T}$, it remains true that,
\begin{equation}
\frac{1}{\sqrt{\psi^{(1)}(S/2)}}T_k(\bs{\xi}_k) \overset{d}{\to} N(0,1) \quad \te{if } k \notin K_{u,T} \label{eq:43}
\end{equation}
where the convergence is uniformly over $k \notin K_{u,T}$, since $\sigma_{uk}^2=0$ implies that $T_k(\bs{\xi}_k)$ is independent of $k$.\\[3mm]
\textbf{\emph{Part (ii)}} \quad By the same argument as in Section \ref{II-sec:3.1}, it follows that $K_{u,T} \subseteq \widehat{K}_u(\bs{\xi})$ with probability tending to 1 as $S, T \to \infty$, using that $\lambda_{u,T} = o(\log(T))$ as in Theorem \ref{thm:2}.\\[3mm]
In order to show $\widehat{K}_u(\bs{\xi}) \subseteq K_{u,T}$ with probability tending to 1 as $S,T \to \infty$, we use a standard argument (see e.g. \cite[Chapter 8]{J11}). By the uniform weak convergence in eq.(\ref{eq:43}) for $k \notin K_{u,T}$,
\begin{equation}
P\left(\frac{T_k(\bs{\xi}_k)}{\sqrt{\psi^{(1)}(S/2)}} \leq x \right) = (1+R_S)P(Z \leq x), \quad \quad Z \sim N(0,1) \nonumber
\end{equation}
where $R_S = o_S(1)$ is uniform over $k \notin K_{u,T}$. By independence across indices $k$ and using Gaussian tail probabilities, we upper bound
\begin{eqnarray}
\lefteqn{P\left( \sup_{k \notin K_{u,T}} \frac{|T_k(\bs{\xi}_k)|}{\sqrt{\psi^{(1)}(S/2)}} \geq \sqrt{2\log(T-k_{u,T})} \right) \leq} \nn
&\hspace{2cm}& 1- \left( 1- 2P\left( \frac{T_k(\bs{\xi}_k)}{\sqrt{\psi^{(1)}(S/2)}} \geq \sqrt{2\log(T-k_{u,T})}\right) \right)^{T-k_{u,T}} \nn
&\hspace{2cm} =& 1-\left( 1- 2(1+R_S) \widetilde{\Phi}\left(\sqrt{2\log(T-k_{u,T})}\right) \right)^{T-k_{u,T}} \nn
&\hspace{2cm} \leq & 2(T-k_{u,T})(1+R_S) \frac{\phi(\sqrt{2\log(T-k_{u,T})})}{\sqrt{2\log(T-k_{u,T})}}\nn
&\hspace{2cm} =& \frac{2(1+R_S)}{\sqrt{\pi \log(T-k_{u,T})}} \to 0, \quad \quad \te{as } S,T \to \infty \nonumber
\end{eqnarray}
from which we conclude that also $\widehat{K}_u(\bs{\xi}) \subseteq K_{u,T}$ with probability tending to 1 as $S,T \to \infty$.
\end{proof}

\section{Proof of Theorem \ref{thm:3}}\label{II-sec:5}
\begin{proof}
For ease of notation we write $\bs{\xi}_k := \bs{\xi}_k^{(2)} \sim N(\bs{h}_k, 2\bs{V}_k)$ with $\bs{h}_k = h_k\bs{1}_S$. In the first part of this proof we show that, conditional on $\hat{\bs{h}} = \hat{\bs{h}}(\bs{\xi}^{(1)})$, the estimators $\widehat{R}(\bs{\xi}, \hat{\bs{h}})$ are asymptotically normal, i.e.
\begin{equation}
\frac{\widehat{R}(\bs{\xi}, \hat{\bs{h}}) - \Vert \bs{h} - \hat{\bs{h}} \Vert^2}{\tau(\bs{h}, \hat{\bs{h}})}\ \overset{d}{\to}\ N(0,1), \quad \quad \te{as } S \to \infty \label{eq:44}
\end{equation}
Asymptotic confidence regions can then be constructed (unconditional on $\hat{\bs{h}}$) based on Gaussian quantiles. \\[3mm]
Conditional on $\hat{\bs{h}}$, we decompose
\begin{eqnarray} \label{eq:45}
\lefteqn{\frac{\widehat{R}(\bs{\xi}, \hat{\bs{h}}) - \Vert \bs{h} - \hat{\bs{h}} \Vert^2}{\tau(\bs{h}, \hat{\bs{h}})} =}\nn
&& \frac{1}{\tau(\bs{h}, \hat{\bs{h}})}\left( \sum_{k=1}^T \left[ \sum_{s=1}^S \left[ w_{sk}(\xi_{sk} - \hat{h}_k)^2 \right] - 2\sigma_{k,T}^2 - (\hat{h}_k - h_k)^2 \right] \right) \\
&=& \underbrace{\sum_{k=1}^T \frac{1}{\tau(\bs{h},\hat{\bs{h}})} \left[ \sum_{s=1}^S w_{sk}(\xi_{sk}-h_k)^2 - 2\sigma_{k,T}^2 \right]}_{\te{(i)}} + \underbrace{\sum_{k=1}^T \frac{2(h_k-\hat{h}_k)}{\tau(\bs{h},\hat{\bs{h}})} \sum_{s=1}^S w_{sk}(\xi_{sk}-h_k)}_{\te{(ii)}} \nonumber
\end{eqnarray}
where we use that $\sum_s w_{sk} = 1$ for each $k \in \{1,\ldots,T\}$. \\[3mm]
First, we derive the asymptotic distribution of term (i). Write $\bs{Z}_k = \frac{1}{\sqrt{2}} \bs{V}_k^{-1/2}(\bs{\xi}_k - \bs{h}_k)$, such that $\bs{Z}_k \sim N(\bs{0}, \te{I}_S)$. Here $\bs{V}_k^{-1/2}$ is a symmetric matrix square root of $\bs{V}_k^{-1}$. We can rewrite,
\begin{eqnarray}
\sum_{s=1}^S w_{sk}(\xi_{sk} - h_k)^2 &=& (\bs{\xi}_k - \bs{h}_k)'\te{diag}(\bs{w}_k)(\bs{\xi}_k - \bs{h}_k) \nn
&=& (\bs{\xi}_k - \bs{h}_k)'\bs{V}_k^{-1/2}\bs{V}_k^{1/2} \te{diag}(\bs{w}_k) \bs{V}_k^{1/2}\bs{V}_k^{-1/2}(\bs{\xi}_k - \bs{h}_k) \nn
&=& \bs{Z}_k' \bs{\Gamma}_k \bs{Z}_k \nn
&=& \bs{Z}_k' \bs{P}_k'\bs{\Lambda}_k \bs{P}_k \bs{Z}_k \nonumber
\end{eqnarray}
with $\bs{P}_k'\bs{\Lambda}_k \bs{P}_k$ the eigendecomposition of $\bs{\Gamma}_k = 2\bs{V}_k^{1/2}\te{diag}(\bs{w}_k)\bs{V}_k^{1/2}$, such that $\bs{P}_k\bs{Z}_k \sim N(\bs{0}, \bs{P}_k'\bs{P}_k) \overset{d}{=} N(\bs{0}, \te{I}_S)$. It follows that, 
\begin{equation}
\sum_{s=1}^S w_{sk}(\xi_{sk} - h_k)^2 \ = \ \bs{Z}_k'\bs{P}_k \bs{\Lambda}_k \bs{P}_k \bs{Z}_k \ \overset{d}{=} \ \sum_{s=1}^S \lambda_{sk} A_{sk}^2 \nonumber
\end{equation}
with $\bs{\lambda}_k = (\lambda_{1k},\ldots,\lambda_{sk})'$ the eigenvalues of $\bs{\Gamma}_k$ and $A_{1k}^2,\ldots, A_{Sk}^2 \overset{\te{iid}}{\sim} \chi_1^2$. Furthermore,
\begin{equation}
\sum_{s=1}^S \lambda_{sk}\ =\ \tr(\bs{\Lambda}_k)\ =\ \tr(\bs{\Gamma_k})\ =\ 2\tr(\te{diag}(\bs{w}_k)\bs{V}_k)\ =\ 2\sigma_{k,T}^2 \nonumber
\end{equation}
since $\sum_s w_{sk} =1$, and
\begin{equation}
\Vert \bs{\lambda}_k \Vert\ =\ \sqrt{\tr(\bs{\Lambda}_k^2)}\ =\ \sqrt{\tr\left( \bs{\Gamma}_k' \bs{\Gamma}_k\right)}\ =\ 2\Vert \te{diag}(\bs{w}_k) \bs{V}_k \Vert_F \nonumber
\end{equation} 
Term (i) in eq.(\ref{eq:45}) can now be rewritten as,
\begin{eqnarray}
\lefteqn{\sum_{k=1}^T \frac{1}{\tau(\bs{h},\hat{\bs{h}})}\left[ \sum_{s=1}^S w_{sk}(\xi_{sk}-h_k)^2 - 2\sigma_{k,T}^2 \right] \overset{d}{=}} \nn 
&\hspace{2cm} & \sum_{k=1}^T \frac{\sqrt{8}\Vert \te{diag}(\bs{w}_k) \bs{V}_k \Vert_F \cdot \sum_{s=1}^S \frac{\lambda_{sk}}{\Vert \bs{\lambda}_k \Vert}(A_{sk}^2 - 1)/\sqrt{2}}{\sqrt{\sum_{k=1}^T 8\Vert \te{diag}(\bs{w}_k) \bs{V}_k \Vert_F^2 + 8(h_k-\hat{h}_k)^2\bs{w}_k'\bs{V}_k\bs{w}_k}} \nn
&\hspace{2cm} :=&  C^{(1)} \sum_{k=1}^T  \Big[ B_k^{(1)} \cdot \sum_{s=1}^S \frac{\lambda_{sk}}{\Vert \bs{\lambda}_k \Vert} (A_{sk}^2-1)/\sqrt{2} \Big] \label{eq:46}
\end{eqnarray}
with, 
\begin{eqnarray}
B_k^{(1)} &:=& \frac{\sqrt{8}\Vert \te{diag}(\bs{w}_k) \bs{V}_k \Vert_F}{\sqrt{\sum_{k=1}^T 8\Vert \te{diag}(\bs{w}_k) \bs{V}_k \Vert_F^2}} \nn C^{(1)} &:=& \frac{\sqrt{\sum_{k=1}^T 8 \Vert \te{diag}(\bs{w}_k) \bs{V}_k \Vert_F^2}}{\sqrt{\sum_{k=1}^T 8 \Vert \te{diag}(\bs{w}_k) \bs{V}_k \Vert_F^2 + 8(h_k-\hat{h}_k)^2 \bs{w}_k' \bs{V}_k \bs{w}_k}} \nonumber
\end{eqnarray}
such that $\Vert \bs{B}^{(1)} \Vert = 1$, and $C^{(1)} \in [0,1]$. We show that $\sum_{s=1}^S \frac{\lambda_{sk}}{\Vert \bs{\lambda}_k \Vert}(A_{sk}^2 - 1)/\sqrt{2} \overset{d}{\to} N(0,1)$ independently for all $k=1,\ldots,T$ by the Lindeberg-Feller central limit theorem. First note that,
\begin{eqnarray}
\mathbb{E}\left[ \sum_{s=1}^S \frac{\lambda_{sk}}{\Vert \bs{\lambda}_k \Vert} (A_{sk}^2-1)/\sqrt{2} \right] &=& 0 \nn
\var\left( \sum_{s=1}^S \frac{\lambda_{sk}}{\Vert \bs{\lambda}_k \Vert} (A_{sk}^2-1)/\sqrt{2} \right) &=& \frac{\sum_{s=1}^S \lambda_{sk}^2}{\Vert \bs{\lambda}_k \Vert^2} = 1 \nonumber
\end{eqnarray}
Writing $X_{sk} := (A_{sk}^2-1)/\sqrt{2}$, the Lindeberg conditions are satisfied if:
\begin{equation}
\lim_{S \to \infty} \sum_{s=1}^S \frac{\lambda_{sk}^2}{\Vert \bs{\lambda}_k \Vert^2} \mathbb{E}\left[  X_{sk}^2 \bs{1}\left\{ \dfrac{|\lambda_{sk}|}{\Vert \bs{\lambda}_k \Vert} |X_{sk}| > \epsilon \right\} \right]  = 0 \quad \quad \te{for all } \epsilon > 0 \nonumber
\end{equation}
Since $\sum_{s=1}^S \frac{ \lambda_{sk}^2}{\Vert \bs{\lambda}_k \Vert^2} = 1$, it suffices to show that $\sup_s \mathbb{E}\left[ X_{sk}^2 \bs{1}\left\{ \frac{|\lambda_{sk}|}{\Vert \bs{\lambda}_k \Vert}|X_{sk}| > \epsilon \right\} \right] \to 0$ as $S \to \infty$, which holds if $\sup_{s} |\lambda_{sk}|/\Vert \bs{\lambda}_k \Vert \to 0$ as $S \to \infty$. This is seen by combining Cauchy-Schwarz's inequality and the fact that $\mathbb{E}[X_{sk}^4] < \infty$ (see also \cite[Ex. 2.28]{VDV00}). Note that by the triangle inequality and the Gershgorin circle theorem, 
\begin{equation}
\sup_{s} |\lambda_{sk}| \leq \sup_s \left\{ | \lambda_{sk} - \bs{\Gamma}_{k[s,s]}| + |\bs{\Gamma}_{k[s,s]}| \right\} \leq \sup_s \sum_{i=1}^S | \bs{\Gamma}_{k[s,i]} | = \Vert \bs{\Gamma}_k \Vert_1 \nonumber
\end{equation}
Therefore, since by assumption $\Vert \bs{\Gamma}_k \Vert_1/\Vert \bs{\Gamma}_k \Vert_F \to 0$,
\begin{equation}
\frac{\sup_s |\lambda_{sk}|}{\Vert \bs{\lambda}_k \Vert} \leq \frac{\Vert \bs{\Gamma}_k \Vert_1}{\Vert \bs{\Gamma}_k \Vert_F} \to 0 \quad \quad \te{as } S \to \infty \nonumber
\end{equation}
By an application of the Lindeberg-Feller central limit theorem, we conclude that $\sum_{s=1}^S \frac{\lambda_{sk}}{\Vert \bs{\lambda}_k \Vert}(A_{sk}^2-1)/\sqrt{2} \overset{d}{\to} N(0,1)$ for all $k=1,\ldots,T$.
\begin{remark} Under the stronger assumption $\kappa(\bs{\Gamma}_k) < \infty$, where $\kappa(\cdot)$ denotes the condition number (the maximum eigenvalue divided by the minimum eigenvalue), all the eigenvalues $\lambda_{sk}$ for $s=1,\ldots,S$ are of the same order, and it follows that,
\begin{equation}
\frac{\sup_s |\lambda_{sk}|}{\Vert \bs{\lambda}_k \Vert} \lesssim \frac{1}{\sqrt{S}} \to 0 \quad  \te{as } S \to \infty \nonumber
\end{equation}
which is also sufficient for the Lindeberg conditions to hold. 
\end{remark}
Next, we derive the distribution of term (ii) in eq.(\ref{eq:45}). We note that for each $k$, $\sum_{s=1}^S w_{sk}(\xi_{sk}-h_k)$ is a mean-zero Gaussian random variable, with variance
\begin{equation}
\var\left(\sum_{s=1}^S w_{sk}(\xi_{sk}-h_k)\right) = \bs{w}_k' \cov(\bs{\xi}_k - \bs{h}_k) \bs{w}_k = 2\bs{w}_k' \bs{V}_k \bs{w}_k \nonumber
\end{equation}
Therefore,
\begin{eqnarray}
\lefteqn{\sum_{k=1}^T \frac{2(h_k-\hat{h}_k)}{\tau(\bs{h},\hat{\bs{h}})} \sum_{s=1}^S w_{sk}(\xi_{sk}-h_k) \overset{d}{=}} \nn
&\hspace{2cm}& \sum_{k=1}^T \frac{\sqrt{8}(h_k - \hat{h}_k)\sqrt{\bs{w}_k'\bs{V}_k\bs{w}_k}\cdot Z_k}{\sqrt{\sum_{k=1}^T 8\Vert \te{diag}(\bs{w}_k) \bs{V}_k \Vert_F^2 + 8(h_k-\hat{h}_k)^2\bs{w}_k'\bs{V}_k\bs{w}_k}}  \nn
&\hspace{2cm} :=& C^{(2)} \sum_{k=1}^T B^{(2)}_k  Z_k  \label{eq:47}
\end{eqnarray}
where $Z_1,\ldots,Z_T \overset{\te{iid}}{\sim} N(0,1)$ and,
\begin{eqnarray}
B^{(2)}_k &:=& \frac{\sqrt{8}(h_k - \hat{h}_k)\sqrt{\bs{w}_k'\bs{V}_k\bs{w}_k}}{\sqrt{\sum_{k=1}^T 8(h_k-\hat{h}_k)^2\bs{w}_k'\bs{V}_k\bs{w}_k}} \nn
C^{(2)} &:=& \frac{\sqrt{\sum_{k=1}^T 8(h_k-\hat{h}_k)^2\bs{w}_k'\bs{V}_k\bs{w}_k}}{\sqrt{\sum_{k=1}^T 8\Vert \te{diag}(\bs{w}_k) \bs{V}_k \Vert_F^2 + 8(h_k-\hat{h}_k)^2\bs{w}_k'\bs{V}_k\bs{w}_k}} \nonumber
\end{eqnarray}
such that $\Vert \bs{B}^{(2)} \Vert =1$, and $C^{(2)} \in [0,1]$.\\[3mm]
Combining eq.(\ref{eq:45}), eq.(\ref{eq:46}) and eq.(\ref{eq:47}) we conclude that, conditional on $\hat{\bs{h}}$,
\begin{eqnarray}
\lefteqn{\frac{\widehat{R}(\bs{\xi}, \hat{\bs{h}}) - \Vert \bs{h} - \hat{\bs{h}} \Vert^2}{\tau(\bs{h}, \hat{\bs{h}})} \overset{d}{=}} \nn
&& C^{(1)} \sum_{k=1}^T  \Big[ B_k^{(1)} \cdot \sum_{s=1}^S \frac{\lambda_{sk}}{\Vert \bs{\lambda}_k \Vert} (A_{sk}^2-1)/\sqrt{2} \Big] + C^{(2)} \sum_{k=1}^T B_k^{(2)} Z_k \nn
&\overset{d}{\to} &  C^{(1)} Z_1 + C^{(2)} Z_2 \sim N(0,1) \quad \te{as } S \to \infty \label{eq:48}
\end{eqnarray}
where we use that $\Vert \bs{B}^{(1)} \Vert = \Vert \bs{B}^{(2)} \Vert = 1$ and also $\Vert \bs{C} \Vert = 1$ with $\bs{C} = (C^{(1)}, C^{(2)})'$, combined with the fact that a standard normal random vector is invariant under rotation by a vector of norm 1.\\[3mm]
By the asymptotic normality result in eq.(\ref{eq:48}), for a given confidence level $1-\alpha$
\begin{equation}
\liminf_{S \to \infty} \inf_{\bs{h} \in \ell_2} P\left( \frac{\widehat{R}(\bs{\xi}, \hat{\bs{h}}) - \Vert \bs{h} - \hat{\bs{h}} \Vert^2}{\tau(\bs{h},\hat{\bs{h}})} \geq -z_{\alpha}\, \Big|\, \hat{\bs{h}} \right) \geq 1-\alpha \nonumber
\end{equation}
with $z_\alpha$ a standard normal quantile. By Fatou's lemma, the asymptotic unconditional coverage probability is also at least $1-\alpha$ (see \cite[Section 2]{RV06}). Therefore, 
\begin{equation}
\liminf_{S \to \infty} \inf_{\bs{h} \in \ell_2} P(\bs{h} \in \widehat{C}_{\alpha}(\bs{\xi})) \geq 1-\alpha \nonumber
\end{equation}
where,
\begin{equation}
\widehat{C}_{\alpha}(\bs{\xi}) = \left\{ \bs{h} \in \ell_2 : \Vert \bs{h} - \hat{\bs{h}} \Vert \leq \sqrt{z_\alpha \tau(\bs{h}, \hat{\bs{h}}) + \widehat{R}(\bs{\xi}, \hat{\bs{h}}}) \right\} \nonumber
\end{equation}
which concludes the proof.
\vspace{-2mm}
\end{proof}


\bibliographystyle{imsart-nameyear}
\bibliography{Draft_paper}

\end{document}